\documentclass[aps,prl,twocolumn,amsmath,amssymb,nofootinbib,superscriptaddress]{revtex4-2}
\usepackage{times}
\usepackage[pdftex]{graphicx}
\usepackage{dcolumn}
\usepackage{bm}
\usepackage{amsmath}
\usepackage{amsthm}
\usepackage{braket}
\usepackage{indentfirst}
\usepackage{xr}
\usepackage{setspace} 
\usepackage[colorlinks]{hyperref}
\usepackage[dvipsnames]{xcolor}
\usepackage{xcolor}
\usepackage{mathtools}

\usepackage[linesnumbered,ruled,vlined]{algorithm2e}
\usepackage{tcolorbox}
\tcbuselibrary{breakable}
\usepackage{verbatim}

\definecolor{aqua}{RGB}{69,139,116}

\usepackage{subfigure}
\usepackage{algpseudocode}
\usepackage{amsmath}
\usepackage{verbatim}
\usepackage{makecell}
\usepackage{float}
\usepackage[normalem]{ulem}
\usepackage{ulem}
\newtheorem{theorem}{Theorem}

\begin{document}

\title{Shadow Engineering of Quantum Processes}
\author{Tian-Ci Tian}
\affiliation{Henan Key Laboratory of Quantum Information and Cryptography, Zhengzhou, Henan 450000, China}
\author{De-Tao Jiang}
\affiliation{Henan Key Laboratory of Quantum Information and Cryptography, Zhengzhou, Henan 450000, China}
\author{Wei-Ming Zhu}
\affiliation{Henan Key Laboratory of Quantum Information and Cryptography, Zhengzhou, Henan 450000, China}
\author{Wei-You Liao}
\affiliation{Henan Key Laboratory of Quantum Information and Cryptography, Zhengzhou, Henan 450000, China}
\author{Hong-Wei Li}
\affiliation{Henan Key Laboratory of Quantum Information and Cryptography, Zhengzhou, Henan 450000, China}
\author{He-Liang Huang}
\email{quanhhl@ustc.edu.cn}
\affiliation{Henan Key Laboratory of Quantum Information and Cryptography, Zhengzhou, Henan 450000, China}
\date{\today}

\begin{abstract}
Characterizing quantum processes is essential for hardware benchmarking, error diagnosis, and algorithm verification. While recent work [PRX QUANTUM \textbf{4}, 040337 (2023)] extended classical shadows from quantum state to quantum process, enabling efficient single-channel $\mathcal{E}$  property prediction, its applicability to composite processes $f(\mathcal{E}_1, \mathcal{E}_2,\cdots, \mathcal{E}_k)$  remains unexplored. We introduce shadow engineering, a framework encoding the classical shadows of processes into sparse transfer matrices to predict $f(\mathcal{E}_1, \mathcal{E}_2,\cdots, \mathcal{E}_k)$ properties with proven polynomial sample complexity, matching single-channel efficiency while exponentially lower than quantum process tomography. Crucially, this approach repurposes existing $\mathcal{E}_m$-shadow data without physical execution of $f(\mathcal{E}_1, \mathcal{E}_2,\cdots, \mathcal{E}_k)$, enabling flexible quantum process characterization with minimal hardware overhead. 
We demonstrate the framework's effectiveness and practicality on a superconducting quantum processor for typical applications such as error mitigation and Hamiltonian dynamical simulation.
This framework unlocks new capabilities for predicting complex quantum behaviors without physical re-execution, with immediate applications in near-term device calibration and quantum simulation.
\end{abstract}
\maketitle

{\it{Introduction}}.---Characterizing quantum processes is fundamental to quantum information science, from device benchmarking to revealing the properties of complex dynamics.
Traditional methods like quantum process tomography (QPT) provide complete descriptions of quantum channels but face exponential resource scaling in both measurements and classical post-processing~\cite{mohseni2008quantum,o2004quantum,scott2008optimizing,PhysRevA.103.062615,torlai2023quantum,PhysRevLett.90.193601}. Recent advances in \textit{classical shadows} have revolutionized quantum state tomography by enabling efficient estimation of multiple observables from randomized measurements~\cite{huang2020predicting,PhysRevLett.127.200501,chen2021robust,struchalin2021experimental,huang2022learning,elben2023randomized,acharya2021informationally,akhtar2023scalable,Koh2022classical,PRXQuantum.5.010324,PhysRevLett.130.230403,Zhou2023performanceanalysis,hu2025demonstration}. 
However, extending this to shadow process tomography—which reconstructs the Choi state~\cite{jamiolkowski1972linear,leung2003choi} of a quantum process using classical shadows—still requires exponential samples to predict $\mathcal{E}(\rho)$ for arbitrary input states $\rho$ and quantum processes $\mathcal{E}$ with constant error~\cite{kunjummen2023shadow,levy2024classical,wang2025reducing}.
To overcome this limitation, a polynomial-sample method predicts $\text{Tr}[O\mathcal{E}(\rho)]$ for arbitrary quantum processes $\mathcal{E}$ by constructing classical shadows from random Pauli (or Clifford) measurements of output states, which are generated by applying $\mathcal{E}$ to random product inputs~\cite{huang2023learning,chen2023efficient}.
While effective for single-step channels $\mathcal{E}$, these methods encounter fundamental limitations for composite operations: predicting outputs under transformed processes, such as adjoints ($\mathcal{E}^\dagger$) or concatenations ($\mathcal{E}_2 \circ \mathcal{E}_1$), requires physically implementing these transformations, incurring prohibitive quantum resource overhead and implementation complexity. This limitation obstructs key capabilities including dynamical system analysis and noise characterization, as existing frameworks lack efficient classical manipulation of classical shadows of processes for operational compositions.

We introduce shadow engineering to bridge this gap. Our framework leverages sparse transfer matrices constructed from the classical shadows of processes, enabling fully classical algebraic manipulation of composite operations. 
Given classical shadows of $\mathcal{E}_1, \mathcal{E}_2,\cdots,\mathcal{E}_k$, this framework accurately and efficiently estimates $\text{Tr}[O\cdot(f(\mathcal{E}_1,\mathcal{E}_2,\cdots,\mathcal{E}_k))(\rho)]$ for any bounded-degree observables $O$ and state $\rho$ sampled from a wide range of distributions over arbitrary $n$-qubit states, and process transformation $f$. 
Theoretically, we prove efficient sample complexity scaling for key cases ($\mathcal{E}_1^\dagger$, $\mathcal{E}_2 \circ \mathcal{E}_1$), matching the efficiency of state-of-the-art single-channel shadow methods while unlocking new functionalities like error mitigation~\cite{huang2023near,PRXQuantum.3.010345,cai2023quantum} and Hamiltonian dynamical simulation~\cite{haah2021quantum,PhysRevLett.129.270502}, which we further demonstrate experimentally on superconducting quantum processor. 
This establishes a scalable paradigm for virtual quantum process control—circumventing physical implementation barriers through classical post-processing of sparse process representations, while extending the applicability of finite-scale classical shadows for quantum system characterization.

\begin{figure*}[t]
  \centering
  {\includegraphics[width=0.92\textwidth]{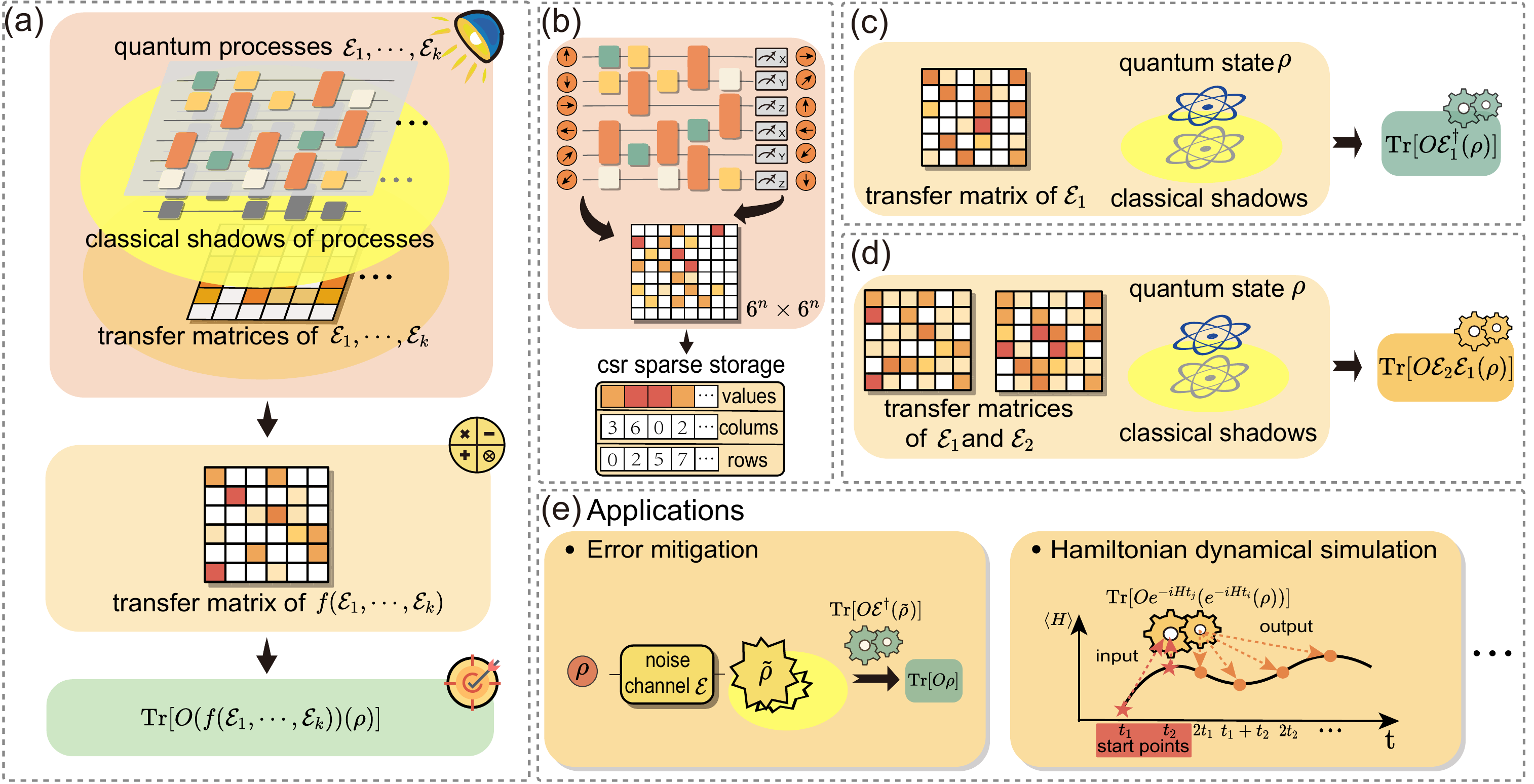}}    
  \caption{{\textbf{Shadow engineering: framework, key cases, and applications.} (a) The framework proceeds in three steps: 
  (1) Obtain classical shadows of single-step channels $\mathcal{E}_1, \dots, \mathcal{E}_k$ and convert them into sparse transfer matrices;
  (2) Combine these matrices to construct the transfer matrix of the composite process $f(\mathcal{E}_1, \dots, \mathcal{E}_k)$;
  (3) Estimate $\text{Tr}[O(f(\mathcal{E}_1,\cdots,\mathcal{E}_k))(\rho)]$ via the obtained composite transfer matrix. 
  (b) Sparse transfer-matrix encoding: First, evolve random product states under under $\mathcal{E}$ and perform randomized Pauli measurements on the outputs~\cite{huang2020predicting,elben2023randomized}.
  Then, map input-outcome index pairs to matrix entries, building a row-normalized $6^n \times 6^n$ sparse matrix (stored in compressed sparse row (CSR) format) from observed frequencies. (c)(d) Shadow engineering for two key cases: using the transfer matrices of $\mathcal{E}_1$ and $\mathcal{E}_2$, shadow engineering can estimate $\text{Tr}[O\mathcal{E}_1^{\dagger}(\rho)]$ (adjoint channel) and $\text{Tr}[O\mathcal{E}_2\mathcal{E}_1(\rho)]$(concatenation), given classical shadows of input states $\rho$. 
  (e) Example applications: (1) Error mitigation: the transfer matrix of noise channel $\mathcal{E}$ is used to compute $\text{Tr}[O\mathcal{E}^{\dagger}(\tilde{\rho})]$, thereby mitigating errors in the noisy state $\tilde{\rho}$;
  (2) Hamiltonian dynamical simulation: from the transfer matrices of a short-time evolution at $t_1$, shadow engineering can predict long-time observables $\text{Tr}[O e^{-iHt_i} (\rho)]$ for multiples $t_i$ of $t_1$, enabling efficient simulation of dynamics.
  }}
  \label{fig1}
\end{figure*}

{\it{Background and notation}}.---We obtain classical shadows of an unknown quantum process $\mathcal{E}$ following the procedure from Ref.~\cite{huang2023learning}. First, uniformly sample an input state ${\mathcal{S}^{\otimes n}_{i}}$ from the $n$-qubit stabilizer basis $\mathcal{S}^{\otimes n}$, where $\mathcal{S} \triangleq \{ \ket{0}\!\bra{0}, \ket{1}\!\bra{1}, \ket{+}\!\bra{+}, \ket{-}\!\bra{-}, \ket{y+}\!\bra{y+}, \ket{y-}\!\bra{y-} \}$. After applying $\mathcal{E}$, a random Pauli measurement on $\mathcal{E}(\mathcal{S}^{\otimes n}_{i})$ yields an outcome $\mathcal{S}^{\otimes n}_{j}$. The following unbiased estimator holds:
\begin{equation}\label{eq0}
\mathbb{E}\big[\mathcal{M}^{-1}(\mathcal{S}^{\otimes n}_{j})\big] = \mathcal{E}(\mathcal{S}^{\otimes n}_{i}),
\end{equation}
where $\mathcal{M}^{-1}\bigl( \bigotimes_{\iota=1}^{n} \rho_\iota \bigr) = \bigotimes_{\iota=1}^{n} (3\rho_\iota - I)$~\cite{huang2020predicting}.
Repeating this procedure $N$ times produces the classical shadow $S_N(\mathcal{E}) \triangleq \left\{ \mathcal{S}^{\otimes n}_{i_{\ell}}, \mathcal{S}^{\otimes n}_{j_{\ell}} \right\}_{\ell=1}^{N}$, with $\ell$ indexing the snapshots.

To efficiently store and post-process the classical shadows of a quantum process, we introduce its \textit{transfer matrix} $T_{\mathcal{E}} \in \mathbb{R}^{6^n \times 6^n}$, defined by
\begin{equation}\label{eq3}
(T_{\mathcal{E}})_{ij} = \left( \frac{1}{3} \right)^n \operatorname{Tr}\bigl[ \mathcal{E}(\mathcal{S}^{\otimes n}_i) \cdot \mathcal{S}^{\otimes n}_j \bigr].
\end{equation}
The entry is the probability of obtaining outcome $\mathcal{S}^{\otimes n}_j$ when measuring $\mathcal{E}(\mathcal{S}^{\otimes n}_i)$ in a random Pauli basis. From classial shadows $S_N(\mathcal{E})$, an unbiased empirical estimator $\hat{T}_{\mathcal{E}}$ is obtained by counting the frequency of each input-outcome index pairs $(i,j)$ and normalizing row-wise, so that $\mathbb{E}[\hat{T}_{\mathcal{E}}] = T_{\mathcal{E}}$.

Therefore, Eq.~\eqref{eq0} can be generalized to the transfer-matrix formulation, which allows the reconstruction of the output states from classical shadows:
\begin{equation}\label{eq1}
\mathcal{E} \circledast (\boldsymbol{\sigma}_{\mathcal{S}^{\otimes n}}) = T_{\mathcal{E}} \; \mathcal{M}^{-1} \circledast (\boldsymbol{\sigma}_{\mathcal{S}^{\otimes n}}),
\end{equation}
where $\boldsymbol{\sigma}_{\mathcal{S}^{\otimes n}}= {[{\mathcal{S}_{1}^{\otimes n}},{\mathcal{S}_{2}^{\otimes n}},...,{\mathcal{S}_{6^n}^{\otimes n}}]^T}$ is the column vector of all basis density operators in ${{\mathcal{S}^{\otimes n}}}$, 
and $\circledast$ denotes applying the superoperator separately to each entry of the vector.

Using the transfer matrix \( T_{\mathcal{E}} \), one can directly predict expectation values \( \mathrm{Tr}[O\mathcal{E}(\rho)] \) for a given observable $O$ and an input state $\rho$. Noting that \( \mathrm{Tr}[O\mathcal{E}(\rho)] = \mathrm{Tr}[\mathcal{E}^{\dagger}(O)\rho] \), the task reduces to finding the Pauli coefficients \( \alpha_P \) of \( \mathcal{E}^\dagger(O)\). 
Using Eq.~\eqref{eq1}, the method for obtaining \( \alpha_P \) descirbed in Ref.~\cite{huang2023learning} can be reformulated in terms of the transfer matrix as:
\begin{align}
 \alpha_P &= 3^{|P|}\,\mathbb{E}_{\rho \in \mathcal{S}^{\otimes n}}\!\bigl[\mathrm{Tr}(O\mathcal{E}(\rho))\,\mathrm{Tr}(P\rho)\bigr] \label{eq:alpha_P_1} \\
 &= \frac{3^{|P|}}{6^n}\sum_{i,j=1}^{6^n} (T_{\mathcal{E}})_{ij}\;
 \mathrm{Tr}[O\mathcal{M}^{-1}(\mathcal{S}^{\otimes n}_j)]\,\mathrm{Tr}[P\mathcal{S}^{\otimes n}_i] \label{eq:alpha_P_2}
\end{align}
for \( P \in \{I,X,Y,Z\}^{\otimes n} \).

{\it{Shadow engineering framework}}.---Given an observable $O$, an input state $\rho$, and classical shadows of $k$ unknown quantum processes $\{\mathcal{E}_m\}_{m=1}^k$, the target of shadow engineering is to predict the expected value $\text{Tr}[O\cdot(f(\mathcal{E}_1,\mathcal{E}_2,\cdots,\mathcal{E}_k))(\rho)]$. As shown in Fig.~\ref{fig1}(a), the framework consists of three stages: 
(1) encoding the transfer matrix of single-step quantum processes, 
(2) constructing the transfer matrix of the transformed process, 
and (3) predicting the target expectation value.

Step (1) converts the classical shadows of each process $\mathcal{E}_m$ into the transfer matrix $\hat{T}_{\mathcal{E}_m}$ (see Fig.~\ref{fig1}(b)). 
Since the shadow data scale polynomially with the qubit number $n$~\cite{huang2023learning}, each $\hat{T}_{\mathcal{E}_m}$ is highly sparse. 
We therefore store these matrices in Compressed Sparse Row (CSR) format~\cite{golub2013matrix} and employ SpGEMM algorithms~\cite{gao2023systematic,islam2025improving} to accelerate subsequent computations.

In step (2), the transfer matrix $\hat{T}_f$ of the transformed process $f(\mathcal{E}_1,\mathcal{E}_2,\cdots,\mathcal{E}_k)$ is constructed from the known matrices $\{\hat{T}_{\mathcal{E}_m}\}_{m=1}^k$. This can be done in two equivalent ways.

The first approach builds $\hat{T}_f$ element-wise from its definition
\((T_f)_{ij} = (1/3)^{n} \text{Tr}[ f(\mathcal{E}_1,\mathcal{E}_2,\cdots,\mathcal{E}_k)(\mathcal{S}^{\otimes n}_i) \, \mathcal{S}^{\otimes n}_j ]\).
The key step is to express \( f(\mathcal{E}_1,\mathcal{E}_2,\cdots,\mathcal{E}_k)(\mathcal{S}^{\otimes n}_i) \) (or its dual) in terms of the outputs \( \mathcal{E}_m(\mathcal{S}^{\otimes n}_l) \). 
For example, for the adjoint channel \( f(\mathcal{E}_1) = \mathcal{E}_1^\dagger \), this directly gives
\begin{equation}
T_{f(\mathcal{E}_1)} = {T_{\mathcal{E}_1}}^{\dagger}.
\end{equation}
Similarly, we can also use the equivalence relationship from Eq.~\eqref{eq1} for $f(\mathcal{E}_1,\mathcal{E}_2,\cdots,\mathcal{E}_k)$, as \begin{equation} T_f \, \mathcal{M}^{-1} \circledast (\boldsymbol{\sigma}_{\mathcal{S}^{\otimes n}}) = f(\mathcal{E}_1,\mathcal{E}_2,\cdots,\mathcal{E}_k) \circledast (\boldsymbol{\sigma}_{\mathcal{S}^{\otimes n}}). \end{equation}
Rewriting the right-hand side as a function of \( \mathcal{E}_m \circledast (\boldsymbol{\sigma}_{\mathcal{S}^{\otimes n}}) \) and substituting each term via Eq.~\eqref{eq1} yields the outcome \( \hat{T}_f \). For instance, concatenation \( f(\mathcal{E}_1,\mathcal{E}_2) = \mathcal{E}_2 \circ \mathcal{E}_1 \) gives
\begin{equation} 
  {T}_{f(\mathcal{E}_1, \mathcal{E}_2)} ={T}_{\mathcal{E}_1} \cdot \mathcal{T} \cdot {T}_{\mathcal{E}_2}. 
\end{equation}
where the integer matrix $\mathcal{T}$ satisfies $\mathcal{M}^{-1}\circledast(\boldsymbol{\sigma}_{\mathcal{S}^{\otimes n}}) = \mathcal{T} \cdot \boldsymbol{\sigma}_{\mathcal{S}^{\otimes n}}$, 
whose extreme sparsity allows for efficient storage and computation in CSR format.

Step (3) is to estimate $\mathrm{Tr}[O f(\mathcal{E}_1,\dots,\mathcal{E}_k)(\rho)]$ using transfer matrix $\hat{T}_f$.
This requires computing the Pauli coefficients $\hat{\alpha}_P(O)$ of $f^{\dagger}(\mathcal{E}_1,\dots,\mathcal{E}_k)(O)$. 
To avoid exponential cost, generally the Pauli truncation of order 
$\kappa = \Theta(\log(1/\epsilon))$ on $f^{\dagger}(\mathcal{E}_1,\dots,\mathcal{E}_k)(O)$ is performed, incurring an error at most $\epsilon$.
Specifically, for each $P$ with $|P|\leq\kappa$, the coefficients $\alpha_P$ is estimated via a sparse version of Eq.~\eqref{eq:alpha_P_2}, as
\begin{equation}
\begin{aligned}
\hat{\alpha}_P(O) 
&= \frac{3^{|P|}}{|\text{supp}(\hat{T}_f)|}
   \sum_{i\in\text{supp}(\hat{T}_f)}
   \mathrm{Tr}\bigl(P\,\mathcal{S}^{\otimes n}_i\bigr) \\
&\phantom{=}\times
   \mathrm{Tr}\Bigl(
   O\sum_{j:(\hat{T}_f)_{ij}\neq 0}
   (\hat{T}_f)_{ij}\,
   \mathcal{M}^{-1}\bigl(\mathcal{S}^{\otimes n}_j\bigr)
   \Bigr)
\end{aligned}
\end{equation}
where \( \operatorname{supp}(\hat{T}_{f}) \) denotes the set of indices of non-zero rows in \( \hat{T}_{f} \). This formulation reduces the computational cost from \( \mathcal{O}(\kappa n^{\kappa} N ) \) to \( \mathcal{O}(\kappa n^{\kappa} |\operatorname{supp}(\hat{T}_{f})|) \) whenever \( |\operatorname{supp}(\hat{T}_{f})| \ll N \), as as it avoids traversing all $N$ non-zero entries of $\hat{T}_f$ for each $P$.

Finally, using the classical shadow $\hat{\rho}$ of the input state $\rho$, the prediction is obtained as
\begin{equation}
h(\rho,O)=\sum_{P:|P|\le\kappa}\hat{\alpha}_P(O)\,\mathrm{Tr}[P\hat{\rho}],
\end{equation}
providing an efficient estimator for $\mathrm{Tr}[O f(\mathcal{E}_1,\dots,\mathcal{E}_k)(\rho)]$.

{\it{Prediction performance}}.---To assess the prediction error of shadow engineering, we analyze its average performance over locally flat distributions \(\mathcal{D}\) of \(n\)-qubit states that are invariant under single-qubit Clifford gates. 
These results are further extended to non-flat distributions in the SM.
Our analysis below focuses on the two representative cases shown in Fig.~\ref{fig1}(c): estimating \(\mathrm{Tr}[O\mathcal{E}_1^{\dagger}(\rho)]\) and \(\mathrm{Tr}[O\mathcal{E}_2\circ\mathcal{E}_1(\rho)]\).

\begin{theorem}[Predicting errors in adjoint channel estimation.]
  Let $n, \epsilon, \epsilon', \delta > 0$. For any unknown $n$-qubit CPTP map $\mathcal{E}_1$, 
  and its classical shadows $S_{N_1}(\mathcal{E}_1)$ obtained by $N_1$ randomized experiments with
\begin{equation}
  \begin{aligned}
    N_1 &= \log(\frac{n}{\delta}) \min(2^{O[\log(1/\epsilon)(\log\log(1/\epsilon) + \log(1/\epsilon^{\prime}))]}, \\ &\times 2^{O\left[\log(1/\epsilon) \log(n)\right]}).
  \end{aligned}
\end{equation}
With probability $\geq 1 - \delta$, 
For any $n$-qubit state $\rho \sim \mathcal{D}$ and any bounded-degree observable $O$ with $\|O\|\leq 1$, 
the prediction error achieved by shadow engineering is
\begin{equation}
    \mathbb{E}_{\rho \sim \mathcal{D}} \left| h(\rho, O) - \text{Tr}[O\mathcal{E}_1^{\dagger}(\rho)] \right|^2\leq  \epsilon + \max(\|O^{(\text{low})}\|^{2 },1)\epsilon' .
\end{equation}
Here, $O^{(\text{low})}$ is the low-degree approximation of $\mathcal{E}_1(O)$.
\end{theorem}

\begin{theorem}[Predicting errors in concatenated channel estimation.]
  Let $n, \epsilon, \epsilon', \delta > 0$. For any unknown $n$-qubit CPTP maps $\mathcal{E}_1$ and $\mathcal{E}_2$, 
  and corresponding classical shadows $S_{N_1}(\mathcal{E}_1)$ and $S_{N_2}(\mathcal{E}_2)$ respectively obtained by $N_1$ and $N_2$ randomized experiments with
\begin{equation}
  \small
  \begin{aligned}
    N_1 =N_2 &= \log(\frac{n}{\delta}) \min(2^{\mathcal{O}[\log(n)+\log(1/\epsilon)(\log\log(1/\epsilon) + \log(1/\epsilon^{\prime}))]}, \\ &\times 2^{O\left[\log(1/\epsilon) \log(n)\right]}).
  \end{aligned}
\end{equation}

With probability $\geq 1 - \delta$, 
For any $n$-qubit state $\rho \sim \mathcal{D}$ and any bounded-degree observable $O$ with $\|O\|\leq 1$, 
the prediction error achieved by shadow engineering is
\begin{equation}
    \mathbb{E}_{\rho \sim \mathcal{D}} \left| h(\rho, O) - \text{Tr}[O\mathcal{E}_2\mathcal{E}_1(\rho)] \right|^2  \leq  \epsilon + \max(\|O^{(\text{low})}\|^{2 },1)\epsilon' .
\end{equation}
Here, $O^{(\text{low})}$ is the low-degree approximation of $\mathcal{E}_1^{\dagger}\mathcal{E}_2^{\dagger}(O)$.
\end{theorem}

The total error comprises Pauli truncation error $\epsilon$ (from approximating $f^{\dagger}(\mathcal{E}_1,\mathcal{E}_2,\cdots,\mathcal{E}_k)(O)$ with $\kappa$ Pauli truncation) and estimation error $\epsilon'$ (from finite-sample estimation of Pauli coefficients via $\hat{T}_{f}$). Detailed proofs are provided in the SM.

\begin{figure}[h]
  \centering
  {\includegraphics[width=0.44\textwidth]{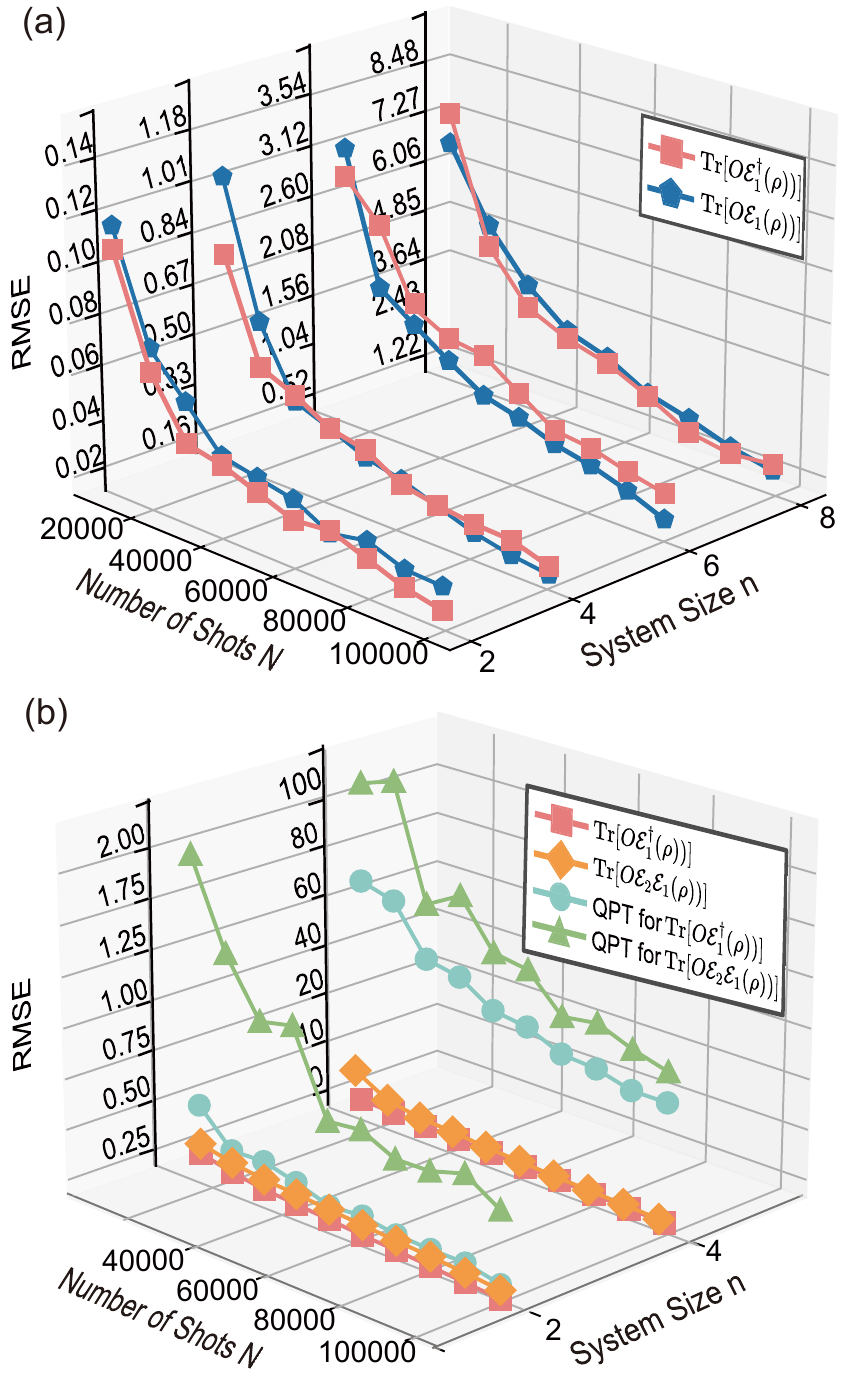}} 
  \caption{{\textbf{Prediction performance of shadow engineering for estimating $\text{Tr}[O\mathcal{E}_1^{\dagger}(\rho)]$ and $\text{Tr}[O\mathcal{E}_2\mathcal{E}_1(\rho)]$.} 
  (a) Using $10^4$-$10^5$ classical shadows for 2-8 qubits, the RMSE for predicting $\text{Tr}[O\mathcal{E}_1^{\dagger}(\rho)]$ with shadow engineering is comparable to that of predicting $\text{Tr}[O\mathcal{E}_1(\rho)]$ using the method in Ref.\cite{huang2023learning}.
  (b) For 2-4 qubits with $10^4$-$10^5$ classical shadows, shadow engineering significantly outperforms QPT in predicting $\text{Tr}[O\mathcal{E}_1^{\dagger}(\rho)]$ and $\text{Tr}[O\mathcal{E}_2\mathcal{E}_1(\rho)]$. 
  }}
  \label{fig2}
\end{figure}

The above theorems indicate that the sample complexity of shadow engineering scales polynomially with the number of qubits, confirming the practical efficiency and scalability of our algorithm for large-scale quantum systems. 

{\it{Numerical simulation and experimental verification}}.---We validate the performance and versatility of shadow engineering through numerical simulations and experiments on a quantum processor

We first benchmark our framework against the method of Ref.\cite{huang2023learning} and standard QPT~\cite{mohseni2008quantum} through numerical simulations, confirming the predicted scaling behavior. Figure~\ref{fig2}(a) presents the root mean square error (RMSE) for predicting $\text{Tr}[O\mathcal{E}_1^{\dagger}(\rho)]$ with shadow engineering (red) versus predicting $\text{Tr}[O\mathcal{E}_1(\rho)]$ with the method of Ref.\cite{huang2023learning} (blue), across 2-8 qubits using $10^4$-$10^5$ measurement shots. The RMSE, computed over 50 experimental repetitions as  $ \sqrt{\sum_{i = 1}^{50}(\langle O\rangle^{\text{est}}_i-\langle O\rangle^{\text{ideal}}_i)^{2}/{50}}$ (``est" indicates estimated values and ``ideal" represents exact theoretical values), shows that both methods achieve comparable accuracy, confirming that shadow engineering extends the direct prediction capability to composite processes.
\begin{figure}[t]
 
  \centering
  {\includegraphics[width=0.44\textwidth]{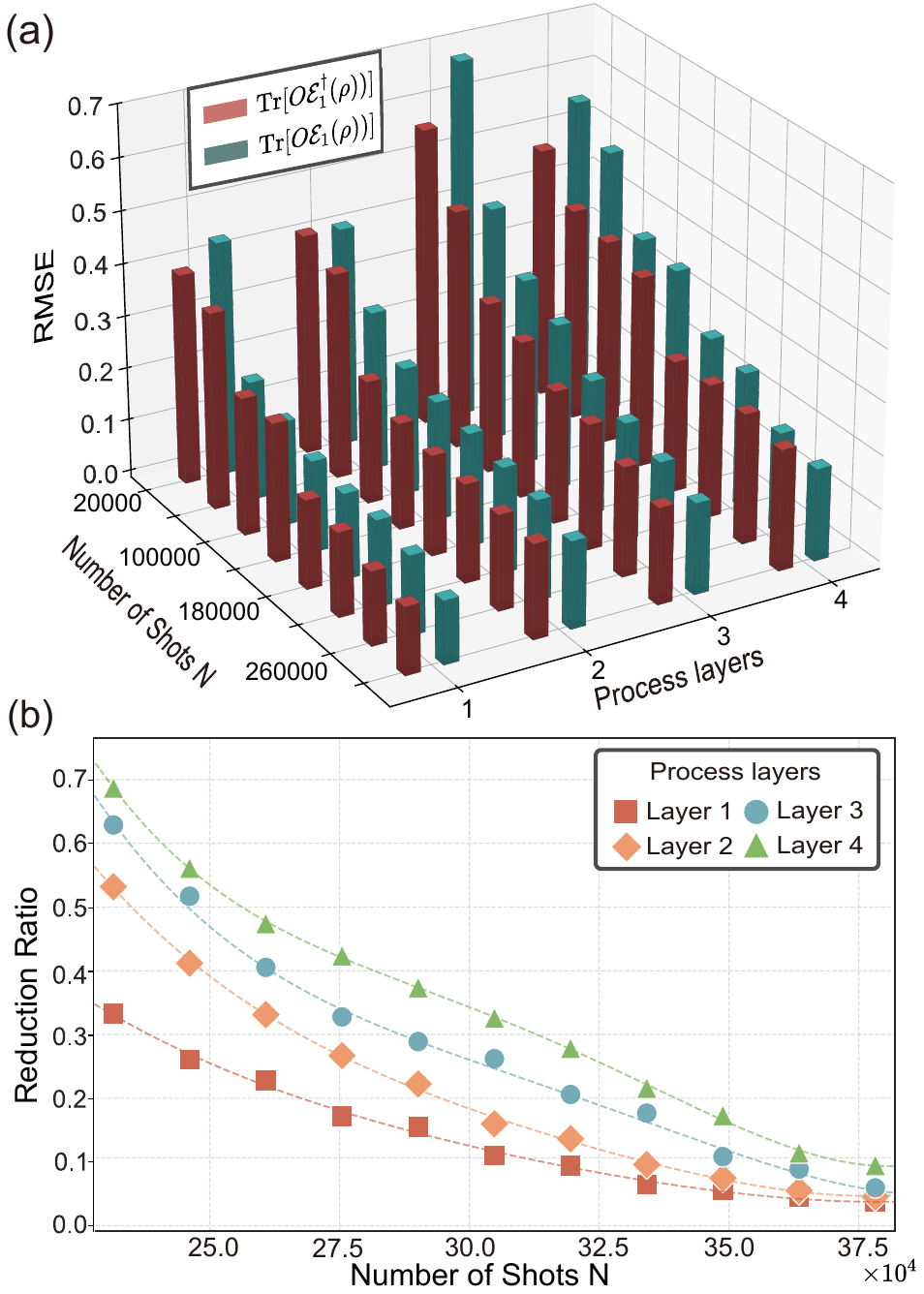}} 
  \caption{{\textbf{Adjoint channel estimation and its application to error mitigation using shadow engineering on a quantum processor.} 
  (a) The 6-qubit noisy process ${E}_M$ is implemented with circuit layers ranging from 1 to 4 (see SM for circuit details); deeper layers introduce stronger noise. Shadow engineering (predicting $\text{Tr}[O\mathcal{E}_M^{\dagger}(\rho)]$) achieves root-mean-square error (RMSE) comparable to the method of Ref.~\cite{huang2023learning} (predicting $\text{Tr}[O\mathcal{E}_M(\rho)]$), with both methods showing a slight increase in error as the layer depth grows. (b) Error-mitigation performance, quantified by the reduction ratio $r = {|h(\rho_{\text{noisy}},O) - \text{Tr}[O\rho]|}/{|\text{Tr}[O\rho_{\text{noisy}}] - \text{Tr}[O\rho]|}$. For each layer (\textit{i.e.}, each ${\mathcal{E}}_M$), $r$ decreases as more classical shadows are used, confirming that the mitigation improves with sample size across all noise levels.}} 
  \label{fig3}
\end{figure}

Furthermore, Fig.~\ref{fig2}(b) compares shadow engineering with QPT for predicting $\text{Tr}[O\mathcal{E}_1^{\dagger}(\rho)]$ and $\text{Tr}[O\mathcal{E}_2\mathcal{E}_1(\rho)]$. 
Shadow engineering uses the classical shadows of $\mathcal{E}_1$ and $\mathcal{E}_2$, whereas QPT requires full tomography of $\mathcal{E}_1^{\dagger}$ (for the adjoint) or separate tomography of $\mathcal{E}_1$ and $\mathcal{E}_2$ followed by composition (for concatenation). Results for 2-4 qubits reveal three key advantages: 1) QPT exhibits substantially larger errors at all system sizes; 2) QPT's composite predictions  $\text{Tr}[O\mathcal{E}_2\mathcal{E}_1(\rho)]$ show severe error propagation from individual channel tomography; 3) Shadow engineering's performance advantage widens with increasing qubit counts. This reflects the fundamental difference in sample complexity: shadow engineering retains polynomial scaling, while QPT requires exponential resources.

Subsequently, we validate the practical performance of shadow engineering on a superconducting quantum processor, which has average readout error 1.255\%, single-qubit gate error 0.122\%, and CZ gate error 1.024\%, and further demonstrate two representative applications: error mitigation and Hamiltonian dynamical simulation.
\begin{figure}[t]

  \centering
  {\includegraphics[width=0.44\textwidth,trim=0 0 0 0, clip]{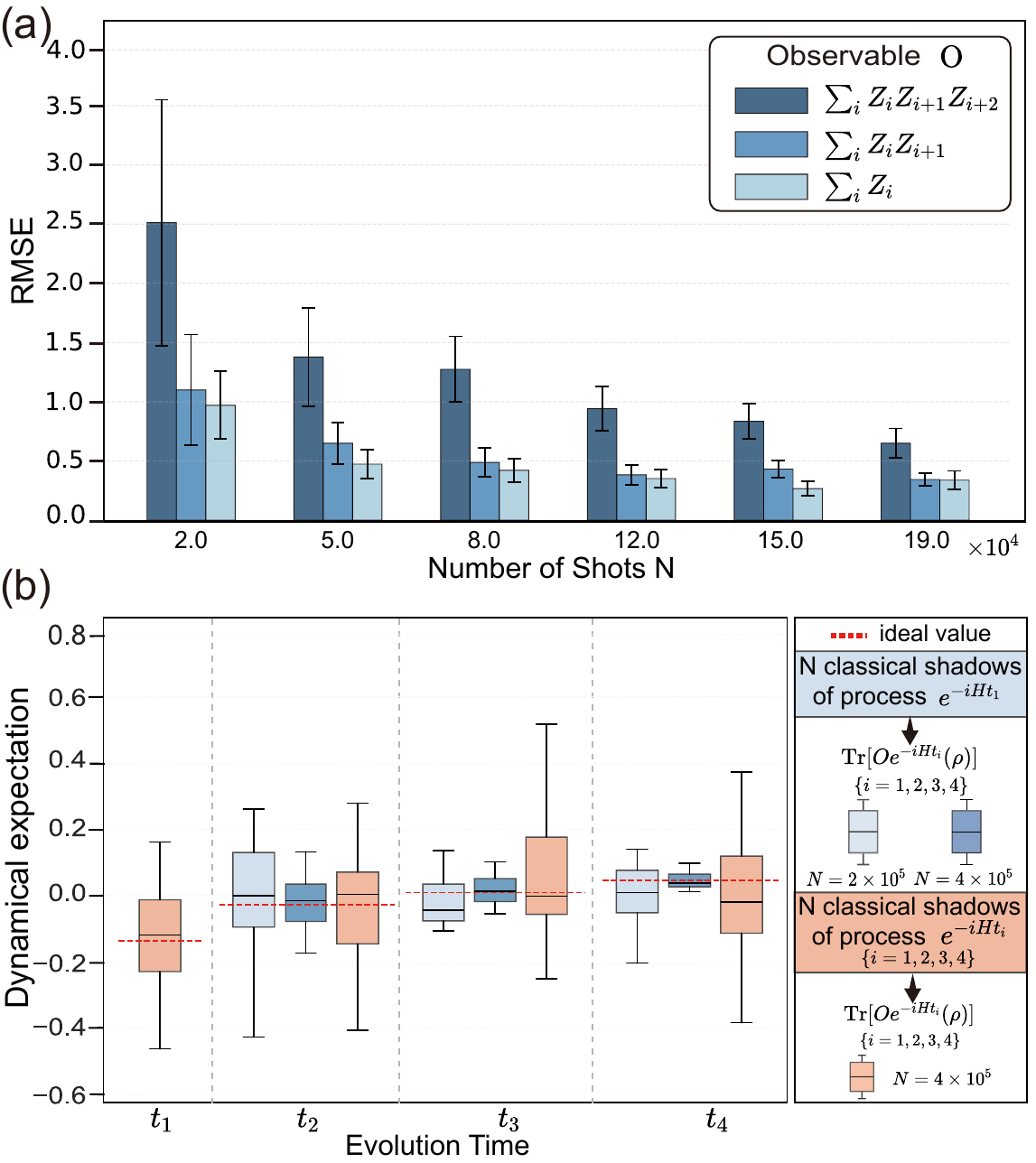}}    
  \caption{{\textbf{Hamiltonian dynamical simulation using shadow engineering on a quantum processor.}
  (a) Predicting $\text{Tr}[O\mathcal{E}_H(t_K)(\rho)] \approx \text{Tr}[Oe^{-iHt_K}(\rho)]$ for  $t_K = K \times t_1$ ($K=1,\cdots,4$) with a 4-qubit Haar random initial state.
  Shadow engineering predictions (blue boxplots), extrapolated from classical shadows acquired at $t_1$, are compared with the method of Ref.~\cite{huang2023learning} (orange boxplots), which uses classical shadows directly acquired at each time point $t_K$.
  Both methods achieve comparable accuracy relative to the ideal evolution (red dashed line), with shadow engineering showing steady improvement with increasing sample size.
  Only one boxplot is shown at $t_1$ since the two methods are equivalent there.
  (b) Prediction error for various properties of the Hamiltonian-evolved state at \(t_2\) using classical shadows from \(t_1\), such as total magnetization and nearest-neighbor correlations. Errors decrease as more shadows are used; for a fixed sample size, lower-locality observables are predicted more accurately, consistent with theoretical scaling. 
  }
   }\label{fig4}
\end{figure}

First, using the classical shadows of a process $\mathcal{E}_M$, we compare the prediction of $\text{Tr}[O\mathcal{E}_M^{\dagger}(\rho)]$ (our framework) with $\text{Tr}[O\mathcal{E}_M(\rho)]$ (method of Ref.~\cite{huang2023learning}). As shown in Fig.~\ref{fig3}(a), the experimental results are in close alignment with numerical simulations: both methods achieve comparable RMSE across different $\mathcal{E}_M$ with circuit layers 1-4, with a slight degradation at deeper layers due to accumulated decoherence. Leveraging this accurate estimation, we made an application example of error mitigation for coherent noise channel: from the noisy state $\rho_{\text{noisy}}=\mathcal{E}_M(\rho)$ affected by noise channel $\mathcal{E}_M$, we compute $h(\rho_{\text{noisy}},O)\approx \text{Tr}[O\mathcal{E}_M^{\dagger}(\rho_{\text{noisy}})]$ to recover \(\text{Tr}[O\rho]\). 
As depicted in Fig.~\ref{fig3}(b), the reduction rate $r = {|h(\rho_{\text{noisy}},O) - \text{Tr}[O\rho]|}/{|\text{Tr}[O\rho_{\text{noisy}}] - \text{Tr}[O\rho]|}$ (averaged over $20$ independent trials) exhibits a steady improvement with the number of classical shadows, converging to small values for all tested $\mathcal{E}_M$ with different circuit layers (which correspond to different noise strengths).

We then assess the performance of shadow engineering for Hamiltonian dynamical simulation and its ability to predict diverse dynamical properties. Specifically, using only shadows of $\mathcal{E}_H(t_1)$ acquired at $t_1=0.5$, shadow engineering predicts $\text{Tr}[O\mathcal{E}_H(t_K)(\rho)] \approx \text{Tr}[Oe^{-iHt_K}(\rho)]$ for $t_K = K \times t_1$ ($K=2,3,4$), where $\mathcal{E}_H(t)=e^{-iHt}$, and \(H=\sum_i Z_i Z_{i+1} + \sum_i X_i\). As shown in Fig.~\ref{fig4}(a), these predictions (blue) match, and in some cases even surpass, the accuracy achieved by the method of Ref.~\cite{huang2023learning} (orange), which requires independent shadows to be collected separately at each $t_K$. Notably, shadow engineering only uses data from $t_1$ to estimate all later time points, whereas the reference method~\cite{huang2023learning} must acquire extensive shadows at every $t_K$ individually. This highlights a key advantage of shadow engineering for long-time dynamical simulation, as it significantly reduces the required quantum data while maintaining predictive accuracy, thereby greatly facilitating the study of many-body dynamical properties. Figure~\ref{fig4}(b) further illustrates predictions for total magnetization, nearest-neighbor correlations, and multi-body correlations at $\text{Tr}[O e^{-iHt_2}(\rho)]$ using shadows at  $t_1$. Accuracy improves with sample size and with decreasing observable locality, in agreement with theoretical scaling.

{\it{Conclusion}}.---We have introduced shadow engineering, a framework enabling classical prediction of transformed quantum process properties through sparse transfer matrices derived from the classcial shadows of single-step processes. Theoretically, we prove polynomial sample complexity for key operations, while numerical experiments demonstrate exponential advantages over quantum process tomography in resource scaling. 
Building on these findings, we further verify the performance of shadow engineering on a superconducting quantum processor, illustrating its practical effectiveness in error mitigation and Hamiltonian dynamical simulation.
Our work establishes a paradigm of virtual quantum process control, where complex transformations are manipulated classically without physical implementation. Future work includes extension to non-Markovian dynamics~\cite{rossini2023single,zhang2012general} and integration with quantum algorithms to reduce implementation overhead, such as simulating $U_1+U_2$ operations in linear combinations of unitaries (LCU)~\cite{PRXQuantum.6.010359,Chakraborty2024implementingany,Loaiza_2025}.
 
\begin{acknowledgments}
H.-L.H. acknowledges support from the National Natural Science Foundation of China (Grant No. 12274464), and Natural Science Foundation of Henan (Grant No. 242300421049)
\end{acknowledgments}
  
\bibliographystyle{apsrev4-2}
\bibliography{b1}

\begin{thebibliography}{5}%
\makeatletter
\providecommand \@ifxundefined [1]{%
 \@ifx{#1\undefined}
}%
\providecommand \@ifnum [1]{%
 \ifnum #1\expandafter \@firstoftwo
 \else \expandafter \@secondoftwo
 \fi
}%
\providecommand \@ifx [1]{%
 \ifx #1\expandafter \@firstoftwo
 \else \expandafter \@secondoftwo
 \fi
}%
\providecommand \natexlab [1]{#1}%
\providecommand \enquote  [1]{``#1''}%
\providecommand \bibnamefont  [1]{#1}%
\providecommand \bibfnamefont [1]{#1}%
\providecommand \citenamefont [1]{#1}%
\providecommand \href@noop [0]{\@secondoftwo}%
\providecommand \href [0]{\begingroup \@sanitize@url \@href}%
\providecommand \@href[1]{\@@startlink{#1}\@@href}%
\providecommand \@@href[1]{\endgroup#1\@@endlink}%
\providecommand \@sanitize@url [0]{\catcode `\\12\catcode `\$12\catcode `\&12\catcode `\#12\catcode `\^12\catcode `\_12\catcode `\%12\relax}%
\providecommand \@@startlink[1]{}%
\providecommand \@@endlink[0]{}%
\providecommand \url  [0]{\begingroup\@sanitize@url \@url }%
\providecommand \@url [1]{\endgroup\@href {#1}{\urlprefix }}%
\providecommand \urlprefix  [0]{URL }%
\providecommand \Eprint [0]{\href }%
\providecommand \doibase [0]{http://dx.doi.org/}%
\providecommand \selectlanguage [0]{\@gobble}%
\providecommand \bibinfo  [0]{\@secondoftwo}%
\providecommand \bibfield  [0]{\@secondoftwo}%
\providecommand \translation [1]{[#1]}%
\providecommand \BibitemOpen [0]{}%
\providecommand \bibitemStop [0]{}%
\providecommand \bibitemNoStop [0]{.\EOS\space}%
\providecommand \EOS [0]{\spacefactor3000\relax}%
\providecommand \BibitemShut  [1]{\csname bibitem#1\endcsname}%
\let\auto@bib@innerbib\@empty
\bibitem [{\citenamefont {Huang}\ \emph {et~al.}(2020)\citenamefont {Huang}, \citenamefont {Kueng},\ and\ \citenamefont {Preskill}}]{huang2020predicting}%
  \BibitemOpen
  \bibfield  {author} {\bibinfo {author} {\bibfnamefont {H.-Y.}\ \bibnamefont {Huang}}, \bibinfo {author} {\bibfnamefont {R.}~\bibnamefont {Kueng}}, \ and\ \bibinfo {author} {\bibfnamefont {J.}~\bibnamefont {Preskill}},\ }\href {\doibase 10.1038/s41567-020-0932-7} {\bibfield  {journal} {\bibinfo  {journal} {Nat. Phys.}\ }\textbf {\bibinfo {volume} {16}},\ \bibinfo {pages} {1050} (\bibinfo {year} {2020})}\BibitemShut {NoStop}%
\bibitem [{\citenamefont {Elben}\ \emph {et~al.}(2023)\citenamefont {Elben}, \citenamefont {Flammia}, \citenamefont {Huang}, \citenamefont {Kueng}, \citenamefont {Preskill}, \citenamefont {Vermersch},\ and\ \citenamefont {Zoller}}]{elben2023randomized}%
  \BibitemOpen
  \bibfield  {author} {\bibinfo {author} {\bibfnamefont {A.}~\bibnamefont {Elben}}, \bibinfo {author} {\bibfnamefont {S.~T.}\ \bibnamefont {Flammia}}, \bibinfo {author} {\bibfnamefont {H.-Y.}\ \bibnamefont {Huang}}, \bibinfo {author} {\bibfnamefont {R.}~\bibnamefont {Kueng}}, \bibinfo {author} {\bibfnamefont {J.}~\bibnamefont {Preskill}}, \bibinfo {author} {\bibfnamefont {B.}~\bibnamefont {Vermersch}}, \ and\ \bibinfo {author} {\bibfnamefont {P.}~\bibnamefont {Zoller}},\ }\href {\doibase 10.1038/s42254-022-00502-2} {\bibfield  {journal} {\bibinfo  {journal} {Nat. Rev. Phys.}\ }\textbf {\bibinfo {volume} {5}},\ \bibinfo {pages} {9} (\bibinfo {year} {2023})}\BibitemShut {NoStop}%
\bibitem [{\citenamefont {Huang}\ \emph {et~al.}(2023)\citenamefont {Huang}, \citenamefont {Chen},\ and\ \citenamefont {Preskill}}]{huang2023learning}%
  \BibitemOpen
  \bibfield  {author} {\bibinfo {author} {\bibfnamefont {H.-Y.}\ \bibnamefont {Huang}}, \bibinfo {author} {\bibfnamefont {S.}~\bibnamefont {Chen}}, \ and\ \bibinfo {author} {\bibfnamefont {J.}~\bibnamefont {Preskill}},\ }\href {\doibase 10.1103/PRXQuantum.4.040337} {\bibfield  {journal} {\bibinfo  {journal} {PRX Quantum}\ }\textbf {\bibinfo {volume} {4}},\ \bibinfo {pages} {040337} (\bibinfo {year} {2023})}\BibitemShut {NoStop}%
\bibitem [{\citenamefont {Mohseni}\ \emph {et~al.}(2008)\citenamefont {Mohseni}, \citenamefont {Rezakhani},\ and\ \citenamefont {Lidar}}]{mohseni2008quantum}%
  \BibitemOpen
  \bibfield  {author} {\bibinfo {author} {\bibfnamefont {M.}~\bibnamefont {Mohseni}}, \bibinfo {author} {\bibfnamefont {A.~T.}\ \bibnamefont {Rezakhani}}, \ and\ \bibinfo {author} {\bibfnamefont {D.~A.}\ \bibnamefont {Lidar}},\ }\href {\doibase 10.1103/PhysRevA.77.032322} {\bibfield  {journal} {\bibinfo  {journal} {Phys. Rev. A}\ }\textbf {\bibinfo {volume} {77}},\ \bibinfo {pages} {032322} (\bibinfo {year} {2008})}\BibitemShut {NoStop}%
\bibitem [{\citenamefont {Layden}(2022)}]{PhysRevLett.128.210501}%
  \BibitemOpen
  \bibfield  {author} {\bibinfo {author} {\bibfnamefont {D.}~\bibnamefont {Layden}},\ }\href {\doibase 10.1103/PhysRevLett.128.210501} {\bibfield  {journal} {\bibinfo  {journal} {Phys. Rev. Lett.}\ }\textbf {\bibinfo {volume} {128}},\ \bibinfo {pages} {210501} (\bibinfo {year} {2022})}\BibitemShut {NoStop}%
\end{thebibliography}%


\begin{thebibliography}{39}%
\makeatletter
\providecommand \@ifxundefined [1]{%
 \@ifx{#1\undefined}
}%
\providecommand \@ifnum [1]{%
 \ifnum #1\expandafter \@firstoftwo
 \else \expandafter \@secondoftwo
 \fi
}%
\providecommand \@ifx [1]{%
 \ifx #1\expandafter \@firstoftwo
 \else \expandafter \@secondoftwo
 \fi
}%
\providecommand \natexlab [1]{#1}%
\providecommand \enquote  [1]{``#1''}%
\providecommand \bibnamefont  [1]{#1}%
\providecommand \bibfnamefont [1]{#1}%
\providecommand \citenamefont [1]{#1}%
\providecommand \href@noop [0]{\@secondoftwo}%
\providecommand \href [0]{\begingroup \@sanitize@url \@href}%
\providecommand \@href[1]{\@@startlink{#1}\@@href}%
\providecommand \@@href[1]{\endgroup#1\@@endlink}%
\providecommand \@sanitize@url [0]{\catcode `\\12\catcode `\$12\catcode `\&12\catcode `\#12\catcode `\^12\catcode `\_12\catcode `\%12\relax}%
\providecommand \@@startlink[1]{}%
\providecommand \@@endlink[0]{}%
\providecommand \url  [0]{\begingroup\@sanitize@url \@url }%
\providecommand \@url [1]{\endgroup\@href {#1}{\urlprefix }}%
\providecommand \urlprefix  [0]{URL }%
\providecommand \Eprint [0]{\href }%
\providecommand \doibase [0]{https://doi.org/}%
\providecommand \selectlanguage [0]{\@gobble}%
\providecommand \bibinfo  [0]{\@secondoftwo}%
\providecommand \bibfield  [0]{\@secondoftwo}%
\providecommand \translation [1]{[#1]}%
\providecommand \BibitemOpen [0]{}%
\providecommand \bibitemStop [0]{}%
\providecommand \bibitemNoStop [0]{.\EOS\space}%
\providecommand \EOS [0]{\spacefactor3000\relax}%
\providecommand \BibitemShut  [1]{\csname bibitem#1\endcsname}%
\let\auto@bib@innerbib\@empty
\bibitem [{\citenamefont {Mohseni}\ \emph {et~al.}(2008)\citenamefont {Mohseni}, \citenamefont {Rezakhani},\ and\ \citenamefont {Lidar}}]{mohseni2008quantum}%
  \BibitemOpen
  \bibfield  {author} {\bibinfo {author} {\bibfnamefont {M.}~\bibnamefont {Mohseni}}, \bibinfo {author} {\bibfnamefont {A.~T.}\ \bibnamefont {Rezakhani}},\ and\ \bibinfo {author} {\bibfnamefont {D.~A.}\ \bibnamefont {Lidar}},\ }\href {https://doi.org/10.1103/PhysRevA.77.032322} {\bibfield  {journal} {\bibinfo  {journal} {Phys. Rev. A}\ }\textbf {\bibinfo {volume} {77}},\ \bibinfo {pages} {032322} (\bibinfo {year} {2008})}\BibitemShut {NoStop}%
\bibitem [{\citenamefont {O'Brien}\ \emph {et~al.}(2004)\citenamefont {O'Brien}, \citenamefont {Pryde}, \citenamefont {Gilchrist}, \citenamefont {James}, \citenamefont {Langford}, \citenamefont {Ralph},\ and\ \citenamefont {White}}]{o2004quantum}%
  \BibitemOpen
  \bibfield  {author} {\bibinfo {author} {\bibfnamefont {J.~L.}\ \bibnamefont {O'Brien}}, \bibinfo {author} {\bibfnamefont {G.~J.}\ \bibnamefont {Pryde}}, \bibinfo {author} {\bibfnamefont {A.}~\bibnamefont {Gilchrist}}, \bibinfo {author} {\bibfnamefont {D.~F.~V.}\ \bibnamefont {James}}, \bibinfo {author} {\bibfnamefont {N.~K.}\ \bibnamefont {Langford}}, \bibinfo {author} {\bibfnamefont {T.~C.}\ \bibnamefont {Ralph}},\ and\ \bibinfo {author} {\bibfnamefont {A.~G.}\ \bibnamefont {White}},\ }\href {https://doi.org/10.1103/PhysRevLett.93.080502} {\bibfield  {journal} {\bibinfo  {journal} {Phys. Rev. Lett.}\ }\textbf {\bibinfo {volume} {93}},\ \bibinfo {pages} {080502} (\bibinfo {year} {2004})}\BibitemShut {NoStop}%
\bibitem [{\citenamefont {Scott}(2008)}]{scott2008optimizing}%
  \BibitemOpen
  \bibfield  {author} {\bibinfo {author} {\bibfnamefont {A.~J.}\ \bibnamefont {Scott}},\ }\href {https://doi.org/10.1088/1751-8113/41/5/055308} {\bibfield  {journal} {\bibinfo  {journal} {J. Phys. A: Math. Theor.}\ }\textbf {\bibinfo {volume} {41}},\ \bibinfo {pages} {055308} (\bibinfo {year} {2008})}\BibitemShut {NoStop}%
\bibitem [{\citenamefont {Sugiyama}\ \emph {et~al.}(2021)\citenamefont {Sugiyama}, \citenamefont {Imori},\ and\ \citenamefont {Tanaka}}]{PhysRevA.103.062615}%
  \BibitemOpen
  \bibfield  {author} {\bibinfo {author} {\bibfnamefont {T.}~\bibnamefont {Sugiyama}}, \bibinfo {author} {\bibfnamefont {S.}~\bibnamefont {Imori}},\ and\ \bibinfo {author} {\bibfnamefont {F.}~\bibnamefont {Tanaka}},\ }\href {https://doi.org/10.1103/PhysRevA.103.062615} {\bibfield  {journal} {\bibinfo  {journal} {Phys. Rev. A}\ }\textbf {\bibinfo {volume} {103}},\ \bibinfo {pages} {062615} (\bibinfo {year} {2021})}\BibitemShut {NoStop}%
\bibitem [{\citenamefont {Torlai}\ \emph {et~al.}(2023)\citenamefont {Torlai}, \citenamefont {Wood}, \citenamefont {Acharya}, \citenamefont {Carleo}, \citenamefont {Carrasquilla},\ and\ \citenamefont {Aolita}}]{torlai2023quantum}%
  \BibitemOpen
  \bibfield  {author} {\bibinfo {author} {\bibfnamefont {G.}~\bibnamefont {Torlai}}, \bibinfo {author} {\bibfnamefont {C.~J.}\ \bibnamefont {Wood}}, \bibinfo {author} {\bibfnamefont {A.}~\bibnamefont {Acharya}}, \bibinfo {author} {\bibfnamefont {G.}~\bibnamefont {Carleo}}, \bibinfo {author} {\bibfnamefont {J.}~\bibnamefont {Carrasquilla}},\ and\ \bibinfo {author} {\bibfnamefont {L.}~\bibnamefont {Aolita}},\ }\href {https://doi.org/10.1038/s41467-023-38332-9} {\bibfield  {journal} {\bibinfo  {journal} {Nat. Commun.}\ }\textbf {\bibinfo {volume} {14}},\ \bibinfo {pages} {2858} (\bibinfo {year} {2023})}\BibitemShut {NoStop}%
\bibitem [{\citenamefont {Altepeter}\ \emph {et~al.}(2003)\citenamefont {Altepeter}, \citenamefont {Branning}, \citenamefont {Jeffrey}, \citenamefont {Wei}, \citenamefont {Kwiat}, \citenamefont {Thew}, \citenamefont {O'Brien}, \citenamefont {Nielsen},\ and\ \citenamefont {White}}]{PhysRevLett.90.193601}%
  \BibitemOpen
  \bibfield  {author} {\bibinfo {author} {\bibfnamefont {J.~B.}\ \bibnamefont {Altepeter}}, \bibinfo {author} {\bibfnamefont {D.}~\bibnamefont {Branning}}, \bibinfo {author} {\bibfnamefont {E.}~\bibnamefont {Jeffrey}}, \bibinfo {author} {\bibfnamefont {T.~C.}\ \bibnamefont {Wei}}, \bibinfo {author} {\bibfnamefont {P.~G.}\ \bibnamefont {Kwiat}}, \bibinfo {author} {\bibfnamefont {R.~T.}\ \bibnamefont {Thew}}, \bibinfo {author} {\bibfnamefont {J.~L.}\ \bibnamefont {O'Brien}}, \bibinfo {author} {\bibfnamefont {M.~A.}\ \bibnamefont {Nielsen}},\ and\ \bibinfo {author} {\bibfnamefont {A.~G.}\ \bibnamefont {White}},\ }\href {https://doi.org/10.1103/PhysRevLett.90.193601} {\bibfield  {journal} {\bibinfo  {journal} {Phys. Rev. Lett.}\ }\textbf {\bibinfo {volume} {90}},\ \bibinfo {pages} {193601} (\bibinfo {year} {2003})}\BibitemShut {NoStop}%
\bibitem [{\citenamefont {Huang}\ \emph {et~al.}(2020)\citenamefont {Huang}, \citenamefont {Kueng},\ and\ \citenamefont {Preskill}}]{huang2020predicting}%
  \BibitemOpen
  \bibfield  {author} {\bibinfo {author} {\bibfnamefont {H.-Y.}\ \bibnamefont {Huang}}, \bibinfo {author} {\bibfnamefont {R.}~\bibnamefont {Kueng}},\ and\ \bibinfo {author} {\bibfnamefont {J.}~\bibnamefont {Preskill}},\ }\href {https://doi.org/10.1038/s41567-020-0932-7} {\bibfield  {journal} {\bibinfo  {journal} {Nat. Phys.}\ }\textbf {\bibinfo {volume} {16}},\ \bibinfo {pages} {1050} (\bibinfo {year} {2020})}\BibitemShut {NoStop}%
\bibitem [{\citenamefont {Zhang}\ \emph {et~al.}(2021)\citenamefont {Zhang}, \citenamefont {Sun}, \citenamefont {Fang}, \citenamefont {Zhang}, \citenamefont {Yuan},\ and\ \citenamefont {Lu}}]{PhysRevLett.127.200501}%
  \BibitemOpen
  \bibfield  {author} {\bibinfo {author} {\bibfnamefont {T.}~\bibnamefont {Zhang}}, \bibinfo {author} {\bibfnamefont {J.}~\bibnamefont {Sun}}, \bibinfo {author} {\bibfnamefont {X.-X.}\ \bibnamefont {Fang}}, \bibinfo {author} {\bibfnamefont {X.-M.}\ \bibnamefont {Zhang}}, \bibinfo {author} {\bibfnamefont {X.}~\bibnamefont {Yuan}},\ and\ \bibinfo {author} {\bibfnamefont {H.}~\bibnamefont {Lu}},\ }\href {https://doi.org/10.1103/PhysRevLett.127.200501} {\bibfield  {journal} {\bibinfo  {journal} {Phys. Rev. Lett.}\ }\textbf {\bibinfo {volume} {127}},\ \bibinfo {pages} {200501} (\bibinfo {year} {2021})}\BibitemShut {NoStop}%
\bibitem [{\citenamefont {Chen}\ \emph {et~al.}(2021)\citenamefont {Chen}, \citenamefont {Yu}, \citenamefont {Zeng},\ and\ \citenamefont {Flammia}}]{chen2021robust}%
  \BibitemOpen
  \bibfield  {author} {\bibinfo {author} {\bibfnamefont {S.}~\bibnamefont {Chen}}, \bibinfo {author} {\bibfnamefont {W.}~\bibnamefont {Yu}}, \bibinfo {author} {\bibfnamefont {P.}~\bibnamefont {Zeng}},\ and\ \bibinfo {author} {\bibfnamefont {S.~T.}\ \bibnamefont {Flammia}},\ }\href {https://doi.org/10.1103/PRXQuantum.2.030348} {\bibfield  {journal} {\bibinfo  {journal} {PRX Quantum}\ }\textbf {\bibinfo {volume} {2}},\ \bibinfo {pages} {030348} (\bibinfo {year} {2021})}\BibitemShut {NoStop}%
\bibitem [{\citenamefont {Struchalin}\ \emph {et~al.}(2021)\citenamefont {Struchalin}, \citenamefont {Zagorovskii}, \citenamefont {Kovlakov}, \citenamefont {Straupe},\ and\ \citenamefont {Kulik}}]{struchalin2021experimental}%
  \BibitemOpen
  \bibfield  {author} {\bibinfo {author} {\bibfnamefont {G.}~\bibnamefont {Struchalin}}, \bibinfo {author} {\bibfnamefont {Y.~A.}\ \bibnamefont {Zagorovskii}}, \bibinfo {author} {\bibfnamefont {E.}~\bibnamefont {Kovlakov}}, \bibinfo {author} {\bibfnamefont {S.}~\bibnamefont {Straupe}},\ and\ \bibinfo {author} {\bibfnamefont {S.}~\bibnamefont {Kulik}},\ }\href {https://doi.org/10.1103/PRXQuantum.2.010307} {\bibfield  {journal} {\bibinfo  {journal} {PRX Quantum}\ }\textbf {\bibinfo {volume} {2}},\ \bibinfo {pages} {010307} (\bibinfo {year} {2021})}\BibitemShut {NoStop}%
\bibitem [{\citenamefont {Huang}(2022)}]{huang2022learning}%
  \BibitemOpen
  \bibfield  {author} {\bibinfo {author} {\bibfnamefont {H.-Y.}\ \bibnamefont {Huang}},\ }\href {https://doi.org/10.1038/s42254-021-00411-5} {\bibfield  {journal} {\bibinfo  {journal} {Nat. Rev. Phys.}\ }\textbf {\bibinfo {volume} {4}},\ \bibinfo {pages} {81} (\bibinfo {year} {2022})}\BibitemShut {NoStop}%
\bibitem [{\citenamefont {Elben}\ \emph {et~al.}(2023)\citenamefont {Elben}, \citenamefont {Flammia}, \citenamefont {Huang}, \citenamefont {Kueng}, \citenamefont {Preskill}, \citenamefont {Vermersch},\ and\ \citenamefont {Zoller}}]{elben2023randomized}%
  \BibitemOpen
  \bibfield  {author} {\bibinfo {author} {\bibfnamefont {A.}~\bibnamefont {Elben}}, \bibinfo {author} {\bibfnamefont {S.~T.}\ \bibnamefont {Flammia}}, \bibinfo {author} {\bibfnamefont {H.-Y.}\ \bibnamefont {Huang}}, \bibinfo {author} {\bibfnamefont {R.}~\bibnamefont {Kueng}}, \bibinfo {author} {\bibfnamefont {J.}~\bibnamefont {Preskill}}, \bibinfo {author} {\bibfnamefont {B.}~\bibnamefont {Vermersch}},\ and\ \bibinfo {author} {\bibfnamefont {P.}~\bibnamefont {Zoller}},\ }\href {https://doi.org/10.1038/s42254-022-00502-2} {\bibfield  {journal} {\bibinfo  {journal} {Nat. Rev. Phys.}\ }\textbf {\bibinfo {volume} {5}},\ \bibinfo {pages} {9} (\bibinfo {year} {2023})}\BibitemShut {NoStop}%
\bibitem [{\citenamefont {Acharya}\ \emph {et~al.}(2021)\citenamefont {Acharya}, \citenamefont {Saha},\ and\ \citenamefont {Sengupta}}]{acharya2021informationally}%
  \BibitemOpen
  \bibfield  {author} {\bibinfo {author} {\bibfnamefont {A.}~\bibnamefont {Acharya}}, \bibinfo {author} {\bibfnamefont {S.}~\bibnamefont {Saha}},\ and\ \bibinfo {author} {\bibfnamefont {A.~M.}\ \bibnamefont {Sengupta}},\ }\href {https://doi.org/10.1103/PhysRevA.104.052418} {\bibfield  {journal} {\bibinfo  {journal} {Phys. Rev. A}\ }\textbf {\bibinfo {volume} {104}},\ \bibinfo {pages} {052418} (\bibinfo {year} {2021})}\BibitemShut {NoStop}%
\bibitem [{\citenamefont {Akhtar}\ \emph {et~al.}(2023)\citenamefont {Akhtar}, \citenamefont {Hu},\ and\ \citenamefont {You}}]{akhtar2023scalable}%
  \BibitemOpen
  \bibfield  {author} {\bibinfo {author} {\bibfnamefont {A.~A.}\ \bibnamefont {Akhtar}}, \bibinfo {author} {\bibfnamefont {H.-Y.}\ \bibnamefont {Hu}},\ and\ \bibinfo {author} {\bibfnamefont {Y.-Z.}\ \bibnamefont {You}},\ }\href {https://doi.org/10.22331/q-2023-06-01-1026} {\bibfield  {journal} {\bibinfo  {journal} {{Quantum}}\ }\textbf {\bibinfo {volume} {7}},\ \bibinfo {pages} {1026} (\bibinfo {year} {2023})}\BibitemShut {NoStop}%
\bibitem [{\citenamefont {Koh}\ and\ \citenamefont {Grewal}(2022)}]{Koh2022classical}%
  \BibitemOpen
  \bibfield  {author} {\bibinfo {author} {\bibfnamefont {D.~E.}\ \bibnamefont {Koh}}\ and\ \bibinfo {author} {\bibfnamefont {S.}~\bibnamefont {Grewal}},\ }\href {https://doi.org/10.22331/q-2022-08-16-776} {\bibfield  {journal} {\bibinfo  {journal} {{Quantum}}\ }\textbf {\bibinfo {volume} {6}},\ \bibinfo {pages} {776} (\bibinfo {year} {2022})}\BibitemShut {NoStop}%
\bibitem [{\citenamefont {Jnane}\ \emph {et~al.}(2024)\citenamefont {Jnane}, \citenamefont {Steinberg}, \citenamefont {Cai}, \citenamefont {Nguyen},\ and\ \citenamefont {Koczor}}]{PRXQuantum.5.010324}%
  \BibitemOpen
  \bibfield  {author} {\bibinfo {author} {\bibfnamefont {H.}~\bibnamefont {Jnane}}, \bibinfo {author} {\bibfnamefont {J.}~\bibnamefont {Steinberg}}, \bibinfo {author} {\bibfnamefont {Z.}~\bibnamefont {Cai}}, \bibinfo {author} {\bibfnamefont {H.~C.}\ \bibnamefont {Nguyen}},\ and\ \bibinfo {author} {\bibfnamefont {B.}~\bibnamefont {Koczor}},\ }\href {https://doi.org/10.1103/PRXQuantum.5.010324} {\bibfield  {journal} {\bibinfo  {journal} {PRX Quantum}\ }\textbf {\bibinfo {volume} {5}},\ \bibinfo {pages} {010324} (\bibinfo {year} {2024})}\BibitemShut {NoStop}%
\bibitem [{\citenamefont {Ippoliti}\ \emph {et~al.}(2023)\citenamefont {Ippoliti}, \citenamefont {Li}, \citenamefont {Rakovszky},\ and\ \citenamefont {Khemani}}]{PhysRevLett.130.230403}%
  \BibitemOpen
  \bibfield  {author} {\bibinfo {author} {\bibfnamefont {M.}~\bibnamefont {Ippoliti}}, \bibinfo {author} {\bibfnamefont {Y.}~\bibnamefont {Li}}, \bibinfo {author} {\bibfnamefont {T.}~\bibnamefont {Rakovszky}},\ and\ \bibinfo {author} {\bibfnamefont {V.}~\bibnamefont {Khemani}},\ }\href {https://doi.org/10.1103/PhysRevLett.130.230403} {\bibfield  {journal} {\bibinfo  {journal} {Phys. Rev. Lett.}\ }\textbf {\bibinfo {volume} {130}},\ \bibinfo {pages} {230403} (\bibinfo {year} {2023})}\BibitemShut {NoStop}%
\bibitem [{\citenamefont {Zhou}\ and\ \citenamefont {Liu}(2023)}]{Zhou2023performanceanalysis}%
  \BibitemOpen
  \bibfield  {author} {\bibinfo {author} {\bibfnamefont {Y.}~\bibnamefont {Zhou}}\ and\ \bibinfo {author} {\bibfnamefont {Q.}~\bibnamefont {Liu}},\ }\href {https://doi.org/10.22331/q-2023-06-29-1044} {\bibfield  {journal} {\bibinfo  {journal} {{Quantum}}\ }\textbf {\bibinfo {volume} {7}},\ \bibinfo {pages} {1044} (\bibinfo {year} {2023})}\BibitemShut {NoStop}%
\bibitem [{\citenamefont {Hu}\ \emph {et~al.}(2025)\citenamefont {Hu}, \citenamefont {Gu}, \citenamefont {Majumder}, \citenamefont {Ren}, \citenamefont {Zhang}, \citenamefont {Wang}, \citenamefont {You}, \citenamefont {Minev}, \citenamefont {Yelin},\ and\ \citenamefont {Seif}}]{hu2025demonstration}%
  \BibitemOpen
  \bibfield  {author} {\bibinfo {author} {\bibfnamefont {H.-Y.}\ \bibnamefont {Hu}}, \bibinfo {author} {\bibfnamefont {A.}~\bibnamefont {Gu}}, \bibinfo {author} {\bibfnamefont {S.}~\bibnamefont {Majumder}}, \bibinfo {author} {\bibfnamefont {H.}~\bibnamefont {Ren}}, \bibinfo {author} {\bibfnamefont {Y.}~\bibnamefont {Zhang}}, \bibinfo {author} {\bibfnamefont {D.~S.}\ \bibnamefont {Wang}}, \bibinfo {author} {\bibfnamefont {Y.-Z.}\ \bibnamefont {You}}, \bibinfo {author} {\bibfnamefont {Z.}~\bibnamefont {Minev}}, \bibinfo {author} {\bibfnamefont {S.~F.}\ \bibnamefont {Yelin}},\ and\ \bibinfo {author} {\bibfnamefont {A.}~\bibnamefont {Seif}},\ }\href {https://doi.org/10.1038/s41467-025-57349-w} {\bibfield  {journal} {\bibinfo  {journal} {Nat. Commun.}\ }\textbf {\bibinfo {volume} {16}},\ \bibinfo {pages} {2943} (\bibinfo {year} {2025})}\BibitemShut {NoStop}%
\bibitem [{\citenamefont {Jamio{\l}kowski}(1972)}]{jamiolkowski1972linear}%
  \BibitemOpen
  \bibfield  {author} {\bibinfo {author} {\bibfnamefont {A.}~\bibnamefont {Jamio{\l}kowski}},\ }\href {https://doi.org/10.1016/0034-4877(72)90070-2} {\bibfield  {journal} {\bibinfo  {journal} {Rep. Math. Phys.}\ }\textbf {\bibinfo {volume} {3}},\ \bibinfo {pages} {275} (\bibinfo {year} {1972})}\BibitemShut {NoStop}%
\bibitem [{\citenamefont {Leung}(2003)}]{leung2003choi}%
  \BibitemOpen
  \bibfield  {author} {\bibinfo {author} {\bibfnamefont {D.~W.}\ \bibnamefont {Leung}},\ }\href {https://doi.org/10.1063/1.1528336} {\bibfield  {journal} {\bibinfo  {journal} {J. Math. Phys.}\ }\textbf {\bibinfo {volume} {44}},\ \bibinfo {pages} {528} (\bibinfo {year} {2003})}\BibitemShut {NoStop}%
\bibitem [{\citenamefont {Kunjummen}\ \emph {et~al.}(2023)\citenamefont {Kunjummen}, \citenamefont {Tran}, \citenamefont {Carney},\ and\ \citenamefont {Taylor}}]{kunjummen2023shadow}%
  \BibitemOpen
  \bibfield  {author} {\bibinfo {author} {\bibfnamefont {J.}~\bibnamefont {Kunjummen}}, \bibinfo {author} {\bibfnamefont {M.~C.}\ \bibnamefont {Tran}}, \bibinfo {author} {\bibfnamefont {D.}~\bibnamefont {Carney}},\ and\ \bibinfo {author} {\bibfnamefont {J.~M.}\ \bibnamefont {Taylor}},\ }\href {https://doi.org/10.1103/PhysRevA.107.042403} {\bibfield  {journal} {\bibinfo  {journal} {Phys. Rev. A}\ }\textbf {\bibinfo {volume} {107}},\ \bibinfo {pages} {042403} (\bibinfo {year} {2023})}\BibitemShut {NoStop}%
\bibitem [{\citenamefont {Levy}\ \emph {et~al.}(2024)\citenamefont {Levy}, \citenamefont {Luo},\ and\ \citenamefont {Clark}}]{levy2024classical}%
  \BibitemOpen
  \bibfield  {author} {\bibinfo {author} {\bibfnamefont {R.}~\bibnamefont {Levy}}, \bibinfo {author} {\bibfnamefont {D.}~\bibnamefont {Luo}},\ and\ \bibinfo {author} {\bibfnamefont {B.~K.}\ \bibnamefont {Clark}},\ }\href {https://doi.org/10.1103/PhysRevResearch.6.013029} {\bibfield  {journal} {\bibinfo  {journal} {Phys. Rev. Res.}\ }\textbf {\bibinfo {volume} {6}},\ \bibinfo {pages} {013029} (\bibinfo {year} {2024})}\BibitemShut {NoStop}%
\bibitem [{\citenamefont {Wang}\ and\ \citenamefont {He}(2025)}]{wang2025reducing}%
  \BibitemOpen
  \bibfield  {author} {\bibinfo {author} {\bibfnamefont {H.}~\bibnamefont {Wang}}\ and\ \bibinfo {author} {\bibfnamefont {K.}~\bibnamefont {He}},\ }\href {https://doi.org/10.1103/13ct-8vtv} {\bibfield  {journal} {\bibinfo  {journal} {Phys. Rev. A}\ }\textbf {\bibinfo {volume} {112}},\ \bibinfo {pages} {012413} (\bibinfo {year} {2025})}\BibitemShut {NoStop}%
\bibitem [{\citenamefont {Huang}\ \emph {et~al.}(2023{\natexlab{a}})\citenamefont {Huang}, \citenamefont {Chen},\ and\ \citenamefont {Preskill}}]{huang2023learning}%
  \BibitemOpen
  \bibfield  {author} {\bibinfo {author} {\bibfnamefont {H.-Y.}\ \bibnamefont {Huang}}, \bibinfo {author} {\bibfnamefont {S.}~\bibnamefont {Chen}},\ and\ \bibinfo {author} {\bibfnamefont {J.}~\bibnamefont {Preskill}},\ }\href {https://doi.org/10.1103/PRXQuantum.4.040337} {\bibfield  {journal} {\bibinfo  {journal} {PRX Quantum}\ }\textbf {\bibinfo {volume} {4}},\ \bibinfo {pages} {040337} (\bibinfo {year} {2023}{\natexlab{a}})}\BibitemShut {NoStop}%
\bibitem [{\citenamefont {Chen}\ \emph {et~al.}()\citenamefont {Chen}, \citenamefont {Yu}, \citenamefont {Zhu},\ and\ \citenamefont {Wang}}]{chen2023efficient}%
  \BibitemOpen
  \bibfield  {author} {\bibinfo {author} {\bibfnamefont {Y.}~\bibnamefont {Chen}}, \bibinfo {author} {\bibfnamefont {Z.}~\bibnamefont {Yu}}, \bibinfo {author} {\bibfnamefont {C.}~\bibnamefont {Zhu}},\ and\ \bibinfo {author} {\bibfnamefont {X.}~\bibnamefont {Wang}},\ }\href@noop {} {}\Eprint {https://arxiv.org/abs/arXiv:2305.04148} {arXiv:2305.04148} \BibitemShut {NoStop}%
\bibitem [{\citenamefont {Huang}\ \emph {et~al.}(2023{\natexlab{b}})\citenamefont {Huang}, \citenamefont {Xu}, \citenamefont {Guo}, \citenamefont {Tian}, \citenamefont {Wei}, \citenamefont {Sun}, \citenamefont {Bao},\ and\ \citenamefont {Long}}]{huang2023near}%
  \BibitemOpen
  \bibfield  {author} {\bibinfo {author} {\bibfnamefont {H.-L.}\ \bibnamefont {Huang}}, \bibinfo {author} {\bibfnamefont {X.-Y.}\ \bibnamefont {Xu}}, \bibinfo {author} {\bibfnamefont {C.}~\bibnamefont {Guo}}, \bibinfo {author} {\bibfnamefont {G.}~\bibnamefont {Tian}}, \bibinfo {author} {\bibfnamefont {S.-J.}\ \bibnamefont {Wei}}, \bibinfo {author} {\bibfnamefont {X.}~\bibnamefont {Sun}}, \bibinfo {author} {\bibfnamefont {W.-S.}\ \bibnamefont {Bao}},\ and\ \bibinfo {author} {\bibfnamefont {G.-L.}\ \bibnamefont {Long}},\ }\href {https://doi.org/10.1007/s11433-022-2057-y} {\bibfield  {journal} {\bibinfo  {journal} {Sci. China-Phys. Mech. Astron.}\ }\textbf {\bibinfo {volume} {66}},\ \bibinfo {pages} {250302} (\bibinfo {year} {2023}{\natexlab{b}})}\BibitemShut {NoStop}%
\bibitem [{\citenamefont {Suzuki}\ \emph {et~al.}(2022)\citenamefont {Suzuki}, \citenamefont {Endo}, \citenamefont {Fujii},\ and\ \citenamefont {Tokunaga}}]{PRXQuantum.3.010345}%
  \BibitemOpen
  \bibfield  {author} {\bibinfo {author} {\bibfnamefont {Y.}~\bibnamefont {Suzuki}}, \bibinfo {author} {\bibfnamefont {S.}~\bibnamefont {Endo}}, \bibinfo {author} {\bibfnamefont {K.}~\bibnamefont {Fujii}},\ and\ \bibinfo {author} {\bibfnamefont {Y.}~\bibnamefont {Tokunaga}},\ }\href {https://doi.org/10.1103/PRXQuantum.3.010345} {\bibfield  {journal} {\bibinfo  {journal} {PRX Quantum}\ }\textbf {\bibinfo {volume} {3}},\ \bibinfo {pages} {010345} (\bibinfo {year} {2022})}\BibitemShut {NoStop}%
\bibitem [{\citenamefont {Cai}\ \emph {et~al.}(2023)\citenamefont {Cai}, \citenamefont {Babbush}, \citenamefont {Benjamin}, \citenamefont {Endo}, \citenamefont {Huggins}, \citenamefont {Li}, \citenamefont {McClean},\ and\ \citenamefont {O'Brien}}]{cai2023quantum}%
  \BibitemOpen
  \bibfield  {author} {\bibinfo {author} {\bibfnamefont {Z.}~\bibnamefont {Cai}}, \bibinfo {author} {\bibfnamefont {R.}~\bibnamefont {Babbush}}, \bibinfo {author} {\bibfnamefont {S.~C.}\ \bibnamefont {Benjamin}}, \bibinfo {author} {\bibfnamefont {S.}~\bibnamefont {Endo}}, \bibinfo {author} {\bibfnamefont {W.~J.}\ \bibnamefont {Huggins}}, \bibinfo {author} {\bibfnamefont {Y.}~\bibnamefont {Li}}, \bibinfo {author} {\bibfnamefont {J.~R.}\ \bibnamefont {McClean}},\ and\ \bibinfo {author} {\bibfnamefont {T.~E.}\ \bibnamefont {O'Brien}},\ }\href {https://doi.org/10.1103/RevModPhys.95.045005} {\bibfield  {journal} {\bibinfo  {journal} {Rev. Mod. Phys.}\ }\textbf {\bibinfo {volume} {95}},\ \bibinfo {pages} {045005} (\bibinfo {year} {2023})}\BibitemShut {NoStop}%
\bibitem [{\citenamefont {Haah}\ \emph {et~al.}(2021)\citenamefont {Haah}, \citenamefont {Hastings}, \citenamefont {Kothari},\ and\ \citenamefont {Low}}]{haah2021quantum}%
  \BibitemOpen
  \bibfield  {author} {\bibinfo {author} {\bibfnamefont {J.}~\bibnamefont {Haah}}, \bibinfo {author} {\bibfnamefont {M.~B.}\ \bibnamefont {Hastings}}, \bibinfo {author} {\bibfnamefont {R.}~\bibnamefont {Kothari}},\ and\ \bibinfo {author} {\bibfnamefont {G.~H.}\ \bibnamefont {Low}},\ }\href {https://doi.org/10.1137/18m1231511} {\bibfield  {journal} {\bibinfo  {journal} {SIAM J. Comput.}\ }\textbf {\bibinfo {volume} {52}},\ \bibinfo {pages} {FOCS18} (\bibinfo {year} {2021})}\BibitemShut {NoStop}%
\bibitem [{\citenamefont {Zhao}\ \emph {et~al.}(2022)\citenamefont {Zhao}, \citenamefont {Zhou}, \citenamefont {Shaw}, \citenamefont {Li},\ and\ \citenamefont {Childs}}]{PhysRevLett.129.270502}%
  \BibitemOpen
  \bibfield  {author} {\bibinfo {author} {\bibfnamefont {Q.}~\bibnamefont {Zhao}}, \bibinfo {author} {\bibfnamefont {Y.}~\bibnamefont {Zhou}}, \bibinfo {author} {\bibfnamefont {A.~F.}\ \bibnamefont {Shaw}}, \bibinfo {author} {\bibfnamefont {T.}~\bibnamefont {Li}},\ and\ \bibinfo {author} {\bibfnamefont {A.~M.}\ \bibnamefont {Childs}},\ }\href {https://doi.org/10.1103/PhysRevLett.129.270502} {\bibfield  {journal} {\bibinfo  {journal} {Phys. Rev. Lett.}\ }\textbf {\bibinfo {volume} {129}},\ \bibinfo {pages} {270502} (\bibinfo {year} {2022})}\BibitemShut {NoStop}%
\bibitem [{\citenamefont {Golub}\ and\ \citenamefont {Van~Loan}(2013)}]{golub2013matrix}%
  \BibitemOpen
  \bibfield  {author} {\bibinfo {author} {\bibfnamefont {G.~H.}\ \bibnamefont {Golub}}\ and\ \bibinfo {author} {\bibfnamefont {C.~F.}\ \bibnamefont {Van~Loan}},\ }\href {https://doi.org/10.1137/1.9781421407944} {\emph {\bibinfo {title} {Matrix Computations}}},\ \bibinfo {edition} {4th}\ ed.\ (\bibinfo  {publisher} {Johns Hopkins University Press},\ \bibinfo {address} {Philadelphia, PA},\ \bibinfo {year} {2013})\BibitemShut {NoStop}%
\bibitem [{\citenamefont {Gao}\ \emph {et~al.}(2023)\citenamefont {Gao}, \citenamefont {Ji}, \citenamefont {Chang}, \citenamefont {Han}, \citenamefont {Wei}, \citenamefont {Liu},\ and\ \citenamefont {Wang}}]{gao2023systematic}%
  \BibitemOpen
  \bibfield  {author} {\bibinfo {author} {\bibfnamefont {J.}~\bibnamefont {Gao}}, \bibinfo {author} {\bibfnamefont {W.}~\bibnamefont {Ji}}, \bibinfo {author} {\bibfnamefont {F.}~\bibnamefont {Chang}}, \bibinfo {author} {\bibfnamefont {S.}~\bibnamefont {Han}}, \bibinfo {author} {\bibfnamefont {B.}~\bibnamefont {Wei}}, \bibinfo {author} {\bibfnamefont {Z.}~\bibnamefont {Liu}},\ and\ \bibinfo {author} {\bibfnamefont {Y.}~\bibnamefont {Wang}},\ }\href {https://doi.org/10.1145/3571157} {\bibfield  {journal} {\bibinfo  {journal} {ACM Comput. Surv.}\ }\textbf {\bibinfo {volume} {55}},\ \bibinfo {eid} {244} (\bibinfo {year} {2023})}\BibitemShut {NoStop}%
\bibitem [{\citenamefont {Islam}\ \emph {et~al.}()\citenamefont {Islam}, \citenamefont {Xu}, \citenamefont {Dai},\ and\ \citenamefont {Bulu{\c{c}}}}]{islam2025improving}%
  \BibitemOpen
  \bibfield  {author} {\bibinfo {author} {\bibfnamefont {A.~A.~R.}\ \bibnamefont {Islam}}, \bibinfo {author} {\bibfnamefont {H.}~\bibnamefont {Xu}}, \bibinfo {author} {\bibfnamefont {D.}~\bibnamefont {Dai}},\ and\ \bibinfo {author} {\bibfnamefont {A.}~\bibnamefont {Bulu{\c{c}}}},\ }\href@noop {} {}\Eprint {https://arxiv.org/abs/arXiv:2507.21253} {arXiv:2507.21253} \BibitemShut {NoStop}%
\bibitem [{\citenamefont {Rossini}\ \emph {et~al.}(2023)\citenamefont {Rossini}, \citenamefont {Maile}, \citenamefont {Ankerhold},\ and\ \citenamefont {Donvil}}]{rossini2023single}%
  \BibitemOpen
  \bibfield  {author} {\bibinfo {author} {\bibfnamefont {M.}~\bibnamefont {Rossini}}, \bibinfo {author} {\bibfnamefont {D.}~\bibnamefont {Maile}}, \bibinfo {author} {\bibfnamefont {J.}~\bibnamefont {Ankerhold}},\ and\ \bibinfo {author} {\bibfnamefont {B.~I.~C.}\ \bibnamefont {Donvil}},\ }\href {https://doi.org/10.1103/PhysRevLett.131.110603} {\bibfield  {journal} {\bibinfo  {journal} {Phys. Rev. Lett.}\ }\textbf {\bibinfo {volume} {131}},\ \bibinfo {pages} {110603} (\bibinfo {year} {2023})}\BibitemShut {NoStop}%
\bibitem [{\citenamefont {Zhang}\ \emph {et~al.}(2012)\citenamefont {Zhang}, \citenamefont {Lo}, \citenamefont {Xiong}, \citenamefont {Tu},\ and\ \citenamefont {Nori}}]{zhang2012general}%
  \BibitemOpen
  \bibfield  {author} {\bibinfo {author} {\bibfnamefont {W.-M.}\ \bibnamefont {Zhang}}, \bibinfo {author} {\bibfnamefont {P.-Y.}\ \bibnamefont {Lo}}, \bibinfo {author} {\bibfnamefont {H.-N.}\ \bibnamefont {Xiong}}, \bibinfo {author} {\bibfnamefont {M.~W.-Y.}\ \bibnamefont {Tu}},\ and\ \bibinfo {author} {\bibfnamefont {F.}~\bibnamefont {Nori}},\ }\href {https://doi.org/10.1103/PhysRevLett.109.170402} {\bibfield  {journal} {\bibinfo  {journal} {Phys. Rev. Lett.}\ }\textbf {\bibinfo {volume} {109}},\ \bibinfo {pages} {170402} (\bibinfo {year} {2012})}\BibitemShut {NoStop}%
\bibitem [{\citenamefont {Zeng}\ \emph {et~al.}(2025)\citenamefont {Zeng}, \citenamefont {Sun}, \citenamefont {Jiang},\ and\ \citenamefont {Zhao}}]{PRXQuantum.6.010359}%
  \BibitemOpen
  \bibfield  {author} {\bibinfo {author} {\bibfnamefont {P.}~\bibnamefont {Zeng}}, \bibinfo {author} {\bibfnamefont {J.}~\bibnamefont {Sun}}, \bibinfo {author} {\bibfnamefont {L.}~\bibnamefont {Jiang}},\ and\ \bibinfo {author} {\bibfnamefont {Q.}~\bibnamefont {Zhao}},\ }\href {https://doi.org/10.1103/PRXQuantum.6.010359} {\bibfield  {journal} {\bibinfo  {journal} {PRX Quantum}\ }\textbf {\bibinfo {volume} {6}},\ \bibinfo {pages} {010359} (\bibinfo {year} {2025})}\BibitemShut {NoStop}%
\bibitem [{\citenamefont {Chakraborty}(2024)}]{Chakraborty2024implementingany}%
  \BibitemOpen
  \bibfield  {author} {\bibinfo {author} {\bibfnamefont {S.}~\bibnamefont {Chakraborty}},\ }\href {https://doi.org/10.22331/q-2024-10-10-1496} {\bibfield  {journal} {\bibinfo  {journal} {{Quantum}}\ }\textbf {\bibinfo {volume} {8}},\ \bibinfo {pages} {1496} (\bibinfo {year} {2024})}\BibitemShut {NoStop}%
\bibitem [{\citenamefont {Loaiza}\ \emph {et~al.}(2025)\citenamefont {Loaiza}, \citenamefont {Sankar~Brahmachari},\ and\ \citenamefont {Izmaylov}}]{Loaiza_2025}%
  \BibitemOpen
  \bibfield  {author} {\bibinfo {author} {\bibfnamefont {I.}~\bibnamefont {Loaiza}}, \bibinfo {author} {\bibfnamefont {A.}~\bibnamefont {Sankar~Brahmachari}},\ and\ \bibinfo {author} {\bibfnamefont {A.~F.}\ \bibnamefont {Izmaylov}},\ }\href {https://doi.org/10.1088/2058-9565/add9c1} {\bibfield  {journal} {\bibinfo  {journal} {Quantum Sci. Technol.}\ }\textbf {\bibinfo {volume} {10}},\ \bibinfo {pages} {035035} (\bibinfo {year} {2025})}\BibitemShut {NoStop}%
\end{thebibliography}%
  
\end{document}


\title{Supplemental Material for ``Shadow Engineering of Quantum Processes"}
\maketitle
\section{Shadow engineering framework}

Given an observable $O$, an input state $\rho$, and classical shadows of $k$ unknown quantum processes $\{\mathcal{E}_m\}_{m=1}^k$, the target of shadow engineering is to predict the expected value $\operatorname{Tr}\left[O \cdot f(\mathcal{E}_1,\mathcal{E}_2,\cdots,\mathcal{E}_k)(\rho)\right]$, where $f(\mathcal{E}_1,\mathcal{E}_2,\cdots,\mathcal{E}_k)$ denotes a transformed quantum process derived from the $k$ unknown processes. 
In this section, we elaborate on the three core stages of shadow engineering: (1) encoding the transfer matrix of single-step quantum processes, (2) constructing the transfer matrix of the transformed process, and (3) predicting the target expectation value via transfer matrix of the transformed process.

\subsection{Encoding the transfer matrix of single-step quantum processes}
The transfer matrix $T_\mathcal{E}$ of an $n$-qubit quantum process $\mathcal{E}$ is a $6^n \times 6^n$ real matrix,
whose elements are defined as
\begin{equation}
    (T_{\mathcal{E}})_{ij} = \left( \frac{1}{3} \right)^n \text{Tr}[ 
      \mathcal{E}\left( \mathcal{S}^{\otimes n}_i \right) \cdot \mathcal{S}^{\otimes n}_j 
  ],
\end{equation}
where $\mathcal{S}^{\otimes n}_i$ and $\mathcal{S}^{\otimes n}_j$ denote the $i$-th and $j$-th states in the $n$-qubit stabilizer basis $\mathcal{S}^{\otimes n} \triangleq \left\{ \bigotimes_{\iota=1}^n s_\iota \mid s_\iota \in {\mathcal{S}} \right\}$, constructed from the single-qubit basis: 
\begin{equation}
    \mathcal{S} \triangleq \{ \ket{0}\!\bra{0}, \ket{1}\!\bra{1}, \ket{+}\!\bra{+}, \ket{-}\!\bra{-}, \ket{y+}\!\bra{y+}, \ket{y-}\!\bra{y-} \}.
\end{equation} 
Each entry $(T_{\mathcal{E}})_{ij}$ represents the probability that, after applying $\mathcal{E}$ to the input state $\mathcal{S}^{\otimes n}_i$ and performing random Pauli measurements (with each qubit measured uniformly in the $X$, $Y$, or $Z$ basis), the measurement outcome corresponds to $\mathcal{S}^{\otimes n}_j$.
The factor $(1/3)^n$ arises from the uniform choice of Pauli measurement basis per qubit, which is necessary to identify $\mathcal{S}^{\otimes n}_j$.
And $\text{Tr}[\mathcal{E}(\mathcal{S}^{\otimes n}_i)\cdot \mathcal{S}^{\otimes n}_j]$ gives the Born rule probability of  the outcome $\mathcal{S}^{\otimes n}_j$ conditioned on this basis choice.

Each row of $T_\mathcal{E}$ thus encodes the complete probability distribution over outcomes for a fixed input stabilizer state, 
enabling the reconstruction of the output state $\mathcal{E}(\mathcal{S}^{\otimes n}_i)$ within the classical shadow tomography~\cite{huang2020predicting,elben2023randomized}. 
Concretely, 
\begin{equation}
    \mathcal{E}(\mathcal{S}^{\otimes n}_i) = \sum_{j=1}^{6^n} (T_{\mathcal{E}})_{ij} \mathcal{M}^{-1}(\mathcal{S}^{\otimes n}_j).
\end{equation}
where $\mathcal{M}^{-1}(\bigotimes_{\iota=1}^{n} \rho_{\iota}) = \bigotimes_{\iota=1}^{n} (3\rho_{\iota} - I)$~\cite{huang2020predicting}.
Collecting these equations for all $i$ yields the compact matrix relation
\begin{equation}\label{eq3}
    \mathcal{E} \circledast (\boldsymbol{\sigma}_{\mathcal{S}^{\otimes n}}) = T_{\mathcal{E}} \mathcal{M}^{-1} \circledast (\boldsymbol{\sigma}_{\mathcal{S}^{\otimes n}})
\end{equation}
where $\boldsymbol{\sigma}_{\mathcal{S}^{\otimes n}}= {[{\mathcal{S}_{1}^{\otimes n}},{\mathcal{S}_{2}^{\otimes n}},...,{\mathcal{S}_{6^n}^{\otimes n}}]^T}$ is a column vector of all $6^n$ states in $\mathcal{S}^{\otimes n}$, and the operator $\circledast$ is defined such that for any channel $\mathcal{C}$, its action on the vector $\boldsymbol{\sigma}_{\mathcal{S}^{\otimes n}}$ is given by: $\mathcal{C} \circledast (\boldsymbol{\sigma}_{\mathcal{S}^{\otimes n}}):= {[\mathcal{C}({\mathcal{S}_{1}^{\otimes n}}),\mathcal{C}({\mathcal{S}_{2}^{\otimes n}}),...,\mathcal{C}({\mathcal{S}_{6^n}^{\otimes n}})]^T}$.
Equation~\eqref{eq3} shows that the action of $\mathcal{E}$ on the enrtire stabilizer basis can be reconstructed by applying the transfer matrix to the classical shadow vector that has been transformed by $\mathcal{M}^{-1}$.

The transfer matrix $T_{\mathcal{E}}$ can be estimated from a dataset of $N$ experimental snapshots, generated via the following pipeline:

\paragraph{Classical shadow dataset preprocessing.}

A classical shadow dataset \(S_{N}(\mathcal{E})\triangleq\left\{\mathcal{S}^{\otimes n}_{i_{\ell}},\mathcal{S}^{\otimes n}_{j_{\ell}}\right\}_{\ell=1}^{N}\) is obtained by repeating the following process for each snapshot \(\ell\): uniformly sample an input state index \(i_{\ell}\) from \(\mathcal{S}^{\otimes n}\), apply the process \(\mathcal{E}\) to this state, and then perform a random Pauli measurement to record the outcome index \(j_{\ell}\) within \(\mathcal{S}^{\otimes n}\).
Each random Pauli measurement is implemented by a suitable single-qubit gate—\(H\) for the \(X\)-basis, \(HS^\dagger\) for the \(Y\)-basis, and \(I\) for the \(Z\)-basis—followed by a computational-basis measurement; the index \(j_\ell\) is determined from the chosen Pauli basis and the measurement outcome bitstring in \(\mathcal{O}(n)\) time. Generating the full dataset \(S_N(\mathcal{E})\) thus requires a total time of \(\mathcal{O}(Nn)\).

\paragraph{Sparse CSR construction.}

Given a classical dataset \(S_N(\mathcal{E})\) consisting of index pairs, the transfer matrix \(T_{\mathcal{E}}\) is estimated in three steps. First, count aggregation: a hash table is used to accumulate the occurrence counts of each index pair \((i_{\ell}, j_{\ell})\), a step that runs in \(\mathcal{O}(N)\) time and yields a set of triplets \((i, j, \text{count}_{ij})\) corresponding to \(N_{\text{non}} \leq N\) unique index pairs. Next, row-wise normalization: for each row index \(i\), the accumulated counts are normalized by the total row-wise count to compute the empirical probabilities \(\hat{p}_{ij} = \text{count}_{ij} / \sum_j \text{count}_{ij}\), which directly estimate the matrix elements \((T_{\mathcal{E}})_{ij}\); this step consumes \(\mathcal{O}(N_{\text{non}})\) time, where \(N_{\text{non}}\) denotes the number of distinct index pairs in \(S_N(\mathcal{E})\). Finally, conversion to compressed sparse row (CSR) format: all normalized triplets are first sorted lexicographically by row index \(i\) and then column index \(j\), a process with time complexity \(\mathcal{O}(N_{\text{non}} \log N_{\text{non}})\), after which the standard CSR arrays are constructed: a \texttt{values} array storing the normalized probabilities \(\hat{p}_{ij}\), a \texttt{columns} array recording the matching column indices \(j\), and a \texttt{rows} array serving as the row pointer, which marks the starting index of each row \(i\) in the \texttt{values} and \texttt{columns} arrays and is computed via cumulative row counts.

\subsection{Constructing the transfer matrix of the transformed process}
As established in the preceding subsection, classical shadows of $k$ unknown single-step quantum processes can be converted into estimated transfer matrices $\{\hat{T}_{\mathcal{E}_m}\}_{m=1}^k$. 
The core goal of this section is to construct the transfer matrix $\hat{T}_f \equiv \hat{T}_{f(\mathcal{E}_1,\mathcal{E}_2,\cdots,\mathcal{E}_k)}$ of the transformed process $f(\mathcal{E}_1,\mathcal{E}_2,\cdots,\mathcal{E}_k)$ directly from these single-channel estimated transfer matrices.

To achieve this, we present two mathematically equivalent but conceptually disinct approaches.
The first approach is a constructive method rooted in the element-wise definition of the transfer matrix, offering intuitive interpretability. 
The second approach leverages global algebraic relations, providing a more efficient route for practical implementation. 
Both methods converge to the same transfer matrix $T_f$, and we elaborate on each approach in the following.

\subsubsection{Element-wise construction from the definition}
This approach constructs \(T_f\) element-wise directly from its definition:
\[
(T_f)_{ij} = \left( \frac{1}{3} \right)^{n} \operatorname{Tr}\left[ f(\mathcal{E}_1,\ldots,\mathcal{E}_k)(\mathcal{S}^{\otimes n}_i) \cdot \mathcal{S}^{\otimes n}_j \right].
\]
The key step consists in explicitly rewriting \( f(\mathcal{E}_1,\ldots,\mathcal{E}_k)(\mathcal{S}^{\otimes n}_i) \) or its adjoint $f^{\dagger}(\mathcal{E}_1,\ldots,\mathcal{E}_k)(\mathcal{S}^{\otimes n}_j)$ in terms of the single-channel output states \(\mathcal{E}_m(\mathcal{S}^{\otimes n}_l)\), using the substitution \(\mathcal{E}_m(\mathcal{S}^{\otimes n}_l) = \sum_p (T_{\mathcal{E}_m})_{lp} \mathcal{S}^{\otimes n}_p\).

To illustrate the element-wise construction in a concrete and rigorous manner, we analyze the important case of the adjoint channel and formalize this result in the following lemma.

\begin{lemma}[Transfer matrix relation of $\mathcal{E}_1$ and $\mathcal{E}^{\dagger}_1$]\label{lemma1} 
Let $T_{\mathcal{E}_1}$ and $T_{\mathcal{E}_1^{\dagger}}$ denote the transfer matrices of quantum processes $\mathcal{E}_1$ and  $\mathcal{E}_1^{\dagger}$, respectively. Then
\begin{equation}
T_{\mathcal{E}_1^{\dagger}}  = {T_{\mathcal{E}_1}}^{\dagger}
\end{equation}
\end{lemma}
\begin{proof}
Let $\mathcal{S}^{\otimes n}_i$ and $\mathcal{S}^{\otimes n}_j  $ represent the \(i\)-th and \(j\)-th elements of \(\mathcal{S}^{\otimes n}\), respectively. 
By definition, the $(i,j)$-th element of transfer matrix for \(\mathcal{E}_1\) is:
\begin{equation}
    (T_{\mathcal{E}_1})_{ij} = \left( \frac{1}{3} \right)^n \text{Tr}[ 
      \mathcal{E}_1\left( \mathcal{S}^{\otimes n}_i \right) \cdot \mathcal{S}^{\otimes n}_j 
  ].
\end{equation}
For \(\mathcal{E}_1^\dagger\), the $(j,i)$-th element of its transfer matrix satisfies:
\begin{equation}
    (T_{\mathcal{E}_1^{\dagger}})_{ji} = \left( \frac{1}{3} \right)^n \text{Tr}[ 
      \mathcal{E}_1^{\dagger}\left( \mathcal{S}^{\otimes n}_j \right) \cdot \mathcal{S}^{\otimes n}_i 
  ].
\end{equation}
Furthermore, the cyclic property of the trace gives:
\begin{equation}
    \left( \frac{1}{3} \right)^n \text{Tr}[ 
      \mathcal{E}_1^{\dagger}\left( \mathcal{S}^{\otimes n}_j \right) \cdot \mathcal{S}^{\otimes n}_i 
  ] = \left( \frac{1}{3} \right)^n \text{Tr}[ 
      \mathcal{E}_1\left( \mathcal{S}^{\otimes n}_i \right) \cdot \mathcal{S}^{\otimes n}_j 
  ].
\end{equation}
This implies:
\begin{equation}
(T_{\mathcal{E}_1})_{ij} = (T_{\mathcal{E}_1^\dagger})_{ji},
\end{equation}
By definition of the adjoint matrix, we thus conclude:
\begin{equation}
T_{\mathcal{E}_1^\dagger} = {T_{\mathcal{E}_1}}^{\dagger}.
\end{equation}
\end{proof}

\subsubsection{Algebraic manipulation}
This approach leverages the global vectorized relation from Eq.~\eqref{eq3} applied to the composite process, yielding
\begin{equation}\label{vectorized_composite_main}
T_f \, \mathcal{M}^{-1} \circledast (\boldsymbol{\sigma}_{\mathcal{S}^{\otimes n}}) = f(\mathcal{E}_1,\cdots,\mathcal{E}_k) \circledast (\boldsymbol{\sigma}_{\mathcal{S}^{\otimes n}}),
\end{equation}
Since the function \(f\) is constructed from standard operations on the individual channels \(\{\mathcal{E}_m\}_{m=1}^k\) such as composition or linear combination, the right-hand side can be fully expressed in terms of the elementary vectors \(\mathcal{E}_m \circledast (\boldsymbol{\sigma}_{\mathcal{S}^{\otimes n}})\). 
Using the substitution \(\mathcal{E}_m \circledast (\boldsymbol{\sigma}_{\mathcal{S}^{\otimes n}}) = T_{\mathcal{E}_m} \, \mathcal{M}^{-1} \circledast (\boldsymbol{\sigma}_{\mathcal{S}^{\otimes n}})\), we obtain
\[
f(\mathcal{E}_1,\cdots,\mathcal{E}_k) \circledast (\boldsymbol{\sigma}_{\mathcal{S}^{\otimes n}}) = H(T_{\mathcal{E}_1},\dots,T_{\mathcal{E}_k}) \, \mathcal{M}^{-1} \circledast (\boldsymbol{\sigma}_{\mathcal{S}^{\otimes n}}),
\]
where \(H\) denotes a matrix function uniquely determined by the form of \(f\).

By equating the two expressions and using the fact that \(\mathcal{M}^{-1} \circledast (\boldsymbol{\sigma}_{\mathcal{S}^{\otimes n}})\) spans the relevant space, we arrive at a general construction rule:
\begin{equation}\label{eq:construction_rule_main}
T_f = H(T_{\mathcal{E}_1},\dots,T_{\mathcal{E}_k}).
\end{equation}
In other words, the transfer matrix of the composite process is obtained by applying the function \(H\) directly to the estimated transfer matrices of the constituent processes.

To illustrate the practical application of this approach, 
we next consider the concatenation of two quantum channels, which yields a similarly clean transfer matrix relation, formalized in the following lemma.

\begin{lemma}[Transfer matrix relation of $\mathcal{E}_1,\mathcal{E}_2$ and $\mathcal{E}_2 \circ \mathcal{E}_1$]\label{lemma2}
Let $\mathcal{E}_1$ and $\mathcal{E}_2$ be quantum channels with transfer matrices $T_{\mathcal{E}_1}$ and $T_{\mathcal{E}_2}$, respectively.
Let $f(\mathcal{E}_1,\mathcal{E}_2)=\mathcal{E}_2\circ\mathcal{E}_1$ denote their concatenation, and let $\mathcal{T}$ be the fixed sparse matrix satisfying
\[
\mathcal{M}^{-1}\circledast(\boldsymbol{\sigma}_{\mathcal{S}^{\otimes n}}) = \mathcal{T}\cdot\boldsymbol{\sigma}_{\mathcal{S}^{\otimes n}}
\]
for state vector $\boldsymbol{\sigma}_{\mathcal{S}^{\otimes n}}$.
Then the transfer matrix of the concatenated channel satisfies
\[
T_{f(\mathcal{E}_1,\mathcal{E}_2)}
=
T_{\mathcal{E}_1}\cdot\mathcal{T}\cdot T_{\mathcal{E}_2}.
\]
\end{lemma}

\begin{proof}
By definition of the concatenated process,
\begin{align*}
f(\mathcal{E}_1, \mathcal{E}_2) \circledast (\boldsymbol{\sigma}_{\mathcal{S}^{\otimes n}})
&= (\mathcal{E}_2 \circ \mathcal{E}_1) \circledast (\boldsymbol{\sigma}_{\mathcal{S}^{\otimes n}}) \\
&= \mathcal{E}_2 \left( \mathcal{E}_1 \circledast (\boldsymbol{\sigma}_{\mathcal{S}^{\otimes n}}) \right)\\
&= \mathcal{E}_2 \left( T_{\mathcal{E}_1} \, \mathcal{M}^{-1} \circledast (\boldsymbol{\sigma}_{\mathcal{S}^{\otimes n}}) \right)
\end{align*}
By linearity of the map $\mathcal{M}^{-1}\circledast(\cdot)$, there exists a fixed sparse matrix $\mathcal{T}$ such that
\[
\mathcal{M}^{-1}\circledast(\boldsymbol{\sigma}_{\mathcal{S}^{\otimes n}}) = \mathcal{T}\cdot\boldsymbol{\sigma}_{\mathcal{S}^{\otimes n}}
\]
Continuing the identity yields
\begin{align*}
f(\mathcal{E}_1, \mathcal{E}_2) \circledast (\boldsymbol{\sigma}_{\mathcal{S}^{\otimes n}})
&= \mathcal{E}_2 \left( T_{\mathcal{E}_1} \, \mathcal{T} \boldsymbol{\sigma}_{\mathcal{S}^{\otimes n}} \right) \\
&= T_{\mathcal{E}_1} \, \mathcal{T} \left( \mathcal{E}_2 \boldsymbol{\sigma}_{\mathcal{S}^{\otimes n}} \right) \\
&= T_{\mathcal{E}_1} \, \mathcal{T} T_{\mathcal{E}_2} \mathcal{M}^{-1} \circledast (\boldsymbol{\sigma}_{\mathcal{S}^{\otimes n}})
\end{align*}
Comparing with the defining vectorized relation~\eqref{vectorized_composite_main} for the composite process $f$, we conclude
\[
T_{f(\mathcal{E}_1,\mathcal{E}_2)}
=
T_{\mathcal{E}_1}\cdot\mathcal{T}\cdot T_{\mathcal{E}_2}.
\]
\end{proof}

Together, these two element-wise and algebraic approaches establish a complete, unified framework for constructing the transfer matrix of arbitrary transformed quantum processes directly from the transfer matrices of their constituent channels.

\subsection{Predicting $\text{Tr}[O\cdot f(\mathcal{E}_1,\mathcal{E}_2,\cdots,\mathcal{E}_k)\rho]$ via transfer matrix of the transformed process}
In this section, we describe how to evaluate the expectation value $\text{Tr}[O\cdot f(\mathcal{E}_1,\mathcal{E}_2,\cdots,\mathcal{E}_k)\rho]$ from the transfer matrix of transformed process $f(\mathcal{E}_1,\mathcal{E}_2,\cdots,\mathcal{E}_k)$, given an observable $O$ and an input state $\rho$,
and provided rigorous theoretical justification for each step of the procedure.

Once the transfer matrix $T_f$ of the composite process $f(\mathcal{E}_1,\mathcal{E}_2,\cdots,\mathcal{E}_k)$ is constructed, shadow engineering enables efficient prediction of the expectation value $\text{Tr}[Of(\mathcal{E}_1,\mathcal{E}_2,\cdots,\mathcal{E}_k)(\rho)]$.
This guarantee holds for bounded-degree observables $O$ with $\|O\|=\mathcal{O}(1)$, as well as for $n$-qubit states $\rho$ drawn from a wide class of distributions.
This class comprises states that are either locally flat--invariant under single-qubit Clifford gates acting on any qubit--or polynomially close to locally flat as quantified by maximum likelihood ratio. 
Examples of such distributions include random product states, ground or thermal states of random local Hamiltonians, and circuit output states with final random single-qubit Clifford layers.

In Alg.~\ref{Alg1}, we detail the procedure for computing the low-degree approximation $h(\rho,O)\approx \text{Tr}[O\cdot f(\mathcal{E}_1,\mathcal{E}_2,\cdots,\mathcal{E}_k)(\rho)]$ from the transfer matrix $T_f$, given $\rho$ and $O$.
\begin{algorithm}[t]
  \SetAlgoLined
  \SetAlgoNoEnd
  \SetAlgoHangIndent{0em}
  \SetInd{1em}{1em}
  \SetKwInOut{Input}{Input}\SetKwInOut{Output}{Output}
  \Input{Transfer matrix \(\hat{T}_{f}\) of quantum process $f(\mathcal{E}_1,\mathcal{E}_2,\cdots,\mathcal{E}_k)$, hyperparameter \(\tilde{\epsilon}\), observable \(O = \sum_{Q \in \{I,X,Y,Z\}^{\otimes n}} a_Q Q\), classical shadow \(\hat{\rho}\) of input state $\rho$}
  \pagebreak[3]
  \Output{Low - degree approximation \(h(\rho, O)\) to expectation $\text{Tr}[O(f(\mathcal{E}_1,\mathcal{E}_2,\cdots,\mathcal{E}_k))(\rho)]$}
  Initialize \(\tilde{\epsilon}\), \(\kappa = \Theta(\log(1/{\epsilon}))\)\;
  \For{each Pauli observable \(P \in \{I,X,Y,Z\}^{\otimes n}\) with \(|P| \leq \kappa\)}{
     \(\hat{x}_P(O) := \frac{1}{|\text{supp}(\hat{T}_f)|}
   \sum_{i\in\text{supp}(\hat{T}_f)}
   \mathrm{Tr}\bigl(P\,\mathcal{S}^{\otimes n}_i\bigr) \phantom{=}\times
   \mathrm{Tr}\Bigl(
   O\sum_{j:(\hat{T}_f)_{ij}\neq 0}
   (\hat{T}_f)_{ij}\,
   \mathcal{M}^{-1}\bigl(\mathcal{S}^{\otimes n}_j\bigr)
   \Bigr)\);

    \(\hat{\alpha}_P(O) := 
    \begin{cases}
    3^{|P|} \hat{x}_P(O),&\text{if } |\hat{x}_P(O)| > 2 \cdot 3^{|P|/2} \sqrt{\tilde{\epsilon}} \sum_{\substack{Q: \\ a_Q \neq 0}} |a_Q|  \quad \text{and } \left(\frac{1}{3}\right)^{|P|} > 2 \tilde{\epsilon}, \\
    0, & \text{otherwise}
    \end{cases}\)
  }
  \(h(\rho, O) := \sum_{P: |P| \leq \kappa} \hat{\alpha}_P(O) \text{Tr}(P\hat{\rho})\);   \footnotemark[1]
  \caption{Post-processing algorithm of shadow engineering}
  \label{Alg1}
\end{algorithm}

\footnotetext[1]{
\begin{spacing}{0.5}
{\footnotesize 
\begin{itemize}
  \setlength{\itemsep}{-0.5ex}
  \item[*] $\|O\| \leq 1$ that is written as a sum of few-body observables, where each qubit is acted on by $O(1)$ of them.
  \item[*] Hyperparameter $\tilde{\epsilon}>0$ scales inverse polynomially in the classical shadows size $N$.
  \item[*] \( \operatorname{supp}(\hat{T}_{f}) \) denotes the set of indices of non-zero rows in \( \hat{T}_{f} \).
  \end{itemize}
}
\end{spacing}}

This approximation is formulated within the Heisenberg picture of quantum evolution, where the prediction is obtained by computing the Pauli coefficients $\alpha_P$ of the observable evolved under the adjoint transformed process:
\begin{equation}
    f^{\dagger}(\mathcal{E}_1,\mathcal{E}_2,\cdots,\mathcal{E}_k)(O) =\sum_{P\in \{I, X, Y, Z\}^{\otimes n}} \alpha_P(O)P, 
\end{equation}
The construction is justified by the fundamental identity $\text{Tr}[O(f(\mathcal{E}_1,\mathcal{E}_2,\cdots,\mathcal{E}_k)(\rho))] = \text{Tr}[f^{\dagger}(\mathcal{E}_1,\mathcal{E}_2,\cdots,\mathcal{E}_k)(O)\cdot\rho]$,
which relates expectation values in the Schr\"odinger and Heisenberg representations.

Directly computing all $\mathcal{O}(4^n)$ Pauli coefficients incurs exponential cost in the system size $n$.
To overcome this issue, we employ a $\kappa$-truncation strategy that drastically reduces the number of coefficients to be evaluated, while bounding the approximation error by leveraging the locally flat structure of the state distribution.
The validity of this truncation is formalized by the following lemma:
\begin{lemma}[Truncation error bound, Corollary 13 of Ref.~\cite{huang2023learning}]\label{Lemma:lowdegree}
Let $O=\sum_{P\in \{I, X, Y, Z\}^{\otimes n}}\alpha_PP$ is an $n$-qubit observable, and let $\mathcal{D}$ be a distribution over quantum states invariant under single-qubit $H$ and $S$ gates. 
For $\kappa \geq 0$, define the truncated observable $O^{(\kappa)}=\sum_{P:|P|\textless \kappa} \alpha_P P$. The squared truncation error satisfies:
\begin{equation}
    \mathbb{E}_{\rho\sim\mathcal{D}}|\text{Tr}(O\rho)-\text{Tr}(O^{(\kappa)}\rho)|^2\leq\left(\frac{2}{3}\right)^{\kappa}\|O\|^2.
\end{equation}
\end{lemma}
Since $\| f^{\dagger}(\mathcal{E}_1,\mathcal{E}_2,\cdots,\mathcal{E}_k)(O)\|\leq \|O\|=\mathcal{O}(1)$, the $\kappa$-truncation yields a constant-error approximation of the evolved observable:
\begin{equation}
    f^{\dagger}(\mathcal{E}_1,\mathcal{E}_2,\cdots,\mathcal{E}_k)(O) \approx \sum_{P\in\{I,X,Y,Z\}^{\otimes n}: |P|\leq \kappa}\alpha_P(O)P
\end{equation}
This reduces the number of Pauli coefficients to compute from $\mathcal{O}(4^n)$ to $\mathcal{O}(n^\kappa)$, enabling efficient prediction of the target expectation value. 

To compute the coefficients $\alpha_P$ for $|P| \leq \kappa$, we utilize the estimated transfer matrix $\hat{T}_{f}$ and the following foundational result:
\begin{lemma}[Lemma 16 of Ref.~\cite{huang2023learning}]\label{alpha_p}
Let $O=\sum_{P\in \{I, X, Y, Z\}^{\otimes n}}\alpha_PP$ be an $n$-qubit observable, and let $\mathcal{D}^0$ denote the uniform distribution over the set $\mathcal{S}^{\otimes n}$. 
For any Pauli observable $P\in \{I, X, Y, Z\}^{\otimes n}$, the expectation value satisfies:
\begin{equation}
    \mathbb{E}_{\rho\sim\mathcal{D}^0}\text{Tr}[O\rho]\text{Tr}[P\rho]=\left(\frac{1}{3}\right)^{|P|}\alpha_P.
\end{equation}
\end{lemma}
By applying this lemma to the Heisenberg-evolved observable $f^{\dagger}(\mathcal{E}_1,\mathcal{E}_2,\cdots,\mathcal{E}_k)(O)$, we derive an explicit expression for its Pauli coefficients:
\begin{equation}
    \alpha_P(O) = 3^{|P|} \times x_P(O),
\end{equation} 
where $x_P(O) = \mathbb{E}_{\rho\sim\mathcal{D}^0}\text{Tr}[f^{\dagger}(\mathcal{E}_1,\mathcal{E}_2,\cdots,\mathcal{E}_k)(O)\cdot\rho]\text{Tr}[P\rho]$.

Using the estimated transfer matrix $\hat{T}_f$--whose rows correspond to input states uniformly sampled from $\mathcal{S}^{\otimes n}$--we approximate $\text{Tr}[f^\dagger(\mathcal{E}_1,\cdots,\mathcal{E}_k)(O) \cdot \rho] = \text{Tr}[O f(\mathcal{E}_1,\cdots,\mathcal{E}_k)(\rho)]$ and reconstruct each coefficient via:
\begin{equation}
\begin{aligned}
    \hat{\alpha}_P(O)=3^{|P|}\times \hat{x}_P(O),
\end{aligned}
\end{equation}
where the empirical estimate $\hat{x}_P(O)$ is given by:
\begin{equation}
    \hat{x}_P(O) = \frac{1}{|\text{supp}(\hat{T}_f)|}\sum_{i\in\text{supp}(\hat{T}_f)}\text{Tr}\big[P\mathcal{S}^{\otimes n}_i\big]\text{Tr}\bigg[O\sum_{j:(\hat{T}_f)_{ij}\neq 0}(\hat{T}_f)_{ij}\mathcal{M}^{-1}(\mathcal{S}^{\otimes n}_j)\bigg]
\end{equation}

To avoid noise amplification from small coefficients with negligible influence, we apply a filtering strategy based on the following lemma:
\begin{lemma}[Coefficient filtering, adapted from Lemma 18 of Ref.~\cite{huang2023learning}]\label{lem:filtering_kappa}
Let $\tilde{\epsilon}, \eta > 0$ and define $S_\kappa = \{P \in \{I, X, Y, Z\}^{\otimes n} : |P| \leq \kappa\}$. 
Suppose for all $P \in S_\kappa$ we have $\alpha_P \in [-\eta, \eta]$, and $x_P = 3^{-|P|} \cdot \alpha_P  \in [-\eta, \eta]$. Assume there exist $A > 0$ and $1 \leq r < 2$ such that:
\begin{equation}
\sum_{P \in S_\kappa} |\alpha_P|^r \leq A^r.
\label{eq:alpha_sum_bound}
\end{equation}

Given estimates $\hat{x}_P$ with $|\hat{x}_P - x_P| < \eta \tilde{\epsilon} $ for all $P \in S_\kappa$, define the filtered coefficients:
\begin{equation}\label{rule}
\hat{\alpha}_P =
\begin{cases} 
0, & 3^{-|P|} \leq 2\tilde{\epsilon} \\
0, & 3^{-|P|} > 2\tilde{\epsilon},\  3^{|P|/2} |\hat{x}_P|   \leq 2\eta \sqrt{\tilde{\epsilon}} \\
3^{|P|} \hat{x}_P , & 3^{-|P|}  > 2\tilde{\epsilon},\  3^{|P|/2} |\hat{x}_P| > 2\eta \sqrt{\tilde{\epsilon}}
\end{cases}.
\end{equation}

Then the following error bounds hold:
\begin{equation}
\sum_{P \in S_\kappa} 3^{-|P|} |\hat{\alpha}_P - \alpha_P|^2 \leq 6A^r \eta^{2-r} \tilde{\epsilon}^{1-(r/2)}.
\label{eq:global_error}
\end{equation}

\end{lemma}

By applying the filtering rule from Lem.~\ref{lem:filtering_kappa} (Eq.~\ref{rule}), we obtain the filtered coefficients $\alpha_P(O)$ of $f^{\dagger}(\mathcal{E}_1,\mathcal{E}_2,\cdots,\mathcal{E}_k)(O)$.

For the $\mathcal{O}(n^{\kappa})$ Pauli operators $P$ with $|P|\leq \kappa$, computing each coeddicient $\alpha_P(O)$ involves two sequential steps: 
First, we evaluate $\text{Tr}(P\mathcal{S}^{\otimes n}_i)$ for each nonzero row $i$ of $\hat{T}_f$, with each evaluation taking an average of $\mathcal{O}(\kappa)$ time. 
Second, for each row $i$, we compute the term $\text{Tr}\bigg[O\sum_{j:(\hat{T}_f)_{ij}\neq 0}(\hat{T}_f)_{ij}\mathcal{M}^{-1}(\mathcal{S}^{\otimes n}_j)\bigg]$, which contributes a total time complexity of $\mathcal{O}(N)$. 

Accordingly, the total computation cost to compute all filtered coefficents $\alpha_P(O)$ (for $|P|\leq \kappa$) is $\mathcal{O}(\kappa n^{\kappa}|\text{supp}(\hat{T}_f)|)$.

Finally, we leverage the classical shadow $\hat{\rho}$ of the state $\rho$--which enables efficient access to its $\kappa$-body reduced density matrices and is generated via random Pauli measurements~\cite{huang2020predicting,elben2023randomized}--to approximate the expectation value as:
\begin{equation}
  h(\rho,O) = \sum_{P\in\{I,X,Y,Z\}^{\otimes n}: |P|\leq \kappa}\hat{\alpha}_P(O)\text{Tr}[P\hat{\rho}]
\end{equation}
This gives an efficient classical predictor $h(\rho,O)$ for $\mathrm{Tr}\left[O\left(f(\mathcal{E}_1,\mathcal{E}_2,\cdots,\mathcal{E}_k)(\rho)\right)\right]$.
With proper implementation, the overall computational complexity of shadow engineering is $\mathcal{O}(\kappa n^{\kappa}|\text{supp}(\hat{T}_f)|)$.

\section{Two key cases of shadow engineering}

This section illustrates the shadow engineering framework through two key cases of quantum process transformation: process adjoint ($\mathcal{E}_1^{\dagger}$) and process concatenation ($\mathcal{E}_2 \circ \mathcal{E}_1$). 
For each case, we elaborate on the construction of the corresponding transfer matrix and analyze the scaling behavior of the prediction error.

\subsection{Shadow engineering for predicting $\text{Tr}[O\mathcal{E}_1^{\dagger}(\rho)]$}
We first consider the case $f(\mathcal{E}_1) = \mathcal{E}_1^{\dagger}$.
By Lem.~\ref{lemma1}, the transfer matrices of $\mathcal{E}_1$ and $\mathcal{E}_1^{\dagger}$ satisfy
\[
T_{\mathcal{E}_1^\dagger} = T_{\mathcal{E}_1}^{\dagger}.
\]

While Lem.~\ref{lemma1} establishes the adjoint relation for ideal transfer matrices, we cannot directly estimate $T_{\mathcal{E}^{\dagger}_1}$ by taking the adjoint of the empirical estimate ${\hat{T}_{\mathcal{E}_1}}^{\dagger}$--this is because ${\hat{T}_{\mathcal{E}_1}}^{\dagger}$ may fail to satisfy the property that the sum of each row equals 1. 
Under the uniform input distribution over $\mathcal{S}^{\otimes n}$, the transfer matrix $T_{\mathcal{E}}$ is doubly stochastic (rows and columns each sum to 1), which guarantees that the distribution of measurement outcomes is also uniform.

To construct a valid estimator for $T_{\mathcal{E}^{\dagger}}$, we first build a counting matrix $T_{(0)}$ from the classical shadows dataset $S_{N_1}(\mathcal{E}_1)\triangleq\left\{\mathcal{S}^{\otimes n}_{i_{\ell}},\mathcal{S}^{\otimes n}_{j_{\ell}}\right\}_{\ell=1}^{N_1}$. 
Sepcifically, we initialize a $6^n \times 6^n$ matrix to zero, then increment the $(i_{\ell},j_{\ell})$-th entry by 1 for each pair $(i_{\ell},j_{\ell})\in S_{N_1}(\mathcal{E}_1)$, yielding $T_{(0)}$.
The matrix ${T_{(0)}}$ empirically captures the row and column probability distributions of $T_{\mathcal{E}_1}$. 
Since ${T_{\mathcal{E}_1}}^{\dagger} = T_{\mathcal{E}_1^{\dagger}}$ by Lem.~\ref{lemma1}, the transposed matrix ${T_{(0)}}^{\dagger}$ aligns with the row and column distributions of $T_{\mathcal{E}_1^{\dagger}}$. 
We therefore obtain a valid estimator $\hat{T}_{\mathcal{E}_1^{\dagger}}$ for $T_{\mathcal{E}_1^{\dagger}}$ by performing row normalization on ${T_{(0)}}^{\dagger}$.

In summary, using a classical shadows $S_{N_1}(\mathcal{E}_1)$ of size $N_1$, we compute the estimated transfer matrix $\hat{T}_{\mathcal{E}_1^{\dagger}}$. 
Within the shadow engineering framework, this enables us to approximate $h(\rho,O)\approx \text{Tr}[O\mathcal{E}_1^{\dagger}(\rho)]$ for any $n$-qubit state $\rho \sim \mathcal{D}$ and any bounded-degree observable $O$.

To quantify the performance of this estimator, we characterize the prediction error explicitly, with the following theorem establishing a rigorous error bound that scales with the dataset size  $N_1$.

\begin{theorem}[Prediction error for estimating $\text{Tr}(O\mathcal{E}_1^{\dagger}(\rho))$]
 Let $n\ge1$ and $\epsilon, \epsilon', \delta > 0$. For any unknown $n$-qubit CPTP map $\mathcal{E}_1$, 
  and its classical shadows $S_{N_1}(\mathcal{E}_1)$ obtained by $N_1$ randomized experiments with
\begin{equation}
    N_1 = \log(\frac{n}{\delta}) \text{min}(2^{O[\log(1/\epsilon)(\log\log(1/\epsilon) + \log(1/\epsilon^{\prime}))]}, 2^{O\left[\log(1/\epsilon) \log(n)\right]}).
\end{equation}
Then, with probability at least $1 - \delta$, 
For any $n$-qubit state $\rho \sim \mathcal{D}$ and any bounded-degree observable $O$, 
the estimate $h(\rho,O)$ satisfies
\begin{equation}
    \mathbb{E}_{\rho \sim \mathcal{D}} \left| h(\rho, O) - \text{Tr}[O\mathcal{E}_1^{\dagger}(\rho)] \right|^2 \leq \left( \epsilon + \epsilon' \left[ \frac{\|O^{(\text{unk},\kappa)}\|}{\|O\|} \right]^{2  \left\lceil \log_{1.5}(1/\epsilon) \right\rceil/ \left[\left\lceil  \log_{1.5}(1/\epsilon)\right\rceil  + 1\right]  } \right) \|O\|^2.
\end{equation}
Here, $O^{(\mathrm{unk},\kappa)}$ denotes the low-degree truncation of $\mathcal{E}_1(O)$.
\end{theorem}
\begin{proof}
    We begin by considering a bounded-degree observable defined as:
    \begin{equation}
    O = \sum_{Q \in \{I, X, Y, Z\}^{\otimes n}, |Q| \leq \nu} a_Q Q,
    \end{equation}
    
    Our goal is to compute the Pauli coefficients of the Heisenberg-evolved observable under $\mathcal{E}_1^{\dagger}$:
    \begin{equation}
    O^{(\mathrm{unk})}  \triangleq \mathcal{E}_1(O) = \sum_{P \in \{I, X, Y, Z\}^{\otimes n}} \alpha_P(O) P.
    \end{equation}

    To circumvent the exponential cost of computing all coefficients, we employ a $\kappa$-truncation to $O^{(\mathrm{unk})}$:
    \begin{equation}
    O^{(\mathrm{unk},\kappa)} \triangleq \sum_{{P \in \{I, X, Y, Z\}^{\otimes n} ,|P| \leq \kappa}} \alpha_P(O) P.
    \end{equation}
    We analyze the prediction error for input states $\rho$ drawn from the $\mathcal{D}$ distribution, aiming to bound:
    \begin{equation}
    \mathop{\mathbb{E}}_{\rho \sim \mathcal{D}} \left|h(\rho,O) - \text{Tr}[O^{(\mathrm{unk})}\rho]\right|^2.
    \end{equation}
    Applying the triangle inequality, we decompose this error into two components:
    \begin{equation}
        \mathop{\mathbb{E}}_{\rho \sim \mathcal{D}} \left|h(\rho,O) - \text{Tr}[O^{(\mathrm{unk})}\rho]\right|^2 \leq   \underbrace{\mathop{\mathbb{E}}_{\rho \sim \mathcal{D}} \left|\text{Tr}[O^{(\mathrm{unk},\kappa)}\rho] - \text{Tr}[O^{(\mathrm{unk})}\rho]\right|^2}_{\text{(i) Truncation error}} + \underbrace{\mathop{\mathbb{E}}_{\rho \sim \mathcal{D}} \left|h(\rho,O) - \text{Tr}[O^{(\mathrm{unk},\kappa)}\rho]\right|^2}_{\text{(ii) Estimation error}}
    \end{equation}
    These correspond to:
    \begin{enumerate}
        \item \textbf{Truncation error:} due to approximating $O^{(\text{unk})}$ by its low-weight part $O^{(\text{unk},\kappa)}$,
        \item \textbf{Estimation error:} arising from the finite-sample estimation of the coefficients $\alpha_P(O)$ for $|P|\leq \kappa$.
    \end{enumerate}

    We use Lem.~\ref{Lemma:lowdegree} to bound the truncation error and the following lemma to bound the estimation error.
    \begin{lemma}[Estimation error bound, adapted from Lemma 15 of Ref.~\cite{huang2023learning}] \label{Lemma:estimation error}
    Given two $n$-qubit observable $O_1,O_2$ with 
    \begin{equation}
        O_1 -O_2 = \sum_{P\in \{I, X, Y, Z\}^{\otimes n}}\Delta\alpha_PP
    \end{equation}and a distribution $\mathcal{D}$ over quantum states that is invariant under single-qubit $H$ and $S$ gates, we have 
     \begin{equation}
        \mathop{\mathbb{E}}_{\rho \sim \mathcal{D}} \left|\text{Tr}[O_1\rho]-\text{Tr}[O_2\rho]\right|^2\leq \sum_{P\in \{I, X, Y, Z\}^{\otimes n}}  \mathop{\mathbb{E}}_{\rho \sim D} [ \gamma^*(\rho_{\mathrm{dom}(P)}) ]\left(\frac{2}{3}\right)^{|P|}|\Delta\alpha_P|^2.
    \end{equation}
    where $\gamma^*(\rho_{\mathrm{dom}(P)})$ denotes the \emph{nonlinear purity} in the subsystem of size $|P|$ (domain of $P$), defined as
    \begin{equation}
        \gamma^*(\rho_{\mathrm{dom}(P)}) \coloneqq \frac{1}{2^{|P|}} \sum_{Q \in \{X, Y, Z\}^{\otimes |P|}} \operatorname{tr}[Q\rho]^2.
    \end{equation}
    Here, $|P|$ is the weight of $P$ (number of non-identity terms) and $\mathrm{dom}(P)$ is the subsystem where $P$ acts non-trivially.
    \end{lemma}
    We first bound the truncation error introduced by approximating $\mathcal{E}_1(O)$ with its low-degree component. 
    Applying Lem.~\ref{Lemma:lowdegree} to $O^{(\mathrm{unk})}$, we obtain for $\rho \sim \mathcal{D}$:
    \begin{equation}
        \mathop{\mathbb{E}}_{\rho \sim \mathcal{D}} \left|\mathrm{Tr}[O^{(\mathrm{unk})}\rho] - \mathrm{Tr}[O^{(\mathrm{unk},\kappa)}\rho]\right|^2 \leq \left(\frac{2}{3}\right)^{\kappa}\|O^{(\mathrm{unk})}\|^2.
    \end{equation}

    Since $\mathcal{E}_1$ is a CPTP map, it is contractive in the operator norm, implying:
    \begin{equation}
        \|O^{(\mathrm{unk})}\| = \|\mathcal{E}_1(O)\| \leq \|O\|.
    \end{equation} 
    
    To control the truncation error, we choose $\kappa = \left\lceil \log_{1.5} \left( {1}/{\epsilon} \right) \right\rceil$ to obtain:
    \begin{equation}\label{leq:truncation_error0}
        \mathop{\mathbb{E}}_{\rho \sim \mathcal{D}} \left|\mathrm{Tr}[O^{(\mathrm{unk})}\rho] - \mathrm{Tr}[O^{(\mathrm{unk},\kappa)}\rho]\right|^2 \leq \epsilon \|O\|^2.
    \end{equation}
    where $\epsilon$ is the truncation error parameter. 
    
    Next, we analyze the estimation error arising from the finite-sample approximation for the Pauli coefficients $\alpha_P$ (for $|P| \leq \kappa$) of $O^{(\mathrm{unk},\kappa)}$.
    
    Building on Lem.~\ref{alpha_p}, we define the core quantity for coefficient estimation:
    \begin{equation}
            x_{P}(O)=\mathbb{E}_{\rho\sim\mathcal{D}^{0}}\text{Tr}[P\rho]\text{Tr}[\mathcal{E}_1(O)\rho],
        \end{equation}
    where $\mathcal{D}^0$ is the uniform distribution over $\mathcal{S}^{\otimes n}$. Since the transfer matrix $\hat{T}_{\mathcal{E}_1^{\dagger}}$ is constructed from rows sampled uniformly from $\mathcal{S}^{\otimes n}$, its enables direct estimation of $x_P(O)$. 
    The Pauli coefficient is then recovered as  $\hat{\alpha}_P(O) = 3^{|P|} \hat{x}_P(O)$,
    where 
    \begin{equation}
        \hat{x}_P(O) = \frac{1}{|\text{supp}(\hat{T}_{\mathcal{E}_1^{\dagger}})|}\sum_{a\in\text{supp}(\hat{T}_{\mathcal{E}_1^{\dagger}})}\text{Tr}\big[P\mathcal{S}^{\otimes n}_a\big]\text{Tr}\bigg[O\sum_{l:(\hat{T}_{\mathcal{E}_1^{\dagger}})_{al}\neq 0}(\hat{T}_{\mathcal{E}_1^{\dagger}})_{al}\mathcal{M}^{-1}(\mathcal{S}^{\otimes n}_l)\bigg]
    \end{equation}
    
    To quantify the estimation error, we first analyze the variance properties of the empirical estimator $\hat{S}_a(Q)$ for Pauli observable $Q \in \{I, X, Y, Z\}^{\otimes n}$ with $|Q|\leq \nu =\mathcal{O}(1)$. 
    For the $a$-th state $\mathcal{S}^{\otimes n}_a $ ($a\in\text{supp}(\hat{T}_{\mathcal{E}_1^{\dagger}}$), the ideal value is $S_a(Q)=\text{Tr}[Q\mathcal{E}_1^{\dagger}(\mathcal{S}^{\otimes n}_a)]$, while the empirical estimator $\hat{S}_a(Q)$ is constructed from the estimated transfer matrix as: 
    \begin{equation}
        \hat{S}_a(Q)=\sum_{l\in \text{supp}((\hat{T}_{\mathcal{E}_1^{\dagger}})_a)}(\hat{T}_{\mathcal{E}_1^{\dagger}})_{al}\text{Tr}[Q\mathcal{M}^{-1}(\mathcal{S}^{\otimes n}_l)]=\frac{1}{N_{1(a)}}\sum_{\mathcal{S}^{\otimes n}_{i_\ell} \in \mathcal{G}_1(\mathcal{S}^{\otimes n}_a)}\text{Tr}[Q\mathcal{M}^{-1}(\mathcal{S}^{\otimes n}_{i_\ell})]
        \end{equation}
    Here, $\text{supp}((\hat{T}_{\mathcal{E}_1^{\dagger}})_a)$ is the set of indices with nonzero elements in the $a$-th row of the transfer matrix $\hat{T}_{\mathcal{E}_1^{\dagger}}$; $ \mathcal{G}_1(S^{\otimes n}_a) = \Big[ S^{\otimes n}_{i_\ell} \;\Big|\; (S^{\otimes n}_{i_\ell}, S^{\otimes n}_{a}) \in S_{N_1}(\mathcal{E}_1), \Big]$ is the collection of the first components in $S_{N_1}(\mathcal{E}_1)$ whose second component is $S^{\otimes n}_a$; and $N_{1(a)}=| \mathcal{G}_1(S^{\otimes n}_a)|$ indicates the number of classical shadows data pairs recorded in the $a$-th row of the transfer matrix $\hat{T}_{\mathcal{E}_1^{\dagger}}$.
    
    A critical bound emerges from the structure of the inverted channel $\mathcal{M}^{-1}$ and the Pauli observable $Q$. Since $Q = \bigotimes_{\iota=1}^n P_\iota$ has only $\mathcal{O}(1)$ non-identity terms, for $\mathcal{S}^{\otimes n}_{i_\ell}=\bigotimes_{\iota=1}^n |s_{i,\iota}\rangle\langle s_{i,\iota}|$, the trace $ \text{Tr}[Q\mathcal{M}^{-1}(\mathcal{S}^{\otimes n}_{i})]$ factorizes as:
    \begin{equation}
        \text{Tr}[Q\mathcal{M}^{-1}(\mathcal{S}^{\otimes n}_{i_\ell})] = \prod_{\iota=1}^n \text{Tr}\left[P_\iota (3|s_{i_\ell,\iota}\rangle\langle s_{i_\ell,\iota}| - I_\iota)\right].
    \end{equation}
    An analysis for different cases of $P_\iota$ is as follows:
    \begin{itemize}
    \item If any $P_\iota$ is $X$ or $Y$, then $\text{Tr}[Q\mathcal{M}^{-1}(\mathcal{S}^{\otimes n}_{i_\ell})] = 0$.
    \item If all non-identity terms are $Z$, then $\left| \text{Tr}[Q\mathcal{M}^{-1}(\mathcal{S}^{\otimes n}_{i_\ell})] \right| = 3^{\mathcal{O}(1)}$.
    \end{itemize}
    Therefore, for any $Q$ with $|Q| = \mathcal{O}(1)$, we have the uniform bound:
    \begin{equation}
        \left|\text{Tr}[Q\mathcal{M}^{-1}(\mathcal{S}^{\otimes n}_{i_\ell})] \right| \leq  3^{\mathcal{O}(1)} = \mathcal{O}(1).
    \end{equation}
    To establish the concentration of the empirical estimator $\hat{S}_a(Q)$, we apply Hoeffding's inequality. For each row $a$ of the transfer matrix $\hat{T}_{\mathcal{E}_1^{\dagger}}$, if the number of samples satisfies
    \begin{equation}
        N_{1(a)}=\Omega\left(\frac{\log(n^\nu/\delta)}{\tilde{\epsilon}^2}\right),
    \end{equation}
    then with probability at least $1-\delta$, for all Pauli observables $Q$ with $|Q| \leq \nu$,
    \begin{equation}
        |\hat{S}_a(Q)-{S}_a(Q)|\leq \tilde{\epsilon}.
    \end{equation}

    To ensure this bound holds over at most $N_1$ nonzero rows of $\hat{T}_{\mathcal{E}_1^{\dagger}}$, we apply a union bound. The total sample size $N_1$ must therefore satisfy
    \begin{equation}
        N_{1}=\Omega\left(\frac{\log(n^\nu N_{1}/\delta)}{\tilde{\epsilon}^2}\right),
    \end{equation}
    Under this condition,
    \begin{equation}
        |\hat{S}_a(Q)-{S}_a(Q)|\leq \tilde{\epsilon}.
    \end{equation}
    holds with a probability $1-\delta$ for at most $N_1$ nonzero rows for the transfer matrix $\hat{T}_{\mathcal{E}_1^{\dagger}}$.
    
    Next, we analyze the error in estimating 
    \begin{equation}
        \hat{x}_P(Q) = \frac{1}{|\text{supp}(\hat{T}_{\mathcal{E}_1^{\dagger}})|}\sum_{a\in \text{supp}(\hat{T}_{\mathcal{E}_1^{\dagger}})}\text{Tr}(P\mathcal{S}^{\otimes n}_a)\hat{S}_a(Q).
    \end{equation}
    
    For any $P$ with $|P|\leq \kappa$, we have $|\text{Tr}(P\mathcal{S}^{\otimes n}_a)|\leq 1$.
    Applying Hoeffding's inequality again, for all $P,Q\in \{I, X, Y, Z\}^{\otimes n} $ with $ |P|\leq \kappa$ and $|Q|\leq \nu$ we establish that the classical shadows size
    \begin{equation}\label{eq:number1}
        N_{1}=\Omega\left(\frac{\log(n^{\nu +\kappa}N_{1}/\delta)}{\tilde{\epsilon}^2}\right),
    \end{equation}
    guarantees
    \begin{equation}
        \left| \hat{x}_P(Q)-x_P(Q)\right|\leq \tilde{\epsilon}
    \end{equation}
    holds with probability at least $1-\delta$.

    Then, using the triangle inequality, we have
    \begin{equation}
        \left| \hat{x}_P(O)-x_P(O)\right|\leq \|O\|_{\text{Pauli},1}\tilde{\epsilon}=\eta \tilde{\epsilon}  
    \end{equation}
    for all $P\in \{I, X, Y, Z\}^{\otimes n} $ with $ |P|\leq \kappa$ .
    
    To apply coefficient filtering Lem.~\ref{lem:filtering_kappa}, we choose $r=2\kappa/(\kappa +1)\in(1,2)$, and seek an upper bound for $\|O^{(\mathrm{unk},\kappa)}\|_{\text{Pauli},r}$. This is provided by the following lemma.
    \begin{lemma}[Norm inequality for a $\kappa$-local observable $O_1$  and Norm inequality for a $\nu$ bounded-degree observable $O_2$]\label{lem:norm inequality}
        Given an $n$-qubit $\kappa$-local observable $O=\sum_{P \in \{I, X, Y, Z\}^{\otimes n}: |P| \leq \kappa} \alpha_P P$, we have
        \begin{equation}
        \sqrt[r]{\sum_{P : |P| \leq \kappa} |\alpha_P|^{r} }\leq \frac{3}{C(\kappa)} \|O\|,
        \label{eq:B11}
        \end{equation}
        where $r = 2\kappa/(\kappa+1) $ and $C(\kappa) = \sqrt{2(\kappa!)}/[2\kappa^{\kappa + 1.5+(\kappa+1)/(2\kappa)}(\sqrt{6} + 2\sqrt{3})^\kappa]$. 
        
        Given an $n$-qubit $\nu$-local observable $O_2$ with bounded degree $d$, we have 
        \begin{equation}
            \|O_2\|_{\text{Pauli,1}}\leq \frac{3}{C(\nu,d)}\|O_2\|
        \end{equation}
        $C(\kappa,d) = \sqrt{2(\kappa!)}/[\sqrt{d}\kappa^{\kappa + 2.5}(2\sqrt{6} + 4\sqrt{3})^\kappa]$. 
        This lemma follows from Corollaries 11 and 12 in Ref.~\cite{huang2023learning}.
    \end{lemma}
    Now, applying Lem.~\ref{lem:norm inequality} to $O^{(\mathrm{unk},\kappa)} = \sum_{P : |P| \leq \kappa} \alpha_P(O) P$, we obtain
    \begin{equation}
    \sum_{P : |P| \leq \kappa} |\alpha_P(O)|^{r} \leq \left(\frac{3}{C(\kappa)} \right)^{r}\|O^{(\mathrm{unk},\kappa)}\|^{r},
    \end{equation}
    where $C(\kappa)$ is defined in the Lemma above.

    Additionally, the lemma gives an upper bound for the Pauli-1 norm of $O$:
    \begin{equation}
    \eta = \|O\|_{\mathrm{Pauli},1} \leq \frac{3}{C(\nu,d)}\|O\|
    \end{equation}

    Combining these bounds via Lem.~\ref{lem:filtering_kappa}, we conclude that
    \begin{equation}
    \sum_{P : |P| \leq \kappa} \left(\frac{1}{3}\right)^{|P|} |\hat{\alpha}_P - \alpha_P|^2 \leq 6 \left(\frac{3}{C(\kappa)} \right)^{r}\|O^{(\mathrm{unk},\kappa)}\|^{r} \left(\frac{3}{C(\nu,d)}\right)^{2-r}\|O\|^{2-r} \tilde{\epsilon}^{1-r/2}
    \end{equation}
    
    To optimize the error scaling, we set the filtering parameter as: 
    \begin{equation}\label{eq:epsilon}
    \tilde{\epsilon} = \left( \frac{\epsilon'}{6 \cdot 2^{\kappa}} \right)^{\kappa+1} \left( \frac{C(\nu, d)}{3} \right)^2 \left( \frac{C(\kappa)}{3} \right)^{2\kappa},
    \end{equation}
    which leads to the refined bound:
    \begin{equation}
    \sum_{P : |P| \leq \kappa} \left(\frac{1}{3}\right)^{|P|} |\hat{\alpha}_P - \alpha_P|^2 \leq \left[ \frac{\|O^{(\mathrm{unk},\kappa)}\|}{\|O\|} \right]^{2\kappa/(\kappa+1)} \frac{\epsilon'}{2^\kappa} \|O\|^2.
    \end{equation}

    According to the definition of nonlinear purity in Lem.~\ref{Lemma:estimation error}, we have:
    \begin{equation}
        \mathbb{E}_{\rho \sim \mathcal{D}}[\gamma^{*}(\rho_{\mathrm{dom}(P)})] {\left(\frac{2}{3}\right)}^{|P|} \leq 2^{\kappa} \left(\frac{1}{3}\right)^{|P|}
    \end{equation}
    holds for all $P$ with $|P|\leq \kappa$ .
    
    Thus, combining this with Lem.~\ref{Lemma:estimation error}, we bound the estimation error as:
    \begin{equation}\label{eq:estimation_error}
    \mathop{\mathbb{E}}_{\rho \sim \mathcal{D}} \left|h(\rho,O) - \mathrm{Tr}(O^{(\mathrm{unk},\kappa)}\rho)\right|^2 \leq \sum_{P : |P| \leq \kappa}\mathbb{E}_{\rho \sim \mathcal{D}}[\gamma^{*}(\rho_{\mathrm{dom}(P)})] {\left(\frac{2}{3}\right)}^{|P|}|\hat{\alpha}_P - \alpha_P|^2 \leq  \epsilon' \left[ \frac{\|O^{(\mathrm{unk},\kappa)}\|}{\|O\|} \right]^{2\kappa/(\kappa+1)} \|O\|^2.
    \end{equation}

    Finally, applying the triangle inequality to combine the truncation error (Eq.~\eqref{leq:truncation_error0}) and estimation error (Eq.~\eqref{eq:estimation_error}), we obtain the overall prediction guarantee under $\mathcal{D}$:
    \begin{equation}
    \mathop{\mathbb{E}}_{\rho \sim \mathcal{D}} \left| h(\rho, O) - \mathrm{Tr}[O \mathcal{E}_1^{\dagger}(\rho)] \right|^2 \leq \left(\epsilon + \epsilon' \left[ \frac{\|O^{(\mathrm{unk},\kappa)}\|}{\|O\|} \right]^{2\kappa/(\kappa+1)}\right) \|O\|^2.
    \end{equation}

    Next, we determine the required size $N_1$ of the classical shadows $S_{N_1}(\mathcal{E}_1)$ to achieve the target error bound.  By the sample–complexity requirement (Eq.~\eqref{eq:number1}), there exists a constant $C_0>0$ such that
    \begin{equation}\label{bigN_1}
    N_1 \leq C_0 \cdot \frac{\log\!\bigl(n^{\nu+\kappa}N_1/\delta\bigr)}{\tilde{\epsilon}^{2}}.
    \end{equation}

    Define
    \begin{equation}\label{eq:T}
    T = 2 \cdot \frac{C_0}{\tilde{\epsilon}^{2}}\log\!\left(2\cdot\frac{C_0 n^{\nu+\kappa}}{\delta\tilde{\epsilon}^{2}}\right).
    \end{equation}

    Assume, for contradiction, that $N_1>T$.  Substituting $N_1>T$ into \eqref{bigN_1} gives
    \begin{equation}\label{eq:step}
    T < N_1 \leq \frac{C_0}{\tilde{\epsilon}^{2}}\log\!\left(\frac{n^{\nu+\kappa}N_1}{\delta}\right)
        < \frac{C_0}{\tilde{\epsilon}^{2}}\log\!\left(\frac{n^{\nu+\kappa}T}{\delta}\right),
    \end{equation}
    where the last inequality uses monotonicity of $\log$.

    Let $A=2C_0 n^{\nu+\kappa}/(\delta\tilde{\epsilon}^{2})$, so that $T=(2C_0/\tilde{\epsilon}^{2})\log A$.  Then \eqref{eq:step} becomes
    \begin{equation}
    2\log A < \log(A\log A)=\log A+\log\log A.
    \end{equation}
    Exponentiating yields
    \begin{equation}
    A^{2} < A\log A \quad\Longrightarrow\quad A<\log A,
    \end{equation}
    which is impossible for $A\ge 2$.  Hence the assumption $N_1>T$ is false; therefore
    \begin{equation}
    N_1 \leq T.
    \end{equation}

    Absorbing constants into $\mathcal{O}$-notation, we obtain
    \begin{equation}\label{totalN1}
    N_1 = \mathcal{O}\!\left(\frac{\log\!\bigl(n^{\nu+\kappa}/(\delta\tilde{\epsilon}^{2})\bigr)}{\tilde{\epsilon}^{2}}\right).
    \end{equation}

    Substituting filtering parameter $\tilde{\epsilon}$ (Eq.~\ref{eq:epsilon}) into Eq.~\eqref{totalN1}, we find that when the classical shadows $S_{N_1}(\mathcal{E}_1)$ is of size
    \begin{equation}\label{eq:dagger1}
        N_1=\log\left(\frac{n}{\delta}\right)2^{\mathcal{O}\left(\log(1/\epsilon)\left[\log\log(1/\epsilon)+\log(1/\epsilon')\right]\right)},
    \end{equation}
    shadow engineering can predict $\mathrm{Tr}[O \mathcal{E}_1^{\dagger}(\rho)]$ with probability at least $1-\delta$ and the prediction error satisfies:
    \begin{equation}
    \mathop{\mathbb{E}}_{\rho \sim \mathcal{D}} \left| h(\rho, O) - \mathrm{Tr}[O \mathcal{E}_1^{\dagger}(\rho)] \right|^2 \leq \left(\epsilon + \epsilon' \left[ \frac{\|O^{(\mathrm{unk},\kappa)}\|}{\|O\|} \right]^{2\kappa/(\kappa+1)}\right) \|O\|^2.
    \end{equation}

    Next, following Ref.~\cite{huang2023learning}, we introduce a second set of parameters: $\kappa = \left\lceil \log_{1.5} \left( {2}/{\epsilon} \right) \right\rceil$ and $\tilde{\epsilon} = \frac{\epsilon}{9 \cdot 2^{\kappa+1}n^{\kappa}} \left( \frac{C(\kappa, d)}{3} \right)^2$. This uses a deeper truncation level $\kappa$ for improved error control. 
    The proof follows the same structure as that for the first parameter set, so we directly state the corresponding conclusion.
    
    Substituting $\tilde{\epsilon}$ into Eq.~\eqref{totalN1}, we find that when the classical shadows $S_{N_1}(\mathcal{E}_1)$ is of size:
    \begin{equation}\label{eq:dagger2}
        N_1=\log\left(\frac{n}{\delta}\right)2^{\mathcal{O}\left(\log(1/\epsilon)\log(n)\right)},
    \end{equation}
    shadow engineering enables prediction of $\mathrm{Tr}[O \mathcal{E}_1^{\dagger}(\rho)]$ with probability $1-\delta$, and the prediction error satisfies:
    \begin{equation}
    \mathop{\mathbb{E}}_{\rho \sim \mathcal{D}} \left| h(\rho, O) - \mathrm{Tr}[O \mathcal{E}_1^{\dagger}(\rho)] \right|^2 \leq \epsilon \|O\|^2.
    \end{equation}
    
    We may then choose the smaller of the two sample sizes given in Eq.~\eqref{eq:dagger1} and Eq.~\eqref{eq:dagger2}, i.e.,
    \begin{equation}
        N_1=\log\left(\frac{n}{\delta}\right)\text{min}\left(2^{\mathcal{O}\left(\log(1/\epsilon)\left[\log\log(1/\epsilon)+\log(1/\epsilon')\right]\right)},2^{\mathcal{O}\left(\log(1/\epsilon)\log(n)\right)}\right),
    \end{equation}
    for any observable $O$ that is a sum of $\nu$-qubit observables (where $\nu = \mathcal{O}(1)$ and each qubit is acted on by $d = \mathcal{O}(1)$ of these $\nu$-qubit observables), and for any $n$-qubit state distribution $\mathcal{D}$ invariant under single-qubit $H$ and $S$ gates, shadow engineering can predict $\mathrm{Tr}[O \mathcal{E}_1^{\dagger}(\rho)]$ with probability at least $1-\delta$. The prediction error satisfies:
    \begin{equation}
    \mathop{\mathbb{E}}_{\rho \sim \mathcal{D}} \left| h(\rho, O) - \mathrm{Tr}[O \mathcal{E}_1^{\dagger}(\rho)] \right|^2 \leq \left(\epsilon + \epsilon' \left[ \frac{\|O^{(\mathrm{unk},\kappa)}\|}{\|O\|} \right]^{2\left\lceil \log_{1.5} \left( 1/\epsilon \right) \right\rceil/\left(\left\lceil \log_{1.5} \left( 1/\epsilon \right) \right\rceil+1\right)}\right) \|O\|^2.
    \end{equation}
    This completes the proof of Theorem 1. 
\end{proof}

\subsection{Shadow engineering for predicting $\text{Tr}[O\mathcal{E}_2\mathcal{E}_1(\rho)]$}
We next consider the second case of $f(\mathcal{E}_1,\mathcal{E}_2) = \mathcal{E}_2 \circ \mathcal{E}_1$.
By Lem.~\ref{lemma2}, the transfer matrices of $\mathcal{E}_2 \circ \mathcal{E}_1$ satisfies

\begin{equation}\label{leq31}
    T_{\mathcal{E}_2 \circ \mathcal{E}_1} = T_{\mathcal{E}_1} \mathcal{T}_n T_{\mathcal{E}_2}. 
\end{equation}
where $\mathcal{T}_n$ is the fixed sparse matrix satisfying
\[
\mathcal{M}^{-1}\circledast(\boldsymbol{\sigma}_{\mathcal{S}^{\otimes n}}) = \mathcal{T}_n\cdot \boldsymbol{\sigma}_{\mathcal{S}^{\otimes n}}.
\]

The mapping matrix $\mathcal{T}_n$ thus plays an essential role in composing transfer matrices by enabling basis conversion between the relevant representations. 
Its structure can be generalized from the single-qubit case to the $n$-qubit case, as detailed below.

For single-qubit case, we have:
\begin{equation}
    \mathcal{M}^{-1} \circledast (\boldsymbol{\sigma}_{\mathcal{S}})=\mathcal{T}_1 \boldsymbol{\sigma}_{\mathcal{S}}
\end{equation}
where $\mathcal{T}_1$ has the explicit form:
\begin{equation}
    \mathcal{T}_1 = 
\begin{pmatrix}
 2 & -1 & 0 & 0 & 0 & 0 \\
 -1 & 2 & 0 & 0 & 0 & 0 \\
 -1 & -1 & 3 & 0 & 0 & 0 \\
 -1 & -1 & 0 & 3 & 0 & 0 \\
 -1 & -1 & 0 & 0 & 3 & 0 \\
 -1 & -1 & 0 & 0 & 0 & 3 
\end{pmatrix}.
\end{equation}

This matrix is derived by expressing each element of $\mathcal{M}^{-1} \circledast (\boldsymbol{\sigma}_{\mathcal{S}})$ as a linear combination of elements in $\boldsymbol{\sigma}_{\mathcal{S}}$. 
Notably, over 50 percent of its entries are zero, rendering it highly sparse--this sparsity allows for efficient storage in the CSR format.

Next, we extend $\mathcal{T}_1$ to the $n$-qubit case by constructing $\mathcal{T}_n$ through the following steps:
\begin{enumerate}
    \item Define \(\mathcal{T}_{\text{ini}}\) as the set of 6 row vectors of $\mathcal{T}_1$:
\begin{equation}
    \mathcal{T}_{\text{ini}} = \{\mathcal{T}_1[0], \mathcal{T}_1[1], \ldots, \mathcal{T}_1[5]\}
\end{equation}
where $\mathcal{T}_1[i]$ denotes the $i$-th row vector of $\mathcal{T}_1$.
    \item Compute the $n$-fold tensor product of $\mathcal{T}_{\text{ini}}$:
    \begin{equation}\label{eq:TN}
        \mathcal{T}_{n} =\mathcal{T}_{\text{ini}}^{\otimes n}
    \end{equation}
    \item Assemble the resulting $6^n \times 6^n$ matrix by treating each element of $\mathcal{T}_{n}$ as a row vector.
\end{enumerate}
The $n$-qubit matrix $\mathcal{T}_n$ inherits exponential sparsity from $\mathcal{T}_1$. Storing $\mathcal{T}_{n}$ in CSR format dramatically reduces both memory requirements and computational complexity.

Using the classical shadows $S_{N_1}(\mathcal{E}_1)$ and  $S_{N_2}(\mathcal{E}_2)$, we computed the estimated transfer matrix $\hat{T}_{\mathcal{E}_2\circ \mathcal{E}_1}$. 
Within the shadow engineering framework, this enables us to approximate $h(\rho,O)\approx \text{Tr}[O\mathcal{E}_2\mathcal{E}_1(\rho)]$ for any $n$-qubit state $\rho \sim \mathcal{D}$ and any bounded-degree observable $O$.

We now present a theorem that quantifies the prediction error as a function of the dataset sizes $N_1,N_2$.

\begin{theorem}[Prediction error for estimating \(\text{Tr}(O\mathcal{E}_2 \circ \mathcal{E}_1(\rho))\)]
 Let $n, \epsilon, \epsilon', \delta > 0$. For any unknown $n$-qubit CPTP map $\mathcal{E}_1$, 
  and its classical shadows $S_{N_1}(\mathcal{E}_1)$ obtained by $N_1$ randomized experiments with
\begin{equation}
        N_1 = N_2 = \log(\frac{n}{\delta}) \text{min}(2^{\mathcal{O}[\log(n)+\log(1/\epsilon)(\log\log(1/\epsilon) + \log(1/\epsilon^{\prime}))]},  2^{O\left[\log(1/\epsilon) \log(n)\right]}),
\end{equation}

With probability at least $  1 - \delta$, 
For any $n$-qubit state $\rho \sim \mathcal{D}$ and any bounded-degree observable $O$, 
the prediction error achieved by the algorithm is
\begin{equation}
    \mathbb{E}_{\rho \sim \mathcal{D}} \left| h(\rho, O) - \text{Tr}[O \mathcal{E}_2\mathcal{E}_1(\rho)] \right|^2\leq \left( \epsilon + \epsilon' \left[ \frac{\|O^{(\mathrm{unk},\kappa)}\|}{\|O\|} \right]^{2  \left\lceil \log_{1.5}(1/\epsilon) \right\rceil/ \left[\left\lceil  \log_{1.5}(1/\epsilon)\right\rceil  + 1\right]  } \right) \|O\|^2.
\end{equation}
where $O^{(\mathrm{unk},\kappa)}$ is the low-degree truncation of $\mathcal{E}_1^{\dagger}\mathcal{E}_2^{\dagger}(O)$.
\end{theorem}

\begin{proof}

    We begin with a bounded-degree observable defined as:
    \begin{equation}
    O = \sum_{Q \in \{I, X, Y, Z\}^{\otimes n}, |Q| \leq \nu} a_Q Q,
    \end{equation}
    
    Our goal is to compute the Pauli coefficients of the Heisenberg-evolved observable under $\mathcal{E}_2\mathcal{E}_1$ :
    \begin{equation}
    O^{(\mathrm{unk})} \triangleq \mathcal{E}_1^\dagger\mathcal{E}_2^\dagger(O) = \sum_{P \in \{I, X, Y, Z\}^{\otimes n}} \alpha_P(O) P.
    \end{equation}
    
    To avoid exponential computational complexity, we apply $\kappa$-truncation to $O^{(\mathrm{unk})}$:
    \begin{equation}
    O^{(\mathrm{unk,\kappa})} \triangleq \sum_{{P \in \{I, X, Y, Z\}^{\otimes n} ,|P| \leq \kappa}} \alpha_P(O) P.
    \end{equation}
    
    We analyze the prediction error for input state $\rho$ drawn from the $\mathcal{D}$ distribution, aiming to bound:
    \begin{equation}
        \mathop{\mathbb{E}}_{\rho \sim \mathcal{D}} \left|h(\rho,O) - \text{Tr}[O^{(\mathrm{unk})}\rho]\right|^2.
    \end{equation}
    Applying the triangle inequality, we decompose this error into two components:
    \begin{equation}
          \mathop{\mathbb{E}}_{\rho \sim \mathcal{D}} \left|h(\rho,O) - \text{Tr}[O^{(\mathrm{unk})}\rho]\right|^2 \leq   \underbrace{\mathop{\mathbb{E}}_{\rho \sim \mathcal{D}} \left|\text{Tr}[O^{(\mathrm{unk,\kappa})}\rho] - \text{Tr}[O^{(\mathrm{unk})}\rho]\right|^2}_{\text{(i) Truncation error}} + \underbrace{\mathop{\mathbb{E}}_{\rho \sim \mathcal{D}} \left|h(\rho,O) - \text{Tr}[O^{(\mathrm{unk,\kappa})}\rho]\right|^2}_{\text{(ii) Estimation error}}
    \end{equation}

    We first bound the truncation error for $\mathcal{E}_1^\dagger\mathcal{E}_2^\dagger(O)$. 
    Applying Lem.~\ref{Lemma:lowdegree} to $O^{(\mathrm{unk})}$ yields:
    \begin{equation}
        \mathop{\mathbb{E}}_{\rho \sim \mathcal{D}} \left|\mathrm{Tr}[O^{(\mathrm{unk})}\rho] - \mathrm{Tr}[O^{(\mathrm{unk},\kappa)}\rho]\right|^2 \leq \left(\frac{2}{3}\right)^{\kappa}\|O\|^2,
    \end{equation}
    since $\|O^{(\mathrm{unk})}\| = \|\mathcal{E}_1^\dagger\mathcal{E}_2^\dagger(O)\| \leq \|O\|$.
    
    To control the truncation error, we select $\kappa = \left\lceil \log_{1.5} \left( {1}/{\epsilon} \right) \right\rceil$ to obtain:
    \begin{equation}\label{leq:truncation_error2}
        \mathop{\mathbb{E}}_{\rho \sim \mathcal{D}} \left|\mathrm{Tr}[O^{(\mathrm{unk})}\rho] - \mathrm{Tr}[O^{(\mathrm{unk},\kappa)}\rho]\right|^2 \leq \epsilon \|O\|^2,
    \end{equation}
    where $\epsilon$ is the truncation error parameter. 
    
    Next, we analyze the estimation error arising from the finite-sample approximation of $\alpha_P(O)$ (for $|P|\leq \kappa$) of $O^{(\text{unk},\kappa)}$.
    According to Lem.~ \ref{alpha_p}, we define
    \begin{equation}
    x_{P}(O) = \mathbb{E}_{\rho\sim\mathcal{D}^{0}} \text{Tr}[P\rho] \text{Tr}[\mathcal{E}_1^\dagger \mathcal{E}_2^\dagger(O) \rho].
    \end{equation}
    The Pauli coefficients are then reconstructed via ${\alpha}_P(O) = 3^{|P|} {x}_P(O)$.
    
    We use the estimated transfer matrices $\hat{T}_{\mathcal{E}_1}$ and $\hat{T}_{\mathcal{E}_2}$ to approximate ${x}_P(O)$, which is equivalent to estimating $\mathbb{E}_{\rho\sim\mathcal{D}^{0}} \text{Tr}[P\rho]\text{Tr}[O\mathcal{E}_2\mathcal{E}_1(\rho)]$.
    To bound the error in this estimation, we consider $\rho=\mathcal{S}^{\otimes n}_c$ ($c \in \text{supp}(\hat{T}_{\mathcal{E}_1})$) and consider the estimation of $\text{Tr}[P\mathcal{S}^{\otimes n}_c]\operatorname{Tr}\!\bigl[O\mathcal{E}_2\mathcal{E}_1(\mathcal{S}^{\otimes n}_c)\bigr]$:

    \begin{align}
    &\Bigl|\text{Tr}[P\mathcal{S}^{\otimes n}_c]\operatorname{Tr}\!\bigl[O\sum_{b=1}^{6^n}(\hat{T}_{\mathcal{E}_1})_{cb}
    \sum_{a=1}^{6^n}(\mathcal{T}_n)_{ba}\sum_{l=1}^{6^n}(\hat{T}_{\mathcal{E}_2})_{al}
    \mathcal{M}^{-1}(\mathcal{S}^{\otimes n}_l)\bigr]
    -\text{Tr}[P\mathcal{S}^{\otimes n}_c]\operatorname{Tr}\!\bigl[O\mathcal{E}_2\mathcal{E}_1(\mathcal{S}^{\otimes n}_c)\bigr]\Bigr|
    \notag\\
    &\quad\le
    \Bigl|\text{Tr}[P\mathcal{S}^{\otimes n}_c]\operatorname{Tr}\!\bigl[O\sum_{b=1}^{6^n}(\hat{T}_{\mathcal{E}_1})_{cb}
    \sum_{a=1}^{6^n}(\mathcal{T}_n)_{ba}\sum_{l=1}^{6^n}(\hat{T}_{\mathcal{E}_2})_{al}
    \mathcal{M}^{-1}(\mathcal{S}^{\otimes n}_l)\bigr]
    -\text{Tr}[P\mathcal{S}^{\otimes n}_c]\operatorname{Tr}\!\bigl[O\sum_{b=1}^{6^n}(\hat{T}_{\mathcal{E}_1})_{cb}
    \sum_{a=1}^{6^n}(\mathcal{T}_n)_{ba}\mathcal{E}_2(\mathcal{S}^{\otimes n}_a)\bigr]\Bigr|
    \label{S93}\\[4pt]
    &\quad+
    \Bigl|\text{Tr}[P\mathcal{S}^{\otimes n}_c]\operatorname{Tr}\!\bigl[O\sum_{b=1}^{6^n}(\hat{T}_{\mathcal{E}_1})_{cb}
    \sum_{a=1}^{6^n}(\mathcal{T}_n)_{ba}\mathcal{E}_2(\mathcal{S}^{\otimes n}_a)\bigr]
    -\text{Tr}[P\mathcal{S}^{\otimes n}_c]\operatorname{Tr}\!\bigl[O\mathcal{E}_2\mathcal{E}_1(\mathcal{S}^{\otimes n}_c)\bigr]\Bigr|
    \label{S94}
    \end{align}
    
    We now analyze the first term (Eq.~\eqref{S93}), which originates from the estimation of $\mathcal{E}_2(\mathcal{S}^{\otimes n}_a)$ via $\hat{T}_{\mathcal{E}_2}$. 
    For a Pauli observable $Q \in \{I, X, Y, Z\}^{\otimes n}$ with $|Q| \leq \nu = \mathcal{O}(1)$, consider the ideal value ${S}^{(2)}_a(Q) = \text{Tr}[Q\mathcal{E}_2(\mathcal{S}^{\otimes n}_a)]~(a\in[1,6^n])$ and its empirical estimator constructed from the transfer matrix:
    \begin{equation}
        \hat{S}^{(2)}_a(Q)=\sum_{l\in \text{supp}((\hat{T}_{\mathcal{E}_2})_a)}(\hat{T}_{\mathcal{E}_2})_{al}\text{Tr}[Q\mathcal{M}^{-1}(\mathcal{S}^{\otimes n}_l)]=\frac{1}{N_{2(a)}}\sum_{\mathcal{S}^{\otimes n}_{j_\ell} \in \mathcal{G}_2(\mathcal{S}^{\otimes n}_a)}\text{Tr}[Q\mathcal{M}^{-1}(\mathcal{S}^{\otimes n}_{j_\ell})]
    \end{equation}
    where $\mathcal{G}_2(S^{\otimes n}_a) = \Big[ S^{\otimes n}_{j_\ell} \;\Big|\; (S^{\otimes n}_{a},S^{\otimes n}_{j_\ell}) \in S_{N_2}(\mathcal{E}_2), \Big]$ is the set of the second components in $S_{N_2}(\mathcal{E}_2)$ whose first component is $S^{\otimes n}_a$, and $N_{2(a)}=|\mathcal{G}_2(S^{\otimes n}_a)|$ indicates the number of classical shadows data pairs recorded in the $a$-th row of the transfer matrix $\hat{T}_{\mathcal{E}_2}$.
    
    Note that $|\text{Tr}[Q\mathcal{M}^{-1}(\mathcal{S}^{\otimes n}{j_\ell})]| = \mathcal{O}(1)$ since $|Q| = \mathcal{O}(1)$.
    For each row $a$ of the transfer matrix $T_{\mathcal{E}_2}$, if the number of samples satisfIes
    \begin{equation}
        N_{2(a)}=\Omega\left(\frac{\log(n^\nu/\delta)}{\tilde{\epsilon}^2}\right),
    \end{equation}
    Hoeffding's inequality guarantees that for all $|Q|\leq \nu$:
    \begin{equation}
        |\hat{S}^{(2)}_a(Q))-{S}^{(2)}_a(Q)|\leq \tilde{\epsilon}_2
    \end{equation}
    holds with probability at least $ 1-\delta$.

    To ensure this bound holds uniformly over $N_2$ nonzero rows $a$ of $\hat{T}_{\mathcal{E}_2}$, we take the full classical shadows data $S_{N_2}(\mathcal{E}_2)$ of size
    \begin{equation}
        N_{2}=\Omega\left(\frac{\log(n^\nu N_{2}/\delta)}{\tilde{\epsilon}_2^2}\right),
    \end{equation}
    By applying Hoeffding’s inequality and a union bound, we obtain that with probability at least $1 - \delta$,
    \begin{equation}
        |\hat{S}^{(2)}_a(Q)-{S}^{(2)}_a(Q)|\leq \tilde{\epsilon}.
    \end{equation}
    holds simultaneously for all $Q$ with $|Q| \leq \nu$.

    Next, we extend the bound to $ \left| \sum_{b=1}^{6^n} (\hat{T}_{\mathcal{E}_1})_{cb} \sum_{a=1}^{6^n} (\mathcal{T}_n)_{ba} \hat{S}^{(2)}_a(Q) - \sum_{b=1}^{6^n} (\hat{T}_{\mathcal{E}_1})_{cb} \sum_{a=1}^{6^n} (\mathcal{T}_n)_{ba} {S}^{(2)}_a(Q) \right|$, 
    which quantifies the error in estimating $\text{Tr}[Q\sum_{b=1}^{6^n}(\hat{T}_{\mathcal{E}_1})_{cb} \sum_{a=1}^{6^n} (\mathcal{T}_n)_{ba}\mathcal{E}_2(\mathcal{S}^{\otimes n}_a)]$, where \(\mathcal{T}_n\) is the basis-transform matrix defined in Eq.~\eqref{eq:TN}.
    
    To bound the above expression, we define a column vector $\overrightarrow{\Delta S}$ whose $a$-th entry is $\left|\hat{S}^{(2)}_a(Q) - {S}^{(2)}_a(Q)\right|$ for each $a \in [1, 6^n]$, and consider the Euclidean norm of $\mathcal{A}\cdot\overrightarrow{\Delta S}$, where $\mathcal{A} = \hat{T}_{\mathcal{E}_1} \mathcal{T}_n$. 
    
    Note that $\hat{T}_{\mathcal{E}_1}$ has at most $N_1$ nonzero rows (each row summing to 1), and remaining $6^n - N_1$ rows are zero.
    Since every row of $\mathcal{T}_n$ also sums to 1, it follows that $\mathcal{A}$ has at most $N_1$ nonzero rows, each summing to 1, with all other rows zero. 
    
    We now bound the Euclidean norm:
    \begin{equation}
        \begin{aligned}
    \|\mathcal{A}\cdot\overrightarrow{\Delta S}\|_2&=\sqrt{\sum_{c=1}^{6^n}\left(\sum_{a=1}^{6^n}\mathcal{A}_{ca}\overrightarrow{\Delta S}_a\right)^2}\\&=\sqrt{\sum_{c=1}^{6^n}\left(\sum_{a=1}^{6^n}\mathcal{A}_{ca}|\hat{S}^{(2)}_a(Q)-{S}^{(2)}_a(Q)|\right)^2}\\&\leq\sqrt{\sum_{c=1}^{6^n}\left(\sum_{a=1}^{6^n}\mathcal{A}_{ca}\tilde{\epsilon}_2\right)^2}\\
            &= \sqrt{N_1}\tilde{\epsilon}_2.
    \end{aligned}
    \end{equation}
    The absolute value of any component of $\mathcal{A}\cdot\overrightarrow{\Delta S}$ is bounded by its Euclidean norm, hence:
    \begin{equation}
        \bigg|\sum_{b=1}^{6^n} (\hat{T}_{\mathcal{E}_1})_{cb} \sum_{a=1}^{6^n} (\mathcal{T}_n)_{ba} \big|\hat{S}^{(2)}_a(Q)-{S}^{(2)}_a(Q)\big|\bigg|\leq \|\mathcal{A}\cdot\overrightarrow{\Delta S}\|_2
    \end{equation}
    Moreover, by the triangle inequality:
    \begin{equation}
            \bigg|\sum_{b=1}^{6^n} (\hat{T}_{\mathcal{E}_1})_{cb} \sum_{a=1}^{6^n} (\mathcal{T}_n)_{ba} \hat{S}^{(2)}_a(Q)-\sum_{b=1}^{6^n} (\hat{T}_{\mathcal{E}_1})_{cb} \sum_{a=1}^{6^n} (\mathcal{T}_n)_{ba} {S}^{(2)}_a(Q)\bigg| \leq \bigg|\sum_{b=1}^{6^n} (\hat{T}_{\mathcal{E}_1})_{cb} \sum_{a=1}^{6^n} (\mathcal{T}_n)_{ba} \big|\hat{S}^{(2)}_a(Q)-{S}^{(2)}_a(Q)\big|\bigg|.
    \end{equation}
    Therefore, we conclude:
    \begin{equation}\label{result1}
        \bigg|\sum_{b=1}^{6^n} (\hat{T}_{\mathcal{E}_1})_{cb} \sum_{a=1}^{6^n} (\mathcal{T}_n)_{ba} \hat{S}^{(2)}_a(Q)-\sum_{b=1}^{6^n} (\hat{T}_{\mathcal{E}_1})_{cb} \sum_{a=1}^{6^n} (\mathcal{T}_n)_{ba} {S}^{(2)}_a(Q)\bigg|\leq \sqrt{N_1}\tilde{\epsilon}_2
    \end{equation}

    Finally, since the observable $O$ has Pauli-1 norm $\|O\|_{\text{Pauli,1}} =\eta$, applying triangle inequality again yields:
    \begin{equation}
            \left|\text{Tr}\left[O\sum_{b=1}^{6^n} (\hat{T}_{\mathcal{E}_1})_{cb} \sum_{a=1}^{6^n} (\mathcal{T}_n)_{ba} \sum_{l=1}^{6^n}(\hat{T}_{\mathcal{E}_2})_{al}\mathcal{M}^{-1}(\mathcal{S}^{\otimes n}_l)\right]-\text{Tr}\left[O\sum_{b=1}^{6^n} (\hat{T}_{\mathcal{E}_1})_{cb} \sum_{a=1}^{6^n} (\mathcal{T}_n)_{ba}\mathcal{E}_2(\mathcal{S}^{\otimes n}_a)\right]\right|\leq \eta \sqrt{N_1}\tilde{{\epsilon}}_2
    \end{equation}
    
    Considering the classical shadows $S_{N_2}(\mathcal{E}_2)$ of size
    \begin{equation}
        N_{2}=\Omega\left(\frac{\log(n^{\nu+\kappa} N_{2}/\delta)}{\tilde{\epsilon}_2^2}\right),
    \end{equation}
    we ensure that for all $|P|\leq \kappa$ the following bound holds:
    \begin{equation}\label{oneparts}
    \begin{aligned}
    &\Bigl|\operatorname{Tr}[P\mathcal{S}^{\otimes n}_c]\,
    \operatorname{Tr}\!\bigl[O\sum_{b=1}^{6^n}(\hat{T}_{\mathcal{E}_1})_{cb}
    \sum_{a=1}^{6^n}(\mathcal{T}_n)_{ba}
    \sum_{l=1}^{6^n}(\hat{T}_{\mathcal{E}_2})_{al}\mathcal{M}^{-1}(\mathcal{S}^{\otimes n}_l)\bigr] \\
    &\quad-\operatorname{Tr}[P\mathcal{S}^{\otimes n}_c]\,
    \operatorname{Tr}\!\bigl[O\sum_{b=1}^{6^n}(\hat{T}_{\mathcal{E}_1})_{cb}
    \sum_{a=1}^{6^n}(\mathcal{T}_n)_{ba}\mathcal{E}_2(\mathcal{S}^{\otimes n}_a)\bigr]\Bigr|
    \leq \eta\sqrt{N_1}\tilde{\epsilon}_2.
    \end{aligned}
    \end{equation}

    Next, we analyze the second term (Eq.~\eqref{S94}).    
    Using the identity
    \begin{equation}
        \begin{aligned}
            \text{Tr}\left[Q\sum_{b=1}^{6^n} (\hat{T}_{\mathcal{E}_1})_{cb} \sum_{a=1}^{6^n} (\mathcal{T}_n)_{ba} \mathcal{E}_2(\mathcal{S}^{\otimes n}_a)\right]
            &= \text{Tr}\left[Q\mathcal{E}_2\left(\sum_{b=1}^{6^n} (\hat{T}_{\mathcal{E}_1})_{cb} \sum_{a=1}^{6^n} (\mathcal{T}_n)_{ba} \mathcal{S}^{\otimes n}_a\right)\right] \\
            &= \text{Tr}\left[Q\mathcal{E}_2\left(\sum_{b=1}^{6^n} (\hat{T}_{\mathcal{E}_1})_{cb}\mathcal{M}^{-1}(\mathcal{S}^{\otimes n}_b)\right)\right],
    \end{aligned}
    \end{equation}
    we aim to bound the absolute error between $\text{Tr}[P\mathcal{S}^{\otimes n}_c]\text{Tr}\left[Q\mathcal{E}_2\left(\sum_{b=1}^{6^n} (\hat{T}_{\mathcal{E}_1})_{cb}\mathcal{M}^{-1}(\mathcal{S}^{\otimes n}_b)\right)\right]$ and $\text{Tr}[P\mathcal{S}^{\otimes n}_c]\text{Tr}\left[Q\mathcal{E}_2\mathcal{E}_1(\mathcal{S}^{\otimes n}_c)\right]$ for $c\in\text{supp}(\hat{T}_{\mathcal{E}_1})$.

    For the state $\mathcal{S}^{\otimes n}_c \in \mathcal{S}^{\otimes n}$, define the ideal value $S^{(1)}_c(Q)=\text{Tr}[Q\mathcal{E}_1(\mathcal{S}^{\otimes n}_c)]$. Its empirical estimator is constructed as:
    \begin{equation}
        \hat{S}^{(1)}_c(Q)=\sum_{l\in \text{supp}((\hat{T}_{\mathcal{E}_1})_c)}(\hat{T}_{\mathcal{E}_1})_{cl}\text{Tr}[Q\mathcal{M}^{-1}(\mathcal{S}^{\otimes n}_l)]=\frac{1}{N_{1(c)}}\sum_{\mathcal{S}^{\otimes n}_{i_\ell} \in \mathcal{G}_3(\mathcal{S}^{\otimes n}_c)}\text{Tr}[Q\mathcal{M}^{-1}(\mathcal{S}^{\otimes n}_{i_\ell})],
        \end{equation}
    where $\mathcal{G}_3(S^{\otimes n}_c) = \Big[ S^{\otimes n}_{j_\ell} \;\Big|\; (S^{\otimes n}_{c},S^{\otimes n}_{j_\ell}) \in S_{N_1}(\mathcal{E}_1), \Big]$ represents the set of the first components in $S_{N_1}(\mathcal{E}_1)$ whose second component is $S^{\otimes n}_c$, and $N_{1(c)}=|\mathcal{G}_3(S^{\otimes n}_c)|$ is the number of classical shadows pairs recorded in the $c$-th row of the transfer matrix $\hat{T}_{\mathcal{E}_1}$.
    
    Note that $|\mathrm{Tr}[Q \mathcal{M}^{-1}(\mathcal{S}^{\otimes n}_{j\ell})]| = \mathcal{O}(1)$ since $|Q| = \mathcal{O}(1)$.
    If the classical shadows $S_{N_1}(\mathcal{E}_1)$ is of size
    \begin{equation}
    N_1 = \Omega\left(\frac{\log(n^{\nu}N_1/\delta)}{\tilde{\epsilon}_1^2}\right)
    \end{equation}  
    then by Hoeffding's inequality and a union bound, for all $Q $ with $|Q| \leq \nu$,   
    \begin{equation}
        |\hat{S}^{(1)}_c(Q) - S^{(1)}_c(Q)| \leq \tilde{\epsilon}_1
    \end{equation}
    holds for all $c\in\text{supp}(\hat{T}_{\mathcal{E}_1})$, where $|\text{supp}(\hat{T}_{\mathcal{E}_1})|\leq N_1$, with probability at least $1 - \delta$.
    
    Then, we perform an error analysis of the following expression, which uses $\hat{T}_{\mathcal{E}_1}$ to estimate $\text{Tr}\left[O \mathcal{E}_2\mathcal{E}_1(S^{\otimes n}_c) \right]$ in Eq.~\eqref{S94}.
    And applying the triangle inequality, we decompose the error as follows:
    \begin{align}
    &\left|\operatorname{Tr}\!\bigl[O\sum_{b=1}^{6^n}(\hat{T}_{\mathcal{E}_1})_{cb}
    \sum_{a=1}^{6^n}(\mathcal{T}_n)_{ba}\mathcal{E}_2(\mathcal{S}^{\otimes n}_a)\bigr]
    -\operatorname{Tr}\!\bigl[O\mathcal{E}_2\mathcal{E}_1(S^{\otimes n}_c)\bigr]\right| \nonumber\\[4pt]
    &\leq  \left|\operatorname{Tr}\!\bigl[\sum_{b}(\hat{T}_{\mathcal{E}_1})_{cb}
    \sum_{a}(\mathcal{T}_n)_{ba}\mathcal{E}_2^\dagger(O)(S^{\otimes n}_a)\bigr]
    -\operatorname{Tr}\!\bigl[\sum_{b}(\hat{T}_{\mathcal{E}_1})_{cb}
    \sum_{a}(\mathcal{T}_n)_{ba}(\mathcal{E}_2^\dagger(O))^{(\nu)}(S^{\otimes n}_a)\bigr]\right|
    \label{S112}\\[4pt]
    & +\left|\operatorname{Tr}\!\bigl[(\mathcal{E}_2^\dagger(O))^{(\nu)}
    \sum_{b}(\hat{T}_{\mathcal{E}_1})_{cb}\mathcal{M}^{-1}(S^{\otimes n}_b)\bigr]
    -\operatorname{Tr}\!\bigl[(\mathcal{E}_2^\dagger(O))^{(\nu)}
    \mathcal{E}_1(S^{\otimes n}_c)\bigr]\right|
    \label{S113}\\[4pt]
    & +\left|\operatorname{Tr}\!\bigl[(\mathcal{E}_2^\dagger(O))^{(\nu)}
    \mathcal{E}_1(S^{\otimes n}_c)\bigr]
    -\operatorname{Tr}\!\bigl[\mathcal{E}_2^\dagger(O)\mathcal{E}_1(S^{\otimes n}_c)\bigr]\right|
    \label{S114}
    \end{align}
    We now bound each of the three error terms separately.
    
    For Eq.~\eqref{S112}: 
    This term quantifies the error introduced by truncating $\mathcal{E}_2^\dagger(O)$ to its low-weight ($\leq \nu$) Pauli components. Define a vector $\overrightarrow{\Delta S}_2 \in \mathbb{R}^{6^n}$ with entries:
    \begin{equation}
        (\overrightarrow{\Delta S}_2)_a = \left|\text{Tr}[(\mathcal{E}_2^\dagger(O) -(\mathcal{E}_2^\dagger(O))^{\nu}) (S^{\otimes n}_a)]\right|,~~~\text{for} ~a=1,\dots,6^n.
    \end{equation}
    
    Since $\mathcal{E}_2^\dagger$ is a CPTP map and $S^{\otimes n}_a$ is a state, we have:
    \begin{equation}
    \left|\text{Tr}[(\mathcal{E}_2^\dagger(O) -(\mathcal{E}_2^\dagger(O))^{\nu}) (S^{\otimes n}_a)]\right|\leq\|\mathcal{E}_2^\dagger(O) -(\mathcal{E}_2^\dagger(O))^{\nu}\|\|S^{\otimes n}_a\|_1 \leq  \|O\|+\|(\mathcal{E}_2^\dagger(O))^{(\nu)}\|
    \end{equation}
    Let $\mathcal{A} = \hat{T}_{\mathcal{E}_1} \mathcal{T}_n$. As argued previously, the matrix $\mathcal{A}$ has at most $N_1$ nonzero rows, each with row sum 1. The Euclidean norm of $\mathcal{A} \overrightarrow{\Delta S}_2$ is bounded by:
    \begin{equation}
        \|\mathcal{A}\cdot\overrightarrow{\Delta S}_2\|_2 \leq  \sqrt{N_1} (\|O\|+\|(\mathcal{E}_2^\dagger(O))^{(\nu)}\|)
    \end{equation}
    This implies a bound on the absolute value of any component of the vector $\mathcal{A} \overrightarrow{\Delta S}_2$, and consequently:
    \begin{equation}
        \bigg|\sum_{b=1}^{6^n} (\hat{T}_{\mathcal{E}_1})_{cb} \sum_{a=1}^{6^n} (\mathcal{T}_n)_{ba} \big|\text{Tr}[\mathcal{E}_2^\dagger(O) (S^{\otimes n}_a)-(\mathcal{E}_2^\dagger(O))^{\nu} (S^{\otimes n}_a)]\big|\bigg|\leq \|\mathcal{A}\cdot\overrightarrow{\Delta S}_2\|_2 \leq  \sqrt{N_1} (\|O\|+\|(\mathcal{E}_2^\dagger(O))^{(\nu)}\|)
    \end{equation}

    For Eq.~\eqref{S113}: 
    We leverage the fact that for all $Q$ with $|Q| \leq \nu$, we have $|\hat{S}^{(1)}_c(Q) - S^{(1)}_c(Q)| \leq \tilde{\epsilon}_1$.
    Express $(\mathcal{E}_2^\dagger(O))^{(\nu)}=\sum_{v} \alpha_v Q_v$ in the Pauli basis $Q_v$. Its bounded by $|\alpha_v| \leq \|(\mathcal{E}_2^\dagger(O))^{(\nu)}\|$ (since \( |\alpha_v| = \left| \frac{1}{2^n} \text{Tr}\left[ (\mathcal{E}_2^\dagger(O))^{(\nu)} Q_v \right] \right| \leq \frac{1}{2^n} \left\| (\mathcal{E}_2^\dagger(O))^{(\nu)} \right\| \left\| Q_v \right\|_1 = \left\| (\mathcal{E}_2^\dagger(O))^{(\nu)} \right\|  \)), and the number of Pauli coefficients is \( \mathcal{O}(n^{\nu}) \).
    \begin{equation}
    \begin{aligned}
    &\left|\operatorname{Tr}\!\left[(\mathcal{E}_2^\dagger(O))^{(\nu)}
    \sum_{b=1}^{6^n}(\hat{T}_{\mathcal{E}_1})_{cb}\mathcal{M}^{-1}(S_b^{\otimes n})\right]
    -\operatorname{Tr}\!\left[(\mathcal{E}_2^\dagger(O))^{(\nu)}\mathcal{E}_1(S_c^{\otimes n})\right]\right|\\
    &\quad=\left|\sum_{v}\alpha_v\bigl(\hat{S}_c^{(1)}(Q)-S_c^{(1)}(Q)\bigr)\right|\\[2pt]
    &\quad\le\sqrt{\sum_{v}|\alpha_v|^2}\sqrt{\sum_{v}\bigl|\hat{S}_c^{(1)}(Q)-S_c^{(1)}(Q)\bigr|^2}\\[2pt]
    &\quad\le\sqrt{\sum_{v}\|(\mathcal{E}_2^\dagger(O))^{(\nu)}\|^2}\sqrt{\sum_{v}\tilde{\epsilon}_1^{2}}\\[2pt]
    &\quad=\|(\mathcal{E}_2^\dagger(O))^{(\nu)}\|\,\tilde{\epsilon}_1\,\mathcal{O}(n^{\nu}).
    \end{aligned}
    \end{equation}
    
    For Eq.~\eqref{S114}:  
    \begin{equation}
        \left| \text{Tr}\left[ (\mathcal{E}_2^\dagger(O))^{(\nu)} \mathcal{E}_1(S^{\otimes n}_c) \right] - \text{Tr}\left[ \mathcal{E}_2^\dagger(O) \mathcal{E}_1(S^{\otimes n}_c) \right] \right|\leq \|(\mathcal{E}_2^\dagger(O))^{(\nu)}-\mathcal{E}_2^\dagger(O)\|\cdot \| \mathcal{E}_1(S^{\otimes n}_c) \|_{1}\leq \|(\mathcal{E}_2^\dagger(O))^{(\nu)}\| + \|O\|
    \end{equation}
    
    Summing the bounds for above three terms yields:
    \begin{equation}\label{S121}
        \left| \text{Tr}\left[O\sum_{b=1}^{6^n} (\hat{T}_{\mathcal{E}_1})_{cb} \sum_{a=1}^{6^n} (\mathcal{T}_n)_{ba} \mathcal{E}_2(\mathcal{S}^{\otimes n}_a)\right]-\text{Tr}\left[ O \mathcal{E}_2\mathcal{E}_1(S^{\otimes n}_c) \right]\right| \leq \mathcal{O}(n^{\nu})\|(\mathcal{E}_2^\dagger(O))^{(\nu)}\| \tilde{\epsilon}_1  +  (\sqrt{N_1}  +1 )(\|O\|+\|(\mathcal{E}_2^\dagger(O))^{(\nu)}\|)
    \end{equation}
    
    Now, considering the required size of the classical shadows $S_{N_1}(\mathcal{E}_1)$:
    \begin{equation}
        N_{1}=\Omega\left(\frac{\log(n^{\nu+\kappa} N_{1}/\delta)}{\tilde{\epsilon}_1^2}\right),
    \end{equation}
    and noting that $|\mathrm{Tr}[P S^{\otimes n}_c]| = \mathcal{O}(1)$ for $|P| \leq \kappa$, we multiply the error bound for Eq.~\eqref{S121} by this factor. 
    Combining this result with the bound from Eq.~\eqref{oneparts} for the first part (which contributes $\eta \sqrt{N_1} \tilde{\epsilon}_2$), the overall error in estimating ${x}_P(O)$ is:
    \begin{equation}
        |\hat{x}_P(O)-x_P(O)|\leq \eta \left(\sqrt{N_1}\tilde{{\epsilon}}_2 +\frac{\|(\mathcal{E}_2^\dagger(O))^{(\nu)}\|\tilde{\epsilon}_1 \mathcal{O}(n^{\nu})}{\eta} + \frac{ (\sqrt{N_1}  +1 )(\|O\|+\|(\mathcal{E}_2^\dagger(O))^{(\nu)}\|)}{\eta}\right)
    \end{equation}
    which holds for all $P$ with $|P| \leq \kappa$. Here, $\hat{x}_P(O)$ is an estimate of ${x}_P(O)$, and
    $\hat{x}_P(O) = \frac{1}{|\text{supp}(\hat{T}_{\mathcal{E}_1})|}\sum_{c \in \text{supp}(\hat{T}_{\mathcal{E}_1})}\text{Tr}\left[P\mathcal{S}^{\otimes n}_c \right]\text{Tr}\left[O\sum_{b=1}^{6^n} (\hat{T}_{\mathcal{E}_1})_{cb} \sum_{a=1}^{6^n} (\mathcal{T}_n)_{ba} \sum_{l=1}^{6^n}(\hat{T}_{\mathcal{E}_2})_{al}\mathcal{M}^{-1}(\mathcal{S}^{\otimes n}_l)\right]$.
    
    Let
    \begin{equation}\label{epsilon1}
        \tilde{\epsilon}_1 = \frac{\eta}{2\|(\mathcal{E}_2^\dagger(O))^{(\nu)}\|\mathcal{O}(n^{\nu})}\left(\frac{\epsilon^{\prime}}{6\cdot 2^{\kappa}}\right)^{\kappa+1}\left(\frac{C(\nu,d)}{3}\right)^2\left(\frac{C(\kappa)}{3}\right)^{2\kappa}-\frac{(\sqrt{N_1}  +1 )(\|O\|+\|(\mathcal{E}_2^\dagger(O))^{(\nu)}\|)}{\|(\mathcal{E}_2^\dagger(O))^{(\nu)}\|\mathcal{O}(n^{\nu})},
    \end{equation}
    \begin{equation}\label{epsilon2}
        \tilde{\epsilon}_2 = \frac{1}{2 \sqrt{N_1}}\left(\frac{\epsilon^{\prime}}{6\cdot 2^\kappa}\right)^{\kappa+1}\left(\frac{C(\nu,d)}{3}\right)^2\left(\frac{C(\kappa)}{3}\right)^{2\kappa}.
    \end{equation}
    
    And we have
    \begin{equation}\label{epsilon_2}
            \tilde{\epsilon} =\sqrt{N_1}\tilde{{\epsilon}}_2 +\frac{\|(\mathcal{E}_2^\dagger(O))^{(\nu)}\|\tilde{\epsilon}_1 \mathcal{O}(n^{\nu})}{\eta} + \frac{ (\sqrt{N_1}  +1 )(\|O\|+\|(\mathcal{E}_2^\dagger(O))^{(\nu)}\|)}{\eta} = \left(\frac{\epsilon^{\prime}}{6\cdot2^{\kappa}}\right)^{\kappa+1}\left(\frac{C(\nu,d)}{3}\right)^2\left(\frac{C(\kappa)}{3}\right)^{2\kappa}.
        \end{equation}
    Applying the norm inequality from Lem.~\ref{lem:norm inequality}, we have:
    \begin{equation}
        \sum_{P:|P|\leq \kappa} \left|\alpha_{P}(O)\right|^{r} \leq \left(\frac{3}{C(\kappa)}\right)^{r} \left\|O^{({\text{unk},\kappa})}\right\|^{r}
    \end{equation}where $r = 2\kappa/(\kappa+1)$.
    
    Thus, by the filtering lemma given in Lem.~\ref{lem:filtering_kappa}, we obtain
    \begin{equation}
    \sum_{P : |P| \leq \kappa} \left(\frac{1}{3}\right)^{|P|} |\hat{\alpha}_P(O) - \alpha_P(O)|^2 \leq 6 \left(\frac{3}{C(\kappa)} \right)^{r}\|O^{(\mathrm{unk},\kappa)}\|^{r} \left(\frac{3}{C(\nu,d)}\right)^{2-r}\|O\|^{2-r} \tilde{\epsilon}^{1-r/2}
    \end{equation}

    Combining with the definition of $\tilde{\epsilon}$ given in Eq.~\eqref{epsilon_2}, we have \begin{equation}
    \sum_{P : |P| \leq \kappa} \left(\frac{1}{3}\right)^{|P|} |\hat{\alpha}_P(O) - \alpha_P(O)|^2 \leq \left[ \frac{\|O^{(\mathrm{unk},\kappa)}\|}{\|O\|} \right]^{2\kappa/(\kappa+1)} \frac{\epsilon'}{2^\kappa} \|O\|^2.
    \end{equation}
    The nonlinear purity property of $\rho$ under the distribution $\mathcal{D}$, we obtain is that for all $P$ with $|P|\leq \kappa$:
    \begin{equation}
        \mathbb{E}_{\rho \sim \mathcal{D}}[\gamma^{*}(\rho_{\mathrm{dom}(P)})] {\left(\frac{2}{3}\right)}^{|P|} \leq 2^{\kappa} \left(\frac{1}{3}\right)^{|P|}.
    \end{equation}
    
    Therefore, by Lem.~\ref{Lemma:estimation error}, we obtain:
    \begin{equation}\label{eq:estimation_error2}
    \mathop{\mathbb{E}}_{\rho \sim \mathcal{D}} \left|h(\rho,O) - \mathrm{Tr}(O^{(\mathrm{unk},\kappa)}\rho)\right|^2 \leq \sum_{P : |P| \leq \kappa}\mathbb{E}_{\rho \sim \mathcal{D}}[\gamma^{*}(\rho_{\mathrm{dom}(P)})] {\left(\frac{2}{3}\right)}^{|P|}|\hat{\alpha}_P(O) - \alpha_P(O)|^2 \leq  \epsilon' \left[ \frac{\|O^{(\mathrm{unk},\kappa)}\|}{\|O\|} \right]^{2\kappa/(\kappa+1)} \|O\|^2.
    \end{equation}

    Finally, combining the truncation error (Eq.~\eqref{leq:truncation_error2}) with the estimation error (Eq.~\eqref{eq:estimation_error2}) via the triangle inequality, we establish the overall prediction guarantee:
    \begin{equation}
    \mathop{\mathbb{E}}_{\rho \sim \mathcal{D}} \left| h(\rho, O) - \mathrm{Tr}[O \mathcal{E}_1^{\dagger}(\rho)] \right|^2 \leq \left(\epsilon + \epsilon' \left[ \frac{\|O^{(\mathrm{unk},\kappa)}\|}{\|O\|} \right]^{2\kappa/(\kappa+1)}\right) \|O\|^2.
    \end{equation}
    
    Next, we determine the required size of the classical shadows to achieve the desired error bound.
    
    From the conclusion:
    \begin{equation}
    N_1 =\mathcal{O}\biggl(\frac{\log(n^{\kappa+\nu}N_1/\delta)}{ {\tilde{\epsilon}_1}^2}\biggr) ,N_2 = \mathcal{O}\biggl(\frac{\log(n^{\kappa +\nu}N_2/\delta)}{{\tilde{\epsilon}_2}^2}\biggr),
    \end{equation}
    we derive the explicit sample complexities:
    \begin{equation}\label{N_total}
    N_1 =\mathcal{O}\left( \frac{\log(n^{\nu+\kappa} / (\delta {\tilde{\epsilon}_1}^2))}{{\tilde{\epsilon}_1}^2} \right),  N_2=\mathcal{O}\left( \frac{\log(n^{\nu+\kappa} / (\delta {\tilde{\epsilon}_2}^2))}{{\tilde{\epsilon}_2}^2} \right)
    \end{equation}
    Substituting the expressions for $\tilde{\epsilon}_1$ and $\tilde{\epsilon}_2$ from Eq.~\eqref{epsilon1} and Eq.~\eqref{epsilon2}, respectively, we obtain the final complexity bounds:
    \begin{equation}
        N_1 = N_2 = \log\left(\frac{n}{\delta}\right)2^{\mathcal{O}\left(\log(n) + \log(1/\epsilon)\left(\log \log(1/\epsilon)+\log(1/\epsilon')\right)\right)}.
    \end{equation}
    Given the classical shadows of size $N_1$ for $\mathcal{E}_1$ and of size $N_2$ for $\mathcal{E}_2$, shadow engineering can predict $\mathrm{Tr}[O \mathcal{E}_2\mathcal{E}_1(\rho)]$ with probability at least $1-\delta$. The prediction error satisfies:
    \begin{equation}
    \mathop{\mathbb{E}}_{\rho \sim \mathcal{D}} \left| h(\rho, O) - \mathrm{Tr}[O \mathcal{E}_2\mathcal{E}_1(\rho)] \right|^2 \leq \left(\epsilon + \epsilon' \left[ \frac{\|O^{(\mathrm{unk},\kappa)}\|}{\|O\|} \right]^{2\left\lceil \log_{1.5} \left( 1/\epsilon \right) \right\rceil/\left(\left\lceil \log_{1.5} \left( 1/\epsilon \right) \right\rceil+1\right)}\right) \|O\|^2.
    \end{equation}
    
    Next, We introduce a secondary parameter $\kappa$ defined as: $\kappa = \left\lceil \log_{1.5} \left( {2}/{\epsilon} \right) \right\rceil$. 
    The proof follows the same lines as for the first parameter set, so we state the corresponding result directly.

    The error parameters $\tilde{\epsilon}_1$ and $\tilde{\epsilon}_2$ are given by:
    \begin{equation}\label{epsilon11}
        \tilde{\epsilon}_1 = \frac{\eta}{2\|(\mathcal{E}_2^\dagger(O))^{(\nu)}\|\mathcal{O}(n^{\nu})}\frac{\epsilon}{9\cdot2^{\kappa+1}\cdot n^{\kappa}}\left(\frac{C(\nu,d)}{3}\right)^2-\frac{(\sqrt{N_1}  +1 )(\|O\|+\|(\mathcal{E}_2^\dagger(O))^{(\nu)}\|)}{\|(\mathcal{E}_2^\dagger(O))^{(\nu)}\|\mathcal{O}(n^{\nu})},
    \end{equation}
    \begin{equation}\label{epsilon12}
        \tilde{\epsilon}_2 = \frac{1}{2 \sqrt{N_1}}\frac{\epsilon}{9\cdot2^{\kappa+1}\cdot n^{\kappa}}\left(\frac{C(\nu,d)}{3}\right)^2.
    \end{equation}
    where $\eta=\|O\|_{\text{Pauli,1}}$ is a constant.

    Substituting these into the expression for the filtering error $\tilde{\epsilon}$ yields:
    \begin{equation}\label{epsilon_12}
            \tilde{\epsilon} =\sqrt{N_1}\tilde{{\epsilon}}_2 +\frac{\|(\mathcal{E}_2^\dagger(O))^{(\nu)}\|\tilde{\epsilon}_1 \mathcal{O}(n^{\nu})}{\eta} + \frac{ (\sqrt{N_1}  +1 )(\|O\|+\|(\mathcal{E}_2^\dagger(O))^{(\nu)}\|)}{\eta} = \frac{\epsilon}{9\cdot2^{\kappa+1}\cdot n^{\kappa}}\left(\frac{C(\nu,d)}{3}\right)^2.
        \end{equation}

    Plugging $\tilde{\epsilon}_1$ and $\tilde{\epsilon}_2$ into Eq.~\eqref{N_total} gives the required sample sizes for the classical shadows of $\mathcal{E}_1$ and $\mathcal{E}_2$:
    \begin{equation}
    N_1 = N_2 =  \log\left(\frac{n}{\delta}\right) \cdot 2^{\mathcal{O}\left(\log(1/\epsilon)\log(n)\right)}.
    \end{equation}
    With these parameters, shadow engineering predicts $\mathrm{Tr}[O \mathcal{E}_2\mathcal{E}_1(\rho)]$ with probability at least $1-\delta$, where the prediction error satisfies:
    \begin{equation}
        \mathop{\mathbb{E}}_{\rho \sim \mathcal{D}} \left| h(\rho, O) - \mathrm{Tr}[O \mathcal{E}_2 \circ \mathcal{E}_1(\rho)] \right|^2 \leq \epsilon \|O\|^2.
    \end{equation}
    
    In conclusion, given the classical shadows of $\mathcal{E}_1$ and $\mathcal{E}_2$ with sizes:
    \begin{equation}
        N_1 = N_2 = \log(\frac{n}{\delta}) \text{min}(2^{\mathcal{O}[\log(n)+\log(1/\epsilon)(\log\log(1/\epsilon) + \log(1/\epsilon^{\prime}))]},  2^{O\left[\log(1/\epsilon) \log(n)\right]}),
    \end{equation}
    for any observable $O$ that is a sum of $\nu$-qubit observables (where $\nu = \mathcal{O}(1)$ and each qubit is acted on by $d = \mathcal{O}(1)$ of these $\nu$-qubit observables), and for any $n$-qubit state distribution $\mathcal{D}$ invariant under single-qubit $H$ and $S$ gates, shadow engineering can predict $\mathrm{Tr}[O \mathcal{E}_2 \circ \mathcal{E}_1(\rho)]$ with probability at least $1-\delta$. The prediction error satisfies:
    \begin{equation}
    \mathop{\mathbb{E}}_{\rho \sim \mathcal{D}} \left| h(\rho, O) - \mathrm{Tr}[O \mathcal{E}_2\circ \mathcal{E}_1(\rho)] \right|^2 \leq \left(\epsilon + \epsilon' \left[ \frac{\|O^{(\mathrm{unk},\kappa)}\|}{\|O\|} \right]^{2\left\lceil \log_{1.5} \left( 1/\epsilon \right) \right\rceil/\left(\left\lceil \log_{1.5} \left( 1/\epsilon \right) \right\rceil+1\right)}\right) \|O\|^2.
    \end{equation}
    This completes the proof of Theorem 2. 
\end{proof}

\section{Details of experiments}
In this section, we assess the performance of the shadow engineering framework using numerical simulations and experimental demonstrations on a superconducting quantum processor. We first benchmark our approach against state-of-the-art methods, including the scheme from Ref.~\cite{huang2023learning} and quantum process tomography (QPT)  \cite{mohseni2008quantum}. We then experimentally implement the framework on programmable superconducting hardware. After specifying the experimental setup, we further demonstrate the efficacy of shadow engineering for error mitigation and Hamiltonian dynamics simulation under realistic hardware noise.

\subsection{Numerical simulation experiments}
Our theoretical analysis establishes that shadow engineering achieves polynomial sample complexity in the number of qubits, verifying its practical efficiency and scalability to large-scale quantum systems. 
To empirically validate this scaling, we perform numerical benchmarks against both the method of Ref.~\cite{huang2023learning} and QPT.

For performance benchmarking, we compare our method against existing approaches, including the method proposed in Ref.~\cite{huang2023learning} and QPT. 
The comparison serves two key purposes: first, to directly comparing the prediction error of $\mathrm{Tr}[O\mathcal{E}_1^{\dagger}(\rho)]$ achieved by shadow engineering with that for $\mathrm{Tr}[O\mathcal{E}_1(\rho)]$ obtained using the method in Ref.~\cite{huang2023learning}; 
second, to benchmark the prediction accuracy of shadow engineering framework for both $\mathrm{Tr}[O\mathcal{E}_1^{\dagger}(\rho)]$ and $\mathrm{Tr}[O\mathcal{E}_2\circ\mathcal{E}_1(\rho)]$ against the corresponding results from QPT.

\noindent\textbf{Experiment 1: Comparison with the method proposed in Ref.~\cite{huang2023learning}.}

To realize the target quantum process $\mathcal{E}_1$, we design a fixed-architecture parameterized quantum circuit with ten layers. Each layer comprises single-qubit rotation gates $RX$, $RY$, and $RZ$ together with all-to-all $\text{CNOT}$ gates. 
Different circuit realizations are generated by uniformly sampling parameter vectors of length $30\times n$ over the interval $[-\pi, \pi)$, where the qubit number $n$ ranges from $2$ to $8$ qubits.

For each system size, we produce $10^5$ classical shadows of $\mathcal{E}_1$ following a standardized pipeline: sampling $10^5$ random input states $\mathcal{S}^{\otimes n}$, executing the parameterized circuit, and performing random Pauli-basis measurements. 
The full dataset is then subsampled into subsets with sizes $[10^4, 2\times10^4, \dots, 10^5]$. 
Both our shadow engineering framework and the method proposed in Ref.~\cite{huang2023learning} adopt identical system configurations and shadow sample for fair comparison. 
A fixed observable $O = \sum_{j=1}^{n-1} (0.27 \sigma_j^x \sigma_{j+1}^x + 0.42 \sigma_j^y \sigma_{j+1}^y + 0.76 \sigma_j^z \sigma_{j+1}^z + 0.6 \sigma_j^z)$ and Haar-random input states $\rho$ are adopted throughout all prediction tasks. 
Our scheme reconstructs the transfer matrix $\hat{T}_{\mathcal{E}_1^\dagger}$ from the classical shadows of $\mathcal{E}_1$ to estimate $\mathrm{Tr}[O\mathcal{E}_1^\dagger(\rho)]$. In contrast, the method proposed in Ref.~\cite{huang2023learning} directly uses classical shadows of the process $\mathcal{E}_1$ to predict $\mathrm{Tr}[O\mathcal{E}_1(\rho)]$.

Prediction accuracy is quantified via the root mean square error (RMSE) over 50 independent repeated trials to ensure statistical robustness:
\begin{equation}
\text{RMSE} = \sqrt{\frac{1}{50}\sum_{i=1}^{50}\left(\langle O \rangle^{\text{est}}_i - \langle O \rangle^{\text{ideal}}_i \right)^2},
\end{equation}
Here, $\langle O \rangle^{\text{est}}_i$ denotes the estimated expectation value in the $i$-trial, while $\langle O \rangle^{\text{ideal}}_i$ represents the exact theoretical values. 
Numerical comparisons show consistent RMSE performance across all tested qubit numbers ($n=2~\text{to}~ 8$). This confirms that our shadow engineering framework achieves prediction accuracy comparable to the method in Ref.~\cite{huang2023learning}.
Furthermore, our method expands the applicable scope of classical shadow resorces: it enables reliable expectation-value prediction for adjoint-channel properties, even with a finite, limited number of shadow snapshots sampled solely from the original quantum process $\mathcal{E}_1$.

\noindent\textbf{Experiment 2: Comparison with QPT.}

We adopt the same target process $\mathcal{E}_1$ defined in Experiment 1, and introduce an additional ten-layer parameterized quantum circuit $\mathcal{E}_2$ with independently randomized parameter set.
For both $n=2$ and $n=4$ qubit systems, we generate classical shadows for $\mathcal{E}_2$ using the same range of sample sizes ($10^4$ to $10^5$) as in the previous experiment. 
For each fixed sample budget, we first reconstruct the transfer matrices of $\mathcal{E}_1$ and $\mathcal{E}_2$ individually from their respective classical shadow datasets. 
Our Shadow engineering framework then directly constructs the effective transfer matrix of the composite channel $\mathcal{E}_2 \circ \mathcal{E}_1$, enabling predictions of $\mathrm{Tr}[O\mathcal{E}_2\mathcal{E}_1(\rho)]$ with the same observable $O$ and Haar-random input states $\rho$ employed throughout all tests.

Numerical results verify that our method yields reliable predictions for both $\mathrm{Tr}[O\mathcal{E}_1^{\dagger}(\rho)]$ and $\mathrm{Tr}[O\mathcal{E}_2\mathcal{E}_1(\rho)]$, with errors scaling polynomially with the qubit number—fully consistent with our theoretical analysis. 
As physically expected, the prediction error for the composite channel prediction is moderately larger than for the single channel case, owing to accumulated statistical fluctuations across two sequential channels.

We further benchmark shadow engineering against QPT, highlighting a fundamental difference in resource requirements.
Shadow engineering relies solely on classical shadows of $\mathcal{E}_1$ and $\mathcal{E}_2$, while QPT demands substantially more resources. Predicting $\mathrm{Tr}[O\mathcal{E}_1^{\dagger}(\rho)]$ with QPT requires full tomography of $\mathcal{E}_1^{\dagger}$, and predicting $\mathrm{Tr}[O\mathcal{E}_2\mathcal{E}_1(\rho)]$ requires full tomography of both $\mathcal{E}_1$ and $\mathcal{E}_2$ followed by exponential-time post-processing to compute the Kraus operators of $\mathcal{E}_2 \circ \mathcal{E}_1$.

To ensure a fair comparison, we allocate identical measurement budgets to both methods at every system size, using the same sample-size sequence $\{10^4, 2\times10^4, \dots, 10^5\}$. We perform full QPT on $\mathcal{E}_1$, $\mathcal{E}_2$, and $\mathcal{E}_1^{\dagger}$, then execute the required post-processing pipelines to compute the two target expectation values.

The results clearly demonstrate that:
\begin{enumerate}
    \item Across all tested system sizes, QPT exhibits significantly larger errors than shadow engineering across all system sizes;
    \item QPT's predictions for the composite observable $\mathrm{Tr}[O\mathcal{E}_2\mathcal{E}_1(\rho)]$ suffer from severe error propagation due to independent tomography of each channel;
    \item The performance gap between shadow engineering and QPT widens as qubit count increases.
\end{enumerate}

This fundamental performance gap originates from their inherent sample complexity scalings: shadow engineering achieves polynomial scaling in both measurement resources and computational cost, whereas standard QPT inevitably incurs exponential overhead with increasing system size.

\subsection{Hardware experiments}
In this section, we validate the effectiveness of our shadow engineering framework on a superconducting quantum processor. We begin with a description of the experimental device, then focus on two core objectives: demonstrating efficient estimation of the adjoint and concatenated channels, and verifying the framework’s practical utility for error mitigation and Hamiltonian dynamical simulation in real quantum systems.
\subsubsection{Device information}
\begin{figure}[h!]

  \centering
  {\includegraphics[width=0.60\textwidth]{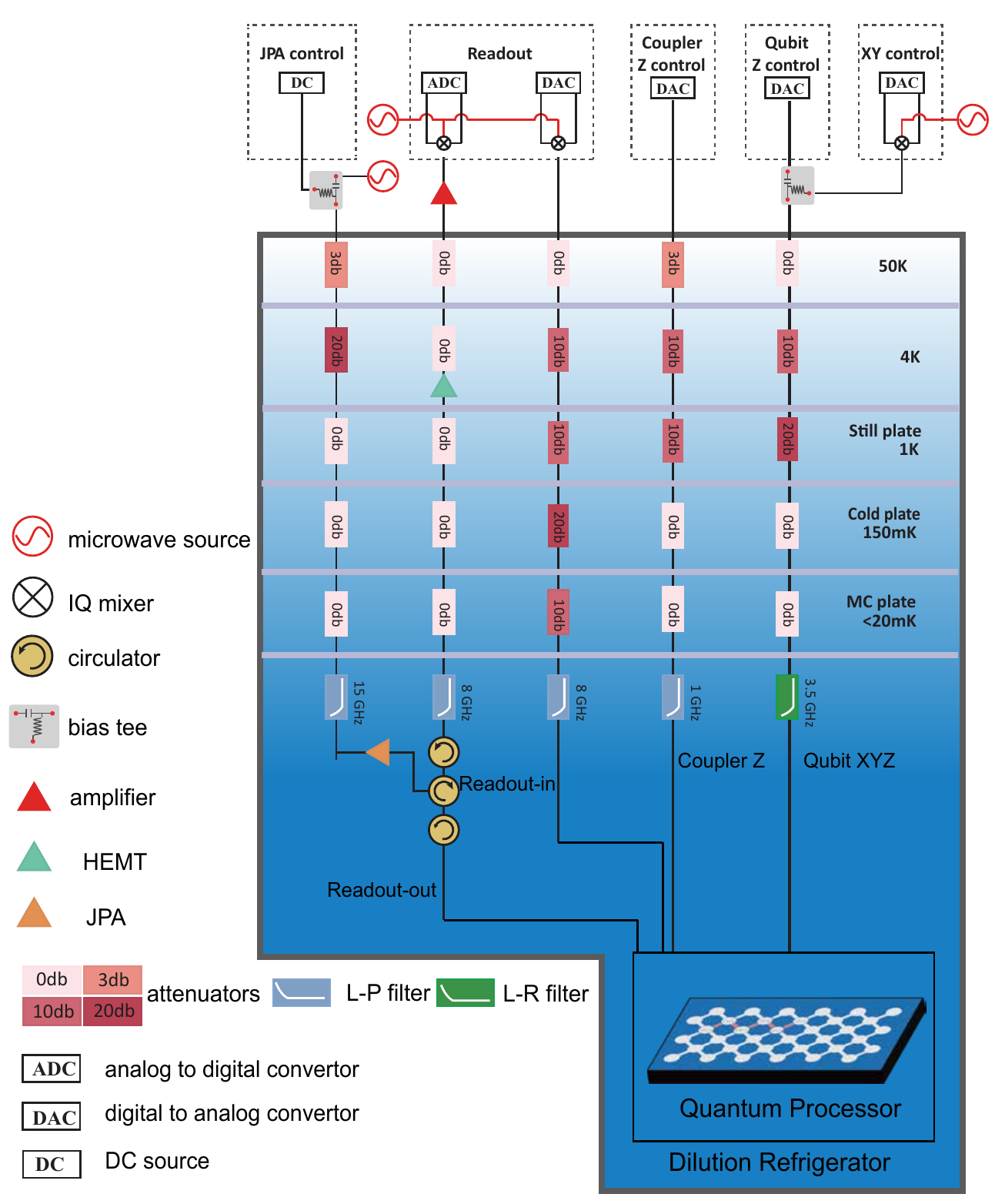}} 
  \caption{{\textbf{Schematic of the superconducting quantum processor.} 
  Each qubit is equipped with independent XY and Z control lines, where room-temperature digital-to-analog converters (DACs) generate the corresponding XY and Z control pulses that are combined via bias tees before being sent to the quantum processor. Microwave sources and mixers produce readout probe pulses and pump signals for the Josephson parametric amplifier (JPA), and attenuators and filters are placed at multiple temperature stages to suppress noise. The readout signals are amplified in three stages: first by a JPA at 20 mK, then by a high-electron-mobility transistor (HEMT) at 4 K, and finally by room-temperature amplifiers, with subsequent digitization and demodulation implemented by analog-to-digital converter (ADC) modules. DC sources provide the static flux bias for JPA operation, and flux bias lines enable precise frequency tuning by combining Z-control pulses with microwave signals.  }}
  \label{device}
\end{figure}

\begin{figure}[h!]
  \centering

  {\includegraphics[width=1\textwidth]{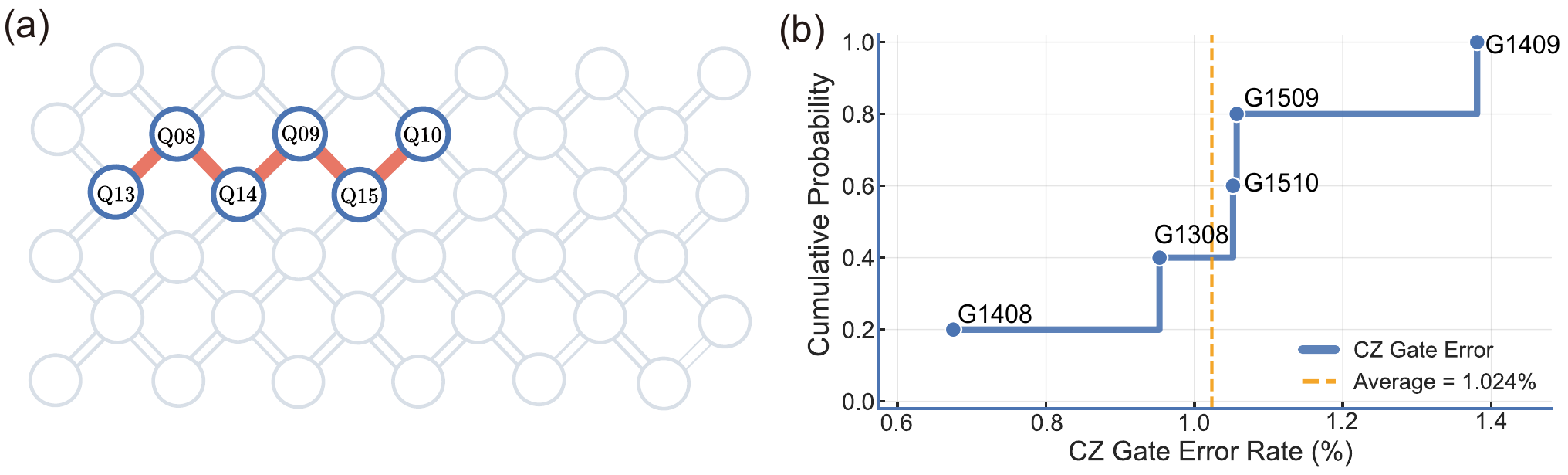}} 
  \caption{{\textbf{Qubit position and CZ gate error rates.} 
  (a) We use a six-qubit chain embedded in a 36-qubit superconducting processor—a $6\times 6$ array of frequency-tunable transmon qubits with tunable couplers. 
  (b) The chain contains five two-qubit gates with an average error rate of $1.024\%$.  }}
  \label{CZ_CDF}
\end{figure}

We perform experiments on a programmable superconducting quantum processor operating below 20 mK in a dilution refrigerator, as shown in Fig.~\ref{device}. The processor consists of a $6\times 6$ two-dimensional array of frequency-tunable transmon qubits in a cross architecture, with nearest-neighbor coupling enabled by dedicated tunable couplers. For our study, we implement a contiguous six-qubit chain within this array (see Fig.~\ref{CZ_CDF}), using layouts optimized for our specific experimental objectives. 
Each qubit is addressed by independent $XY$ and $Z$ control lines: room-temperature digital-to-analog converters (DACs) generate the corresponding control pulses, which are combined via bias tees prior to input to the processor. Microwave sources and mixers produce readout probe pulses and Josephson parametric amplifier (JPA) pump signals, while attenuators and filters are strategically placed across multiple temperature stages to minimize noise and optimize performance.
Readout signals undergo three-stage amplification: initial boosting via a JPA at 20 mK, followed by amplification with a high-electron mobility transistor (HEMT) at 4 K, and final amplification with room-temperature amplifiers. Subsequent digitization and demodulation are carried out using analog-to-digital converter (ADC) modules. DC sources supply the static flux bias required for JPA operation, and flux bias lines enable precise frequency tuning via the combination of Z-control pulses and microwave signals.
After calibration and optimization, key performance metrics of the six-qubit linear chain are summarized in Tab.~\ref{tab:qubit_metrics}: the median qubit idle frequency is 5.262 GHz, with median T1 relaxation and T2 coherence times of 58.5 µs and 6.65 µs, respectively. The median readout error across qubits is $1.255\%$, and the median single-qubit gate error is $0.115\%$. The average CZ gate error is $1.024\%$, whose distribution is shown in Fig.~\ref{CZ_CDF}. These parameters are determined through rigorous characterization to ensure the reliability and accuracy of our experimental results.

\begin{table}[h!]
\centering

\footnotesize
\renewcommand{\arraystretch}{1.5}
\renewcommand{\tabcolsep}{4pt}
\makebox[0.98\columnwidth][c]{
\begin{tabular}{||c c c c c c c c c||}  
 \hline  
 Metric & median & mean & Q13 & Q08 & Q14 & Q09 & Q15 & Q10 \\
 \hline\hline
 $f_{01}$ ($\mathrm{GHz}$) & 5.262 & $5.243 \pm 0.074$ & 5.120 & 5.195 & 5.281 & 5.332 & 5.243 & 5.285 \\
 $f_{01} - f_{12}$ ($\mathrm{MHz}$) & 234.00 & $234.83 \pm 5.62$ & 246 & 234 & 234 & 230 & 235 & 230 \\
 $T_1$ ($\mu\mathrm{s}$) & 58.50 & $56.07 \pm 6.66$ & 62.9 & 58.7 & 53.0 & 44.2 & 58.3 & 59.3 \\
$T_2$ ($\mu\mathrm{s}$) & 6.65 & $6.867 \pm 1.552$ & 4.6 & 8.6 & 8.8 & 7.0 & 5.9 & 6.3 \\
 1Q error (\%) & 0.115 & $0.122 \pm 0.033$ & 0.19 & 0.11 & 0.10 & 0.12 & 0.09 & 0.12 \\
 readout error (\%) & 1.255 & $1.362 \pm 0.443$ & 1.20 & 1.36 & 2.33 & 0.97 & 1.00 & 1.31 \\
 \hline
\end{tabular}
}
\caption{ \textbf{Performance summary of the single qubits used on our superconducting quantum computer.}  Reported T1, T2, 1Q error, readout error were obtained from daily measurements over a 15 day period}
\label{tab:qubit_metrics}
\end{table}

{\subsubsection{Implementation details for adjoint channel estimation}
In this section, we benchmark the shadow engineering framework against the method of Ref.~\cite{huang2023learning} for characterizing coherent noise channels.  
Using quantum circuits of tunable depths to emulate noise channels with controlled strengths, we quantitatively compare the characterization accuracy of both approaches. 
Furthermore, we extend shadow engineering to error mitigation, demonstrating reliable performance across the tested noise range.
}

{\noindent\textbf{Circuit implementation details.} }

\begin{figure}[h]

  \centering
  {\includegraphics[width=0.6\textwidth]{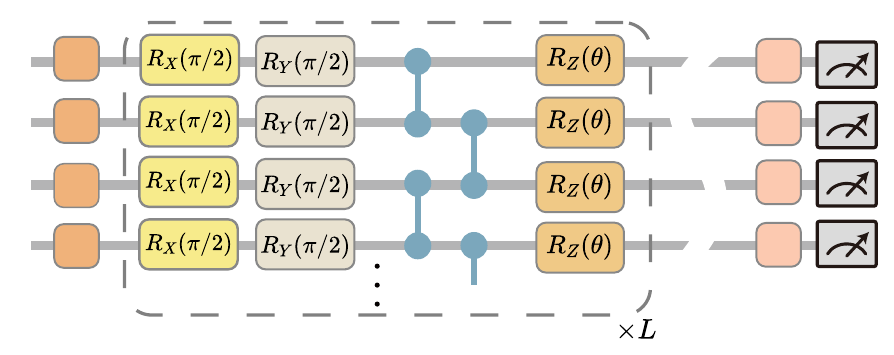}} 
  \caption{{\textbf{Quantum circuit implementation of adjoint channel estimation and error mitigation application.} 
  The $N$-qubit circuit used in noise channel characterization consists of three stages: 
  (i) stabilizer state preparation;
  (ii) $L$ layers of the noise channel $\mathcal{E}_M$, each containing $N$ single-qubit rotations $R_X, R_Y, R_Z$ and $(N-1)$ CZ gates;
  (iii) $N$ single-qubit Pauli measurements implemented via rotation gates and $Z$-basis measurements.
  }}
  \label{EM_circuit}
\end{figure}

{We use parameterized quantum circuits of variable depth to simulate coherent noise channels with tunable strength.
The noise channel $\mathcal{E}_M$ is constructed as follows: a layer of single-qubit $R_X(\pi/2)$ gates is applied, followed by a layer of $R_Y(\pi/2)$ gates; two layers of parallel controlled-$Z$ (CZ) gates are then implemented, and the sequence is completed with a layer of random $R_Z$ rotations.
The sampling procedure, illustrated in Fig.~\ref{EM_circuit}, starts with the preparation of an initial state $\mathcal{S}^{\otimes n}_i$ chosen from the $n$-qubit stabilizer basis $\mathcal{S}^{\otimes n}$ using single-qubit rotations. The noise channel $\mathcal{E}_M$ is subsequently applied.
By varying the circuit depth---that is, the number of repetitions of $\mathcal{E}_M$---we realize a range of effective noise strengths. A single-qubit unitary $U\in\{H,HS^\dagger,\mathbb{I}\}$ is then applied to each qubit, followed by measurement in the computational ($Z$) basis.
From repeated experimental realizations, we reconstruct the classical shadow of the noise channel $\mathcal{E}_M$.
}

{\noindent\textbf{Experiment 3: Comparison with the method in Ref.~\cite{huang2023learning} for noise channel characterization.}
}

{We use the six qubits listed in Tab.~\ref{tab:qubit_metrics} as our target system, chosen for their favorable coherence properties and regular nearest-neighbor connectivity.
As illustrated in Fig.~\ref{EM_circuit}, the sampling protocol is implemented by sequentially applying single-qubit rotations and CZ gates to construct the noise channel $\mathcal{E}_M$ with depths ranging from 1 to 4 layers.
Each additional layer increases the effective noise strength, enabling a systematic investigation of the framework's robustness across different noise regimes.
}

{The experimental procedure is as follows.
We randomly select 200 initial states $\mathcal{S}^{\otimes n}_i$ from the $n$-qubit stabilizer set $\mathcal{S}^{\otimes n}$ and 400 random Pauli measurement bases, generating 400 distinct sampling circuits for each circuit depth.
Each circuit is executed with 1000 shots on the superconducting quantum processor, yielding a total of $4 \times 10^5$ classical shadows across all depths.
}

{In post-processing, we analyze subsampled datasets of sizes $[2, 6, \dots, 30] \times 10^4$.
For each subsample, we reconstruct the transfer matrix $\hat{T}_{\mathcal{E}_M^\dagger}$ that characterizes the adjoint noise channel using Lem.~\ref{lemma1}.
We employ the observable $O = \sum_{i=1}^6 Z_i$ and generate 6-qubit Haar-random states $\rho$ as test input states.
The shadow engineering framework estimates $\operatorname{Tr}[O\mathcal{E}_M^\dagger(\rho)]$ by directly learning the transfer matrix of the adjoint channel.
For benchmarking, we apply the method from Ref.~\cite{huang2023learning} to estimate $\operatorname{Tr}[O\mathcal{E}_M(\rho)]$ using the same classical shadow data, observable, and input states.
To ensure statistical reliability, we perform 20 independent repetitions for each configuration with randomized subsampling and estimation, and compute the RMSE between ideal theoretical values (from noiseless simulations) and experimental estimates.
}

{Experimental results demonstrate that shadow engineering achieves predictive accuracy comparable to the method in Ref.~\cite{huang2023learning} for noise channels of similar complexity, validating the efficacy of directly learning the adjoint channel $\mathcal{E}_M^\dagger$ from classical shadows of $\mathcal{E}_M$.
At fixed depth and sample size, the RMSE values for $\operatorname{Tr}[O\mathcal{E}_M^\dagger(\rho)]$ and $\operatorname{Tr}[O\mathcal{E}_M(\rho)]$ are statistically indistinguishable over the tested parameter range.
We also observe a mild monotonic increase in prediction error with rising circuit depth, consistent with accumulated decoherence and gate imperfections that enhance effective noise, affecting the characterization fidelity of both methods similarly.
}

{\noindent\textbf{Experiment 4: Error mitigation capability under varying noise strengths.}
}

{In this experiment, we extend the noise channel characterization developed in Experiment 3 to the task of quantum error mitigation.
Specifically, given a noisy state $\rho_{\rm noisy} = \mathcal{E}_M(\rho)$, shadow engineering allows us to estimate
\[
\operatorname{Tr}[O\mathcal{E}_M^\dagger(\mathcal{E}_M(\rho))] = \operatorname{Tr}[O\rho],
\]
thus recovering the ideal expectation value and realizing practical error mitigation.
}

{We use a six-qubit Haar-random state as the initial pure state $\rho$ and subject it to the noise channel $\mathcal{E}_M$ with depths from 1 to 4 layers.
For each noisy output state, we perform 500 random single-qubit Pauli measurements with 1000 shots per measurement setting, and reconstruct $\rho_{\rm noisy} = \mathcal{E}_M(\rho)$ using the classical shadows of it.
}

{Using the same observable $O = \sum_{i=1}^6 Z_i$ as in Experiment 3 and the previously acquired classical shadows of the noise channel (totaling $5\times10^5$), we subsample data at the same intervals $[2, 6, \dots, 30] \times 10^4$.
For each subsample, we reconstruct the corresponding transfer matrix and use shadow engineering to estimate $\operatorname{Tr}[O\mathcal{E}_M^\dagger(\rho_{\rm noisy})]$.
To quantify mitigation performance across different sample sizes, we define the error mitigation ratio:
\begin{equation}
    r = \frac{|h(\hat{\rho}_{\text{noisy}}, O) - \mathrm{Tr}[O\rho]|}{|\mathrm{Tr}[O\rho_{\text{noisy}}] - \mathrm{Tr}[O\rho]|}.
    \label{ratio}
\end{equation}
This ratio measures the residual error after mitigation relative to the original noise-induced error, where a smaller 
$r$ indicates more effective mitigation.
In addition, using Eq.~\ref{ratio}, we compute the error mitigation ratio $r$ from the ideal expectation value $\operatorname{Tr}[O\rho]$ (obtained via noiseless simulation) and the noisy expectation value $\operatorname{Tr}[O\rho_{\rm noisy}]$.
Under identical experimental conditions, we conduct 20 independent trials with randomly subsampled classical shadows to obtain the average error mitigation ratio $r$.
}

{The results show that at small sample sizes, the error mitigation ratio for $L=4$ (corresponding to the strongest noise) is higher than those for shallower circuit depths.
As the sample size increases, the mitigation ratios across all depths converge to lower values, indicating improved error mitigation performance with increased sampling.
}

{\subsubsection{Implementation details for concatenated channel estimation}
}

{In this section, we extend the capability of the shadow engineering framework for concatenated channel estimation to Hamiltonian dynamical simulation, demonstrating that our framework enables efficient prediction of quantum many-body dynamics without requiring explicit physical implementation of longer-time evolution.
We further demonstrate high-fidelity prediction of Hamiltonian-evolved properties, including total magnetization, nearest-neighbor correlations, and multi-body correlations.
}

{In these experiments, we implement the time-evolution operator for the one-dimensional transverse-field Ising model (TFIM) Hamiltonian via first-order Trotter-Suzuki decomposition.
The system Hamiltonian is given by
\begin{equation}
    \mathsf{H}_{\mathsf{TFIM}} = -J\sum_i Z_iZ_{i+1} - h\sum_i X_i,
\end{equation}
where we set the coupling strength $J=-1$ and transverse field strength $h=-1$.
The antiferromagnetic interaction $J<0$ favors antiparallel spin alignment between neighboring qubits, while the transverse field $h$ induces single-qubit rotations along the $X$-axis, creating competing dynamics.
The unit ratio $|h/J|=1$ places the system precisely at the quantum critical point, where quantum fluctuations are maximized and correlation functions exhibit algebraic decay.
}

{\noindent\textbf{Circuit implementation details.}}
\begin{figure}[h]
  \centering
 
  \includegraphics[width=0.8\textwidth]{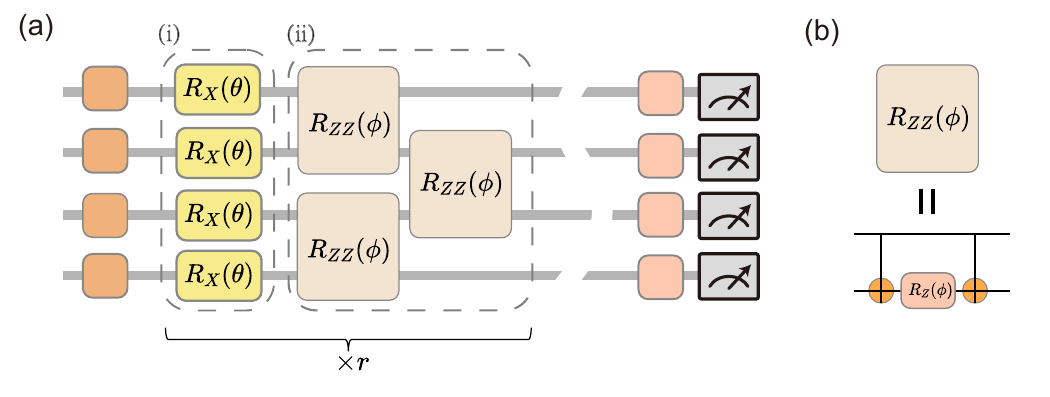} 
  \caption{\textbf{Quantum circuit implementations of Hamiltonian dynamical simulation.}
  (a) The $N$-qubit circuit for concatenated channel characterization consists of three stages:
  (1) stabilizer state preparation;
  (2) $r$ layers of Trotter-decomposed Hamiltonian time evolution, where each layer contains (i) $N$ single-qubit $R_X$ rotations and (ii) $(N-1)$ two-qubit $R_{ZZ}$ entangling gates;
  (3) $N$ single-qubit Pauli measurements implemented via basis rotation gates followed by $Z$-basis readout.
  (b) Realization of the two-qubit $R_{ZZ}(\phi)$ gate on the superconducting quantum processor, decomposed into two CNOT gates and one single-qubit $R_Z(\phi)$ rotation.}
  \label{HD_circuit}
\end{figure}

{For evolution time $t_1=0.5$ and Trotter step number $r=1$~\cite{PhysRevLett.128.210501}, the single-step evolution channel is approximated as
\begin{equation}
    \mathcal{E}_H(t) = e^{-i\mathsf{H}_{\mathsf{TFIM}}t} \approx e^{-iht\sum_i X_i} \cdot e^{-iJt\sum_i Z_iZ_{i+1}}.
\end{equation}
Leveraging the commutativity of terms within the first-order Trotter decomposition, the corresponding quantum circuit takes the form
\begin{equation}
    \mathcal{E}_H(t_1) = \left[\prod_{i=1}^N R_{X_i}(2ht)\right] \cdot \left[\prod_{i=1}^{N-1} R_{ZZ_{i,i+1}}(2Jt)\right],
\label{H(t_1)}
\end{equation}
where $R_{X_i}(\theta) = e^{-i\theta X_i/2}$ denotes single-qubit rotation gates and $R_{ZZ_{i,i+1}}(\phi) = e^{-i\phi Z_iZ_{i+1}/2}$ denotes two-qubit entangling gates.
Substituting $J=-1$, $h=-1$, and $t=0.5$ yields rotation angles $2ht = 2Jt = -1$.
}

{The $R_{ZZ}$ gate is implemented via the standard gate decomposition
\begin{equation}
\begin{aligned}
    R_{ZZ_{i,i+1}}(\phi) &= e^{-i\phi Z_iZ_{i+1}/2} \\
    &= \mathrm{CNOT}_{i,i+1} \cdot R_{Z_{i+1}}(\phi) \cdot \mathrm{CNOT}_{i,i+1},
\end{aligned}
\end{equation}
as shown in Fig.~\ref{HD_circuit}(b).
The full evolution circuit, illustrated in Fig.~\ref{HD_circuit}(a), thus consists of (i) a layer of $N$ $R_X$ rotations, followed by (ii) $N-1$ $R_{ZZ}$ entangling gates.
}

{\noindent\textbf{Experiment 5: Hamiltonian dynamical simulation on a quantum processor.}
}

{In this experiment, we again select the last four qubits listed in Tab.~\ref{tab:qubit_metrics} as our target system. We extend the evolution channel $\mathcal{E}_H(t_1)$ (Eq.~\ref{H(t_1)}, $t_1 = 0.5$, $r = 1$) to $t_2 = 1$ with $r = 2$ Trotter steps~\cite{PhysRevLett.128.210501}, yielding the composite channel
\begin{equation}
    \mathcal{E}_H(t_2) = \left[\prod_{i=1}^N R_{X_i}(2ht_2)\right] \left[\prod_{i=1}^{N-1} R_{ZZ_{i,i+1}}(2Jt_2)\right] \left[\prod_{i=1}^N R_{X_i}(2ht_2)\right] \left[\prod_{i=1}^{N-1} R_{ZZ_{i,i+1}}(2Jt_2)\right]
    = \mathcal{E}_H(t_1) \circ \mathcal{E}_H(t_1),
\end{equation}
with $\mathcal{E}_H(t_3)~(t_3 = 1.5, r = 3)$ and $\mathcal{E}_H(t_4)~(t_4 = 2, r = 4)$ defined analogously.
}

{We establish a control group that employs the method in Ref.~\cite{huang2023learning} to predict properties of the time-evolved state at each time point $t_i$ directly from classical shadows of $\mathcal{E}_H(t_i)$. In contrast, our experimental group applies shadow engineering using only classical shadows of the single-step Hamiltonian channel $\mathcal{E}_H(t_1)$ (at time $t_1$). This approach enables us not only to predict dynamics at $t_1$ (matching the control group at this time step) but also to extrapolate and predict expectation values at subsequent time points $t_i$ ($i = 2, 3, 4$) without additional measurements. We vary the number of classical shadows in the experimental group to evaluate performance across different sample sizes.
}

{For the experimental group, following the circuit layout shown in Fig.~\ref{HD_circuit}, we randomly select 500 initial stabilizer states $S^{\otimes n}_i \in S^{\otimes n}$ and 500 random Pauli measurement bases. These are combined with $\mathcal{E}_H(t_1)$ to construct sampling circuits, each executed with 1000 shots on the quantum processor, resulting in a total of $5 \times 10^5$ classical shadows. In post-processing, we consider subsample sizes of $2 \times 10^5$ and $4 \times 10^5$. For each subsample, we first reconstruct the transfer matrix $\hat{T}_{\mathcal{E}_H(t_1)}$, then use Eq.~\ref{leq31} to derive the transfer matrix $\hat{T}_{\mathcal{E}_H(t_2)}$ for the composite channel $\mathcal{E}_H(t_2) = \mathcal{E}_H(t_1) \circ \mathcal{E}_H(t_1)$. We similarly estimate $\hat{T}_{\mathcal{E}_H(t_3)}$ by composing $\hat{T}_{\mathcal{E}_H(t_2)}$ and $\hat{T}_{\mathcal{E}_H(t_1)}$ via Eq.~\ref{leq31}, and extend this procedure to $\hat{T}_{\mathcal{E}_H(t_4)} = \hat{T}_{\mathcal{E}_H(t_2)} \cdot \mathcal{T} \cdot\hat{T}_{\mathcal{E}_H(t_2)}$. In this manner, shadow engineering yields estimates of the transfer matrices for all $\mathcal{E}_H(t_i)$ ($i = 1, 2, 3, 4$) purely from classical shadows of $\mathcal{E}_H(t_1)$. Using the observable $O = \sum_i Z_i$ and a 4-qubit Haar-random state $\rho$, we then estimate the expectation value $\operatorname{Tr}[O \mathcal{E}_H(t_i)(\rho)]$ at each time step via Alg.~\ref{Alg1}.
}

{For the control group, we follow the identical classical shadow acquisition protocol to obtain $5 \times 10^5$ classical shadows for each of the four time-evolution channels $\mathcal{E}_H(t_i)$ ($i = 1, 2, 3, 4$). Using a fixed subsample of $4 \times 10^5$ shadows, we apply the method from Ref.~\cite{huang2023learning} to estimate $\operatorname{Tr}[O \mathcal{E}_H(t_i)(\rho)]$. Both groups are run with 20 independent repetitions, and results are presented as boxplots. Notably, at $t_1$, the two methods are mathematically equivalent; both utilize shadows of $\mathcal{E}_H(t_1)$ and yield statistically consistent results, so only one box plot is shown for this time step.
}

{
Experimental results demonstrate that shadow engineering only requires initial-time \(t_1\) classical shadows to estimate all subsequent time points, whereas the reference method~\cite{huang2023learning} collects dedicated data at every \(t_K\). This reduces measurement overhead for long-time dynamics while preserving accuracy, enabling efficient characterization of many-body properties. Under identical sample budgets, our approach delivers more stable predictions throughout Hamiltonian evolution than the scheme in Ref.~\cite{huang2023learning}, and larger shadow datasets consistently further improve precision across all tested configurations.
}

{\noindent\textbf{Experiment 6: Prediction errors for various dynamical properties in Hamiltonian dynamical simulation.}
}

{In this experiment, we adopt the same system configuration as in Experiment 5, where shadow engineering leverages the classical shadow of $\mathcal{E}_H(t_1)$ to predict dynamical properties at evolution time $t_2$.
We investigate prediction errors across observables with varying locality.
Specifically, we employ the 1-local observable $O_1 = \sum_i Z_i$, the 2-local observable $O_2 = \sum_i Z_i Z_{i+1}$, and the 3-local observable $O_3 = \sum_i Z_i Z_{i+1} Z_{i+2}$ to estimate the total magnetization, nearest-neighbor correlations, and multi-body correlations of the Hamiltonian dynamics at time $t_2$, respectively.
}

{The experimental protocol is structured as follows.
Following the circuit architecture shown in Fig.~\ref{HD_circuit}, we randomly sample 300 initial stabilizer states $\mathcal{S}^{\otimes n}_i \in \mathcal{S}^{\otimes n}$ and 300 random Pauli measurement bases.
These are combined with the Hamiltonian channel $\mathcal{E}_H$ to construct the sampling circuits for classical shadow acquisition.
Each circuit is executed with 1000 shots on the superconducting quantum processor, yielding $3 \times 10^5$ classical shadows of $\mathcal{E}_H$ in total.
}

{In post-processing, we analyze subsampled datasets with sizes $[2, 5, \dots, 19] \times 10^4$.
For each subsample size:
(1) we first reconstruct the transfer matrix $\hat{T}_{\mathcal{E}_H(t_1)}$ of the channel $\mathcal{E}_H(t_1)$;
(2) we then compute the transfer matrix $\hat{T}_{\mathcal{E}_H(t_1) \circ \mathcal{E}_H(t_1)}$ for the concatenated channel $\mathcal{E}_H(t_1) \circ \mathcal{E}_H(t_1)=\mathcal{E}_H(t_2)$ via Eq.~\eqref{leq31}.
Using these transfer matrices $\hat{T}_{\mathcal{E}_H(t_1) \circ \mathcal{E}_H(t_1)}$ with different estimation precisions and a 4-qubit Haar-random state $\rho$ as the initial state, we predict the Hamiltonian evolution properties $\operatorname{Tr}[O_i \mathcal{E}_H(t_2)(\rho)]$ at time $t_2$ via Alg.~\ref{Alg1} for the three observables of different locality.
To ensure statistical robustness, we perform 20 independent repetitions of the subsampling and estimation procedure for each configuration, and compute the RMSE between the theoretically expected values (from noiseless simulations) and the experimentally estimated values.
This enables a direct comparison of prediction performance across observables with different locality.
}

{Experimental results demonstrate that, for concatenated channel characterization at a fixed sample size, shadow engineering produces lower prediction errors for observables with smaller locality (i.e., 1-local $<$ 2-local $<$ 3-local).
This trend is consistent with the expected sample complexity scaling, where lower-locality observables require fewer samples to achieve accurate estimation due to weaker entanglement and fewer multi-body correlations.
}
\bibliographystyle{apsrev4-1}
\bibliography{b1}